\documentclass[12pt]{revtex4}
\usepackage{graphicx}
\usepackage{amsmath}
\usepackage{amsfonts}
\usepackage{wasysym}
\usepackage{float}
\usepackage{color}
\usepackage{mathrsfs}
\begin{document}
\title{Integrable turbulence and formation of rogue waves}
\author{D.S. Agafontsev$^{(a)}$, V.E. Zakharov$^{(a),(b),(c),(d)}$}

\affiliation{\small \textit{ $^{(a)}$ P.P. Shirshov Institute of Oceanology, 36 Nakhimovsky prosp., Moscow 117218, Russia.\\
$^{(b)}$ Department of Mathematics, University of Arizona, Tucson, AZ, 857201, USA.\\
$^{(c)}$ P.N. Lebedev Physical Institute, 53 Leninsky ave., 119991 Moscow, Russia.\\
$^{(d)}$ Novosibirsk State University, 2 Pirogova, 630090 Novosibirsk, Russia.}}

\begin{abstract}
In the framework of the focusing Nonlinear Schr{\"o}dinger (NLS) equation we study numerically the nonlinear stage of the modulation instability (MI) of the condensate. The development of the MI leads to formation of ``integrable turbulence'' [Zakharov V.\,E., Stud. Appl. Math. 122, 219-234 (2009)]. We study the time evolution of it's major characteristics averaged across realizations of initial data -- the condensate solution seeded by small random noise with fixed statistical properties.

We observe that the system asymptotically approaches to the stationary integrable turbulence, however this is a long process. During this process momenta, as well as kinetic and potential energies, oscillate around their asymptotic values. The amplitudes of these oscillations decay with time $t$ as $t^{-3/2}$, the phases contain the nonlinear phase shift that decays as $t^{-1/2}$, and the frequency of the oscillations is equal to the double maximum growth rate of the MI. The evolution of wave-action spectrum is also oscillatory, and characterized by formation of power-law region $\sim\,|k|^{-\alpha}$ in the small vicinity of the zeroth harmonic $k=0$ with exponent $\alpha$ close to 2/3. The corresponding modes form ``quasi-condensate'', that acquire very significant wave action and macroscopic potential energy.

The probability density function (PDF) of wave amplitudes asymptotically approaches to Rayleigh distribution in oscillatory way. Nevertheless, in the beginning of the nonlinear stage the MI slightly increases the occurrence of rogue waves. This takes place at the moments of potential energy modulus minima, where the PDF acquires ``fat tales'' and the probability of rogue waves occurrence is by about two times larger than in the asymptotic stationary state.

Presented facts need theoretical explanation.
\end{abstract}

\maketitle


\section{Introduction.}

Today the total amount of experimental evidence of rogue waves emergence on the surface of fluid and in optical fibers is huge \cite{kharif2003physical, dysthe2008oceanic, onorato2013rogue, solli2007optical, mussot2009observation}. Thus, the development of a consistent theory of these events is urgently needed. The simplest nonlinear mathematical model for the description of rogue waves phenomenon is the modulation instability (MI) developing from the condensate solution in the framework of the focusing one-dimensional Nonlinear Schrodinger (NLS) equation \cite{kharif2003physical, dysthe2008oceanic, onorato2013rogue}. Without loss of generality we will use the NLS equation in following form:
\begin{equation}
i\Psi_t -\Psi +\Psi_{xx}+|\Psi|^2 \Psi = 0. \label{nlse_condensate}
\end{equation}
The simplest "condensate" solution of this equation $\Psi=1$ is unstable. If we consider modulations to the condensate as
\begin{equation}\label{MI_condensate}
\Psi = 1 + \kappa\exp(ikx+i\Omega t),\quad |\kappa|\ll 1,
\end{equation}
and linearize Eq. (\ref{nlse_condensate}) against the condensate, we obtain
\begin{equation}\label{MI_condensate2}
\Omega^{2}=k^{4}-2k^{2}.
\end{equation}
The modulations with $k\in (-\sqrt{2}, \sqrt{2})$ turn out to be unstable, and the maximum growth rate of the instability,
\begin{equation}\label{max_growth_rate}
\gamma_{0}=\max_{k}\mathrm{Im}\,\Omega=1,
\end{equation}
is realized at $k=\pm 1$. Thus, the characteristic length of the instability is $\ell=2\pi$, and the characteristic time is $1/\gamma_{0}=1$.

To study the nonlinear stage of the MI, one has to solve Eq. (\ref{nlse_condensate}) with the initial data in the form 
\begin{equation}
\Psi|_{t=0} = 1+\epsilon(x),\quad |\epsilon(x)|\ll 1. \label{ensemble}
\end{equation}
It should be noted that the problem of the MI development on the background of any condensate solution $\Psi|_{t=0} = C+\epsilon(x)$, $|\epsilon(x)|\ll |C|$, and for the focusing NLS equation
$$
i\Psi_t + B\,\Psi_{xx} + G\,|\Psi|^2 \Psi = 0
$$
with arbitrary dispersion $B>0$ and nonlinearity $G>0$ coefficients renormalizes to Eqs.~(\ref{nlse_condensate}), (\ref{ensemble}), as can be seen after the scaling and gauge transformations $x=\tilde{x}\sqrt{B/(G|C|^{2})}$, $t=\tilde{t}/(G|C|^{2})$, $\Psi=\tilde{\Psi}\,C e^{i\tilde{t}}$ and $\epsilon=\tilde{\epsilon}\,C e^{i\tilde{t}}$.

If $|\epsilon(x)|\to 0$ at $|x|\to +\infty$, then the problem can be solved analytically with the help of the inverse scattering transformation \cite{shabat1972exact, zakharov1984theory, zakharov2011soliton, zakharov2013nonlinear}. The MI in this case leads to formation of different types of solitonic solutions. The scenario of the MI essentially depends on the fine details of the initial perturbation $\epsilon(x)$. The pure real perturbation leads to formation of homoclinic solutions of Peregrine type \cite{peregrine1983water}, while the pure imaginary one generates "superregular" solitonic solutions described in the publications of V.E. Zakharov and A.A. Gelash \cite{zakharov2011soliton, zakharov2013nonlinear}. The general case when both imaginary and real parts of the perturbation are present is not properly studied yet.

In spite of the apparent significance of these results, they do not answer to the main question -- what happens if the perturbation $\epsilon(x)$ is not localized? To study it, we solve the NLS equation (\ref{nlse_condensate}) numerically in the box $x\in[-L/2, L/2]$ with periodic boundary. Theoretically speaking, this problem can also be solved analytically. Any periodic solution of the NLS equation can be expressed explicitly in terms of Jacobi theta-functions over a certain hyperbolic curve \cite{zakharov1984theory}. However, this beautiful mathematical result can hardly be used for practical purposes. In our numerical experiments $\epsilon(x)$ is a small random noise, and we use the number of harmonics of order $10^{5}$. Then, to model it's evolution in terms of Jacobi functions, we have to make the genus of the curve of order $10^{5}$. It is unrealistic so far to follow this evolution by the use of the exact analytical methods. 

Therefore, we rely completely on numerical experiments. We use integrability of the NLS equation only in the weakest sense. Integrability implies conservation of infinite number of integrals of motion. The first three of these invariants are wave action, 
\begin{equation}\label{wave_action}
N = \frac{1}{L}\int_{-L/2}^{L/2}|\Psi(x,t)|^{2}\,dx,
\end{equation}
momentum,
\begin{equation}\label{momentum}
P=\frac{i}{2L}\int_{-L/2}^{L/2}(\Psi_{x}^{*}\Psi-\Psi_{x}\Psi^{*})\,dx,
\end{equation}
and total energy,
\begin{equation}\label{energy}
E=H_{2}+H_{4},\quad\quad H_{d}=\frac{1}{L}\int_{-L/2}^{L/2} |\Psi_{x}|^2\,dx,\quad\quad H_{4} = -\frac{1}{2L}\int_{-L/2}^{L/2} |\Psi|^4\,dx.
\end{equation}
Here $H_{d}$ is kinetic and $H_{4}$ is potential energy. We define these integrals with the prefactor $1/L$ for further convenience. We use method of numerical simulations that conserves very well the first 12 invariants. 

Our study has two main goals. First, we expect that after a very long evolution the result of the MI of the condensate should be the stationary ``integrable turbulence'' \cite{zakharov2009turbulence} -- thermodynamically equilibrium state defined by infinite number of invariants. In our experiments we indeed observe that the system asymptotically approaches towards it's stationary turbulent state. The investigation of this state has fundamental importance. Note that the similar research has recently been made for the focusing NLS equation, but with incoherent wave field initial conditions \cite{walczak2015optical} (see also \cite{onorato2001freak, janssen2003nonlinear, onorato2004observation, onorato2005modulational, onorato2006extreme} for the dependence of the final state on the Benjamin-Feir index and \cite{randoux2014intermittency} for the defocusing NLS equation). Since we study integrable system which ``remembers'' it's initial state through infinite number of integrals of motion, it is 
not surprising that our asymptotic turbulent state differs from that of  \cite{walczak2015optical}. Second, we examine the beginning of the nonlinear stage of the MI and the subsequent evolution towards the asymptotic turbulent state in order to understand the characteristic features of rogue waves emergence in the framework of the focusing NLS equation. In this sense our study is in line with the intensive modern research on the statistics of waves in different nonlinear systems, and many of these systems fall under the category of generalized 
one-dimensional NLS equation \cite{solli2007optical, mussot2009observation, dudley2008harnessing, kibler2009soliton, montina2009non, 
genty2010collisions, taki2010third, shemer2010applicability, chung2011strong, slunyaev2012stochastic}.

Since we perform our simulations in the finite box $L< +\infty$, after a very long time we will encounter with Fermi - Pasta - Ulam (FPU) recurrence phenomenon \cite{fermi1955studies, infeld1981quantitive}. This means that at some point of time the evolution towards the stationary integrable turbulence will stop, and the system will move back towards the condensate state. Thus, the system of finite size $L$ does not have the asymptotic stationary turbulent state in it's true sense. However, it has quasi-asymptotic state in which it spends most of the time between the development of the MI and the FPU recurrence. Since the time of the FPU recurrence tends to infinity with the system size $L$, the quasi-asymptotic state approaches to the stationary integrable turbulence as $L\to +\infty$. 

We stop our simulations before we observe tendency towards the FPU recurrence. Technically, we perform convergence study comparing our results obtained in the computational boxes $L$ and $2L$. As soon as we observe deviations between these results, we stop simulations in the smaller box $L$ and switch to simulations in the box $2L$, comparing the results with that from the box $4L$. We repeat this procedure until our computational resources allow. In this sense our results can be considered as the subsequent approximations of the stationary integrable turbulence. Taking this into account we will continue to use term "asymptotic state" in it's initial meaning.

One of the important characteristics of the turbulence is wave-action spectrum
\begin{equation}\label{Ik}
I_{k}(t)=\langle|\Psi_{k}(t)|^{2}\rangle.
\end{equation}
Here and below $\langle ...\rangle$ stands for arithmetic average across ensemble of initial data and $\Psi_{k}(t)=\mathscr{F}[\Psi(x,t)]$ is Fourier transform of $\Psi(x,t)$. We define forward $\mathscr{F}$ and backward $\mathscr{F}^{-1}$ Fourier transformations as
\begin{eqnarray}
\Psi_{k}(t)&=&\mathscr{F}[\Psi(x,t)]=\frac{1}{L}\int_{-L/2}^{L/2}\Psi(x,t)e^{-ikx}\,dx, \label{Fourier1}\\
\Psi(x,t)&=&\mathscr{F}^{-1}[\Psi_{k}(t)]=\sum_{k}\Psi_{k}(t)e^{ikx},\label{Fourier2}
\end{eqnarray}
where $k=2\pi n/L$ is wavenumber and $n$ is integer. In our simulations we use $L=2\pi m$ where $m$ is integer, so that our spectral band contains exact wavenumbers $k=\pm 1$ where the maximum growth rate of the MI is achieved. Wave-action spectrum is the spectral density of wave action, since
\begin{equation}\label{Ndensity}
\langle N\rangle = \sum_{k}I_{k}(t).
\end{equation}
Thus, the right-hand side of Eq. (\ref{Ndensity}) is conserved by the motion. According to (\ref{Fourier1}), all wave-action of the condensate $\Psi=1$ is concentrated in the zeroth harmonic $k=0$,
\begin{equation}\label{spectra_condensate}
I_{k} =\left\{ \begin{array}{ll} 
1, & k=0, \\ 
0, & k\neq 0.
\end{array}\right.
\end{equation}

During the development of the MI we observe that wave action disperse across other harmonics. This happens in the form of the oscillatory exchange of wave action between the zeroth harmonic $I_{0}(t)$ from one hand, and the rest of the spectrum from the other. In the result of this process wave-action spectrum approaches towards the asymptotic spectrum. In the beginning of the MI the spectrum has discontinuity at $k=0$ in the form of a high peak occupying the zeroth harmonic only. This peak appears from the initial data (\ref{ensemble}). The remarkable result of our experiments is that the peak does not disappear with the arrival to the nonlinear stage of the MI, but instead decays in oscillatory way and remains detectable for a long time after the beginning of the nonlinear stage. After the peak finally disappears, the singularity in the spectrum at $k=0$ transforms to power-law behavior $\sim\,|k|^{-\alpha}$ at $|k|\le 0.15$ with exponent $\alpha$ close to 2/3. The corresponding modes have very large 
scales 
in the physical space and can be called "quasi-condensate", that in the asymptotic turbulent state has about 40\% of wave 
action, less than 1\% of kinetic energy and about 10\% of potential energy. The asymptotic spectrum decays monotonically as $|k|\to +\infty$; this decay is slower at $0.4\lesssim |k|\lesssim 1$, very fast near $|k|=0$ and $|k|=\sqrt{2}$, and close to exponential from $|k|>1.5$.

Another important characteristic of the turbulence is the (simultaneous) spatial correlation function,
\begin{equation}\label{gx}
g(x,t) = \bigg\langle\frac{1}{L}\int_{-L/2}^{L/2}\Psi(y,t)\Psi^{*}(y-x,t)\,dy\bigg\rangle.
\end{equation}
It is connected with wave-action spectrum by the relation
$$
g(x,t) = \mathscr{F}^{-1}[I_{k}(t)],
$$
that follows from Eqs. (\ref{Fourier1})-(\ref{Fourier2}). Spatial correlation function is fixed to unity $g(0,t)\approx 1$ at $x=0$, since
\begin{equation}\label{corr_x_0}
g(0,t) = \langle N\rangle,
\end{equation}
and for the ensemble of initial data (\ref{ensemble}) wave action $N$ almost coincides with unity (for our experiments $[\langle N\rangle-1]\sim 10^{-9}$).

In the nonlinear stage of the MI we observe that spatial correlation function also evolves in oscillatory way approaching towards the asymptotic correlation function. While the peak at $k=0$ in wave-action spectrum is present, $g(x,t)$ decays with $|x|\to +\infty$ to some nonzero level that is determined by the magnitude of the peak. When the peak disappears, the correlation function decays to zero as $1/|x|$. At lengths $|x|<x_{corr}/2$ the asymptotic spatial correlation is close to Gaussian. Here $x_{corr}\approx 4$ is it's full width at half maximum.

\begin{figure}[t]\centering
\includegraphics[width=8.0cm]{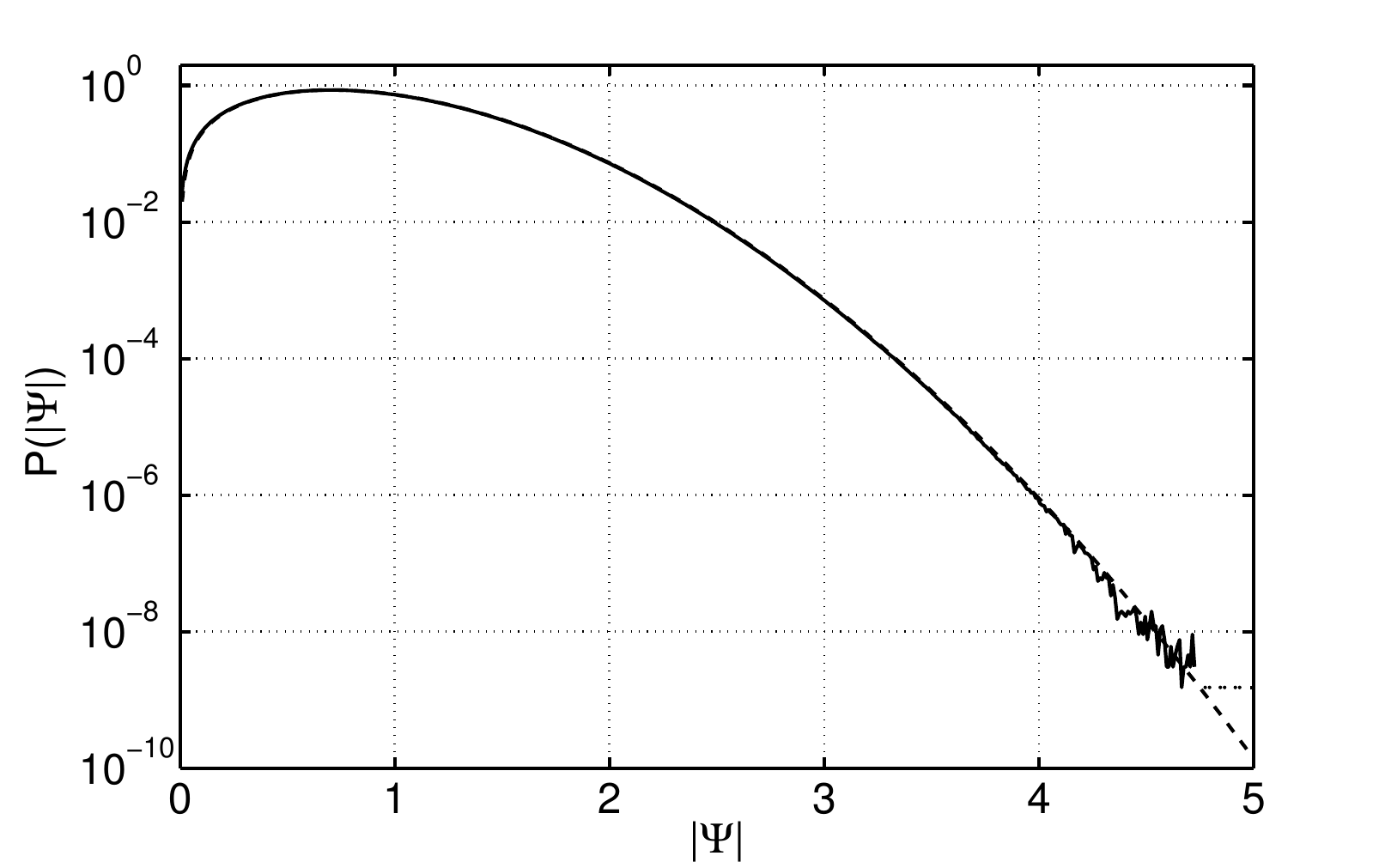}

\caption{\small {\it The PDF $P(|\Psi|)$ (solid line) for linear waves $\Psi(x) = (\sqrt{8\pi}/\theta L)^{1/2}\,\mathscr{F}^{-1}[A_{0} e^{-k^{2}/\theta^{2} + i\xi_{k}}]$, $\theta=5$, $A_{0}=1$, calculated in the periodic box $x\in[-L/2, L/2]$, $L=256\pi$, using $10^{6}$ different realizations of random uncorrelated phases $\xi_{k}$. The average squared amplitude for such linear waves is $\overline{|\Psi|^{2}}\approx A_{0}^{2}=1$ (see Eqs. (\ref{noise})-(\ref{noise_amplitude})). Dashed line is Rayleigh distribution (\ref{Rayleigh}) with $\sigma=1$.}}
\label{fig:linearPDF}
\end{figure}

We also measure the probability density function (PDF) of wave amplitudes $P(|\Psi|,t)$. Let us suppose that the current state of a system consists of multitude of uncorrelated linear waves,
\begin{equation}\label{DFT}
\Psi(x) = \sum_{k}|\Psi_{k}|\, e^{i(kx+\phi_{k})}.
\end{equation}
If phases $\phi_{k}$ are random and uncorrelated, the number of waves $\{k\}$ is large enough, and amplitudes $|\Psi_{k}|$ fall under the conditions of central limit theorem, then real $\textrm{Re}\,\Psi(x)$ and imaginary $\textrm{Im}\,\Psi(x)$ parts of field $\Psi(x)$ are Gaussian-distributed, and the PDF of wave amplitudes coincides with Rayleigh distribution \cite{nazarenko2011wave} (see example on FIG.~\ref{fig:linearPDF}),
\begin{equation}\label{Rayleigh}
P_{R}(|\Psi|) = \frac{2|\Psi|}{\sigma^{2}}e^{-|\Psi|^{2}/\sigma^{2}}.
\end{equation}
For a system that conserves wave action and has Rayleigh PDF, the parameter $\sigma$ can be readily calculated as
$$
\langle N\rangle = \langle |\Psi|^{2}\rangle = \int_{0}^{+\infty}|\Psi|^{2}\,P_{R}(|\Psi|)\,d|\Psi| = \sigma^{2}.
$$
Here $\langle |\Psi|^{2}\rangle$ is ensemble and space average of squared amplitude. For the ensemble of initial data (\ref{ensemble}) this leads to conclusion $\sigma\approx 1$.

Since $\int F(x)x\,dx = (1/2)\int F(x)\,d\,x^{2}$, the PDF $P(|\Psi|^{2},t)$ of squared amplitudes is exponential if the corresponding amplitude PDF $P(|\Psi|,t)$ is Rayleigh one, and vice versa; for $\sigma=1$ this leads to
\begin{equation}\label{Rayleigh_3}
P_{R}(|\Psi|^{2}) = e^{-|\Psi|^{2}}.
\end{equation}
It is more convenient to examine exponential dependencies than Rayleigh ones, and thereby we measure the PDF of squared amplitudes and compare the results with exponential dependency (\ref{Rayleigh_3}) that we call Rayleigh one for simplicity. Throughout the publication we use term "PDF" only in relation to PDF of wave amplitudes. We measure the PDF $P(|\Psi|^{2},t)$ for entire field $\Psi(x,t)$, in contrast to PDFs for local maximums or absolute maximums, and use normalization,
$$
\int_{0}^{+\infty} P(|\Psi|^{2})\,d|\Psi|^{2}=1.
$$
The knowledge of the PDF gives us the probability of occurrence $W(Y,t)$ of waves exceeding certain threshold $|\Psi|^{2}>Y$, 
\begin{equation}\label{WY}
W(Y,t) = \int_{Y}^{+\infty}P(|\Psi|^{2},t)\,d|\Psi|^{2}.
\end{equation}
In case of Rayleigh PDF (\ref{Rayleigh_3}) this probability takes the simple form 
\begin{equation}\label{WYR}
W_{R}(Y,t) = e^{-Y}.
\end{equation}

In addition to the PDF we measure ensemble average kinetic $\langle H_{d}\rangle$ and potential $\langle H_{4}\rangle$ energies, and also the moments
\begin{eqnarray}\label{Mn}
M^{(n)}(t)=\bigg\langle\frac{1}{L}\int_{-L/2}^{+L/2} |\Psi(x,t)|^{n}\,dx\bigg\rangle.
\end{eqnarray}
The moments are connected to the PDF as
\begin{equation}\label{MnPDF}
M^{(n)}(t)=\int_{0}^{+\infty} |\Psi|^{n} P(|\Psi|,t)\,d|\Psi|.
\end{equation}
Thus, for a system with Rayleigh PDF (\ref{Rayleigh_3}) the moments can be easily calculated,
\begin{equation}\label{MnR1}
M_{R}^{(n)}=\Gamma\bigg(\frac{n}{2}+1\bigg),
\end{equation}
where $\Gamma(m)$ is gamma-function. The moment $M^{(2)}(t)$ does not change with time since $M^{(2)}(t)=\langle N\rangle\approx 1$. 

In the nonlinear stage of the MI we observe that kinetic $\langle H_{d}\rangle$ and potential $\langle H_{4}\rangle$ energies, as well as the moments $M^{(n)}(t)$, $n\neq 2$, oscillate with time around their asymptotic values. The amplitudes of these oscillations decay with time as $t^{-3/2}$, the phases contain the nonlinear phase shift that decays as $t^{-1/2}$, and the period of the oscillations is equal to $\pi$. Thus, the frequency of the oscillations is equal to the double maximum growth rate of the MI. The asymptotic values of kinetic and potential energies are 0.5 and -1 respectively, while the asymptotic moments coincide with Rayleigh predictions (\ref{MnR1}).

The PDF in the asymptotic turbulent state coincides with Rayleigh one (\ref{Rayleigh_3}). Thus, the probability of occurrence of waves turns out to be the same as in a wave field described by linear equations (\ref{WYR}). The level of nonlinearity of the turbulence can be estimated by the parameter
$$
Q = \frac{|\langle H_{4}\rangle|}{|\langle H_{d}\rangle|}.
$$
For weak turbulence $|Q|\ll 1$ the Rayleigh PDF would be a natural result. However, for the NLS equation we observe "moderately strong" turbulence with $Q=2$ in the asymptotic state.

Nevertheless, we confirm that in the beginning of the nonlinear stage the MI moderately increases the occurrence of rogue waves. According to the standard definition \cite{kharif2003physical, dysthe2008oceanic, onorato2013rogue}, a rogue wave is a wave that exceeds at least two times the significant wave height $h_{s}$. The significant wave height is calculated as the average wave height of the largest 1/3 waves. It is easy to calculate that for Rayleigh PDF (\ref{Rayleigh_3}) the significant wave amplitude is $h_{s}\approx 1.42$, and rogue waves must exceed $|\Psi|>2.8$ in amplitude or $|\Psi|^{2}>8$ in squared amplitude. 

In our experiments we observe that in the beginning of the nonlinear stage of the MI the PDF evolves significantly with the oscillations of kinetic and potential energies. At the points of time corresponding to local maximums and minimums of potential energy modulus $|\langle H_{4}\rangle|$, the PDF acquires "fat tails" and significantly exceeds Rayleigh PDF (\ref{Rayleigh_3}) in the two regions of squared amplitudes $3\lesssim |\Psi|^{2}\lesssim 7$ and $10\lesssim |\Psi|^{2}\lesssim 15$ respectively. It is interesting that the evolution of the PDF goes in such a way that in the beginning of the nonlinear stage the "standard" rogue waves $|\Psi|^{2}>8$ appear even less frequently than predicted by Rayleigh PDF (\ref{WYR}). 

The waves from the first region $3\lesssim |\Psi|^{2}\lesssim 7$ are "imperfect" rogue waves, since they do not match the criterion for the "standard" rogue waves $|\Psi|^{2}>8$. The "imperfect" rogue waves are the typical outcome of the MI, and can be seen at the first several local maximums of $|\langle H_{4}\rangle|$. In space these waves form a modulated lattice of large waves with distance between them close to the characteristic length $\ell=2\pi$ of the MI. In the beginning of the nonlinear stage the probability of occurrence of such waves with $|\Psi|^{2}>4$ is by about three times larger than Rayleigh one (\ref{WYR}). The crests of the "imperfect" rogue waves are mostly composed of the imaginary part of wave field $\Psi(x)$, $|\mathrm{Re}\,\Psi|\ll|\mathrm{Im}\,\Psi|$. At the first, third, and so on, local maximums of $|\langle H_{4}\rangle|$ it is positive $\mathrm{Im}\,\Psi>0$, and at the second, fourth, and so on, local maximums -- negative $\mathrm{Im}\,\Psi<0$. 

The similar scenario is realized in case of the Akhmediev breather \cite{akhmediev1987exact, akhmediev2009extreme, akhmediev2009waves} that corresponds to the maximum growth rate of the MI. At the time of it's maximal elevation this solution is purely imaginary, and at it's maximums the imaginary part is positive $\mathrm{Im}\,\Psi>0$. After the decay this solution changes the phase of the condensate by $e^{i\pi}=-1$. Thus, the following Akhmediev breather -- if it appears -- should have negative imaginary part $\mathrm{Im}\,\Psi<0$ at it's crests, the third Akhmediev breather -- positive, and so on. 

However, it is unclear how these solutions may appear from random statistically homogeneous in space noise one after another with the short interval between them. This interval is equal to the period of the oscillations of the moments, as well as kinetic and potential energies, is close to 4 in the beginning of the nonlinear stage of the MI and approaches to $\pi$ with time. Also, spatial correlation function of the "imperfect" rogue waves significantly decreases after a few characteristic lengths $\ell$ of the MI, and takes (locally in time) minimal values. For the Akhmediev breather it remains periodic. Wave-action spectrum of the "imperfect" rogue waves also is not very similar to that of the Akhmediev breather.

In the beginning of the nonlinear stage of the MI and at the local minimums of potential energy modulus $|\langle H_{4}\rangle|$ the waves from the second region $10\lesssim |\Psi|^{2}\lesssim 15$ appear by about two times more frequently for $|\Psi|^{2}>12$ than predicted by Rayleigh PDF (\ref{WYR}). These waves are very rare events, and represent in space a singular high peak with full width at half maximum of about $x_{FW}\sim 1$ and duration in time of about $\Delta T\sim 1$. These "large" rogue waves appear on the background of perturbed wave field that is usually less than $|\Psi|<1.5$ in amplitude. Statistically at this time wave field is significantly correlated, spatial correlation function takes (locally in time) maximal values, and wave-action spectrum has (locally in time) maximal zeroth harmonic with the rest of the spectrum minimally excited.

The crests of the "large" rogue waves are composed mostly of real part of wave field $\Psi(x)$, $|\mathrm{Im}\,\Psi|\ll|\mathrm{Re}\,\Psi|$. At the first, third, and so on, local minimums of $|\langle H_{4}\rangle|$ it is negative $\mathrm{Re}\,\Psi<0$, and at the second, fourth, and so on, local minimums -- positive $\mathrm{Re}\,\Psi>0$. It is interesting, that the Peregrine solution \cite{peregrine1983water} has similar property: at the time of it's maximal elevation this solution is purely real and negative at it's maximum amplitude. However, the Peregrine solution has slightly smaller maximal squared amplitude $\max|\Psi|^{2}=9$.

We also observe extremely large waves with amplitudes of up to  $|\Psi|\sim 6$. However, the accuracy of our simulations is insufficient to study the time evolution of the PDF and the probability of occurrence of such waves.

The paper is organized as follows. In the next Section we describe the numerical methods that we used in the framework of the current study. Section 3 is devoted to the investigation of the asymptotic state of the integrable turbulence, while in Section 4 we describe how the temporal evolution towards this state is arranged. Section 5 contains conclusions and acknowledgements. In the Appendix A we provide detailed graphs for wave-action spectrum and spatial correlation function in the beginning of the nonlinear stage of the MI. Detailed graphs for the evolution of the probability of large waves occurrence are given in the Appendix B.


\section{Numerical methods.}

We integrate Eq. (\ref{nlse_condensate}) numerically on the time interval $t\in[0, 1000]$ in the box $x\in[-L/2, L/2]$, $L=1024\pi$, with periodic boundary. We use $L=2\pi m$, where $m$ is integer, in order to have in our spectral band exact wavenumbers $k=\pm 1$ where the maximum growth rate of the MI is achieved. We continue integration for so long time $t\le 1000$ in order to approach to the asymptotic turbulent state as close as our computational resources allow. We do not integrate beyond $t=1000$ since starting from $t\sim 1200$ we observe tendency towards the FPU recurrence: kinetic and potential energies, as well as the moments $M^{(n)}(t)$, significantly deviate from their asymptotics, as well as from the results obtained on larger computational box $1.5L$ (for that large $L$ we cannot allow ourselves comparison with the box $2L$). We also performed simulations on smaller boxes and found the same phenomenon starting from $t\sim 600$ for $L=512\pi$, $t\sim 300$ for $L=256\pi$, and so on. 

We would like to stress, that we have a very good quantitative agreement of our results obtained on different computational boxes $L$ before the tendency towards the FPU phenomenon appears. Thus, results for $L=256\pi$ coincide with that obtained on larger boxes up to $t\sim 300$, for $L=512\pi$ -- up to $t\sim 600$ and for $L=1024\pi$ -- up to $t\sim 1200$. Note that in order to compare wave-action spectrum obtained on different boxes $L$, we need to use it in the form $I_{k}/\Delta k$, since for finite $L$ wavenumber $k$ actually models area $[k-\Delta k/2, k+\Delta k/2]$ in the spectrum. Here $\Delta k=2\pi/L$ is the distance between the subsequent wavenumbers. Below we will continue to use the spectrum in the definition (\ref{Ik}), where the spectrum of the condensate takes the convenient form (\ref{spectra_condensate}).

We use Runge-Kutta 4th-order method, and calculate spatial derivatives and wave-action spectrum with the help of Fast Fourier transformations (FFT) routines. We perform our simulations on uniform grid with the spatial grid size $\Delta x = L/M$, where $M$ is the number of nodes on the grid. Thus, all summation over wavenumbers $k=2\pi n/L$ in (\ref{Fourier2}), (\ref{Ndensity}), and so on, where $n$ is integer, is performed in the spectral band $k\in[-\pi/\Delta x, \pi/\Delta x]$.

We change the spatial grid size $\Delta x$ adaptively after the analysis of Fourier components of the solution $\Psi_{k}$: we reduce $\Delta x$ when $\Psi_{k}$ at large wave numbers $k$ exceed $10^{-13}\max|\Psi_{k}|$ and increase $\Delta x$ when this criterion allows. The distance between the subsequent wavenumbers $2\pi/L=2^{-9}\approx 0.002$ is fixed by the length of the computational box $L=1024\pi$, and the range of wavenumbers $k\in[-\pi/\Delta x, \pi/\Delta x]$ is determined by $\Delta x$. Thereby, the modification of $\Delta x$ only adds or removes harmonics with large wavenumbers. This allows us to perform interpolation from one uniform grid to another by simply transferring the shared part of the spectrum to the new grid. We checked that the error of such interpolation is comparable with the round-off errors. In order to prevent appearance of numerical instabilities, time step $\Delta t$ also changes with $\Delta x$ as $\Delta t = h\Delta x^{2}$, $h \le 0.1$. 

We start our simulations on the grid with $M=65536$ nodes. In order to calculate the ensemble average characteristics, we interpolate the solution $\Psi(x,t)$ from the current grid (determined by the spectrum $\Psi_{k}$ at the current time $t$) to fixed grid with $M=131072$ nodes. We checked that such grids are sufficient for our computational box, comparing our results with that obtained on larger grids. We start from the initial data (\ref{ensemble}) and use statistically homogeneous in space initial noise that can be written symbolically as
\begin{equation}\label{noise}
\epsilon(x)=A_{0}\bigg(\frac{\sqrt{8\pi}}{\theta L}\bigg)^{1/2} \mathscr{F}^{-1}\bigg[e^{-k^{2}/\theta^{2}+i\xi_{k}}\bigg].
\end{equation}
Here $A_{0}$ is noise amplitude, $\theta$ is noise width in k-space and $\xi_{k}$ are arbitrary phases for each $k$ and each noise realization within the ensemble of initial data. The average squared amplitude of noise in x-space can be calculated as
\begin{eqnarray}
&&\overline{|\epsilon|^{2}}=\frac{1}{L}\int_{-L/2}^{L/2} |\epsilon(x)|^{2}\, dx=\frac{\sqrt{8\pi}}{\theta L}\frac{A_{0}^{2}}{L}\sum_{k_{1},k_{2}}e^{-(k_{1}^{2}+k_{2}^{2})/\theta^{2}+i(\xi_{k_{1}}-\xi_{k_{2}})}\int_{-L/2}^{+L/2} e^{i(k_{1}-k_{2})x}\,dx = \nonumber\\
&&= \frac{\sqrt{8\pi}}{\theta L}A_{0}^{2}\sum_{k}e^{-2k^{2}/\theta^{2}}\approx\frac{\sqrt{8\pi}}{\theta L}A_{0}^{2}\bigg(\frac{2\pi}{L}\bigg)^{-1}\int_{-\infty}^{+\infty}e^{-2k^{2}/\theta^{2}}\,dk = A_{0}^{2}.\label{noise_amplitude}
\end{eqnarray}

We performed several experiments for several ensembles of initial data that differed from each other by noise parameters $A_{0}$ and $\theta$. We didn't find significant dependence of our results on noise amplitude $A_{0}$, except that the system arrives to the nonlinear stage of the MI faster for larger $A_{0}$. We also changed noise width in k-space $\theta$ in the broad range from $\theta=5$ to $\theta=1$, and obtained the same results for these experiments. In the next two Sections we present the results obtained for the most broad in k-space noise distribution $A_{0}=10^{-5}$, $\theta=5$, that we studied. Inside the range of the instability $k\in(-\sqrt{2}, \sqrt{2})$ such noise can be treated as a white noise.

We also tested the following initial noise distribution, 
\begin{eqnarray}
\epsilon_{2}(x)=A_{0}\bigg(\frac{\sqrt{8\pi}}{\theta L}\bigg)^{1/2} \mathscr{F}^{-1}\bigg[10^{-v_{k}}\times e^{-k^{2}/\theta^{2}+i\xi_{k}}\bigg],\label{noise_k_space2}
\end{eqnarray}
where $v_{k}$ is uniformly distributed over $[0, 10]$ random value for each $k$, $A_{0}=10^{-5}$ and $\theta=5$. The multiplier $10^{-v_{k}}$ introduces the detuning between the amplitudes of noise in $k$-space by up to 10 orders of magnitude. However, we came to very similar results, though the oscillatory evolution of the system in the nonlinear stage of the MI, that we will demonstrate below, became slightly less regular. In our opinion this means that our results should be visible for a very wide variety of statistical distributions of noise.

The NLS equation (\ref{nlse_condensate}) has an infinite number of integrals of motion \cite{zakharov1984theory}. The first three of these integrals are wave action (\ref{wave_action}), momentum (\ref{momentum}) and total energy (\ref{energy}), the fourth one is 
$$
c_{4}[\Psi(x)]=\frac{1}{L}\int_{-\infty}^{+\infty} \bigg[\Psi\Psi_{xxx}^{*} + \frac{1}{2}\Psi\frac{d}{dx}(|\Psi|^{2}\Psi^{*}) + |\Psi|^{2}\Psi\Psi_{x}^{*}\bigg] \,dx,
$$
and so on. Our scheme provides very good conservation of the first 12 invariants with accuracy better than $10^{-6}$. We measure absolute errors for integrals $c_{n}[\Psi(x)]$ with even numbers $\mathrm{mod}(n,2)=0$, and relative errors for integrals $c_{n}[\Psi(x)]$ with odd numbers $\mathrm{mod}(n,2)=1$, since for our initial data integrals with even numbers are very close to zero. The first three invariants are conserved by our scheme with accuracy better than $10^{-10}$.

In our experiments we use ensembles of 1000 initial distributions each. We checked our statistical results against the size of the ensembles, the parameters of our numerical scheme and the implementation of other numerical methods (Runge-Kutta 5th-order, Split-Step 2nd- and 4th-order methods \cite{muslu2005higher, mclachlan1995numerical}), and found no difference. 

We also compared our results with that for the MI of the condensate in the framework of the Ablowitz--Ladik (AL) equation \cite{ablowitz1976nonlinear, agafontsev2014extreme},
\begin{equation}\label{AL}
i\frac{d\Psi_{n}}{dt} +\frac{\Psi_{n+1}-2\Psi_{n}+\Psi_{n-1}}{h^{2}} - \Psi_{n} + |\Psi_{n}|^{2}\frac{\Psi_{n+1}+\Psi_{n-1}}{2}=0.
\end{equation}
The AL system is defined on the grid $-M/2\le n \le M/2-1$ with periodic boundary, where $n$ is (integer) node number and $M$ is the total number of nodes, and is analogous to the NLS equation. As shown in \cite{agafontsev2014extreme}, the problem of the MI of the condensate for the AL system has one free parameter $h$ that has the meaning of the constant of coupling between the nodes. The NLS equation can be obtained after the substitution $x=nh$ and in the limit $h\to 0$. Therefore, for $h\ll 1$ the AL system (\ref{AL}) may be considered as the scheme of numerical integration of the NLS equation with fixed discretization along spatial dimension. Using this scheme together with Runge-Kutta 4th order method, we arrived to exactly the same results as in case of the described above numerical scheme for the integration of the NLS equation.

Comparison of our results with that for the AL system is also important from the point of view of integrability. Indeed, the AL system is also completely integrable with the help of the inverse scattering transformation. The schemes of numerical integration of the NLS equation break the integrability due to discretization along both spatial and temporal dimensions. However, during numerical simulations of the AL system the integrability is broken due to discretization along temporal dimension only. Therefore, since our results for the NLS equation and the AL system with $h\ll 1$ coincide, we may conclude that violation of integrability due to spatial discretization does not affect our results.


\section{Integrable turbulence: the asymptotic state.}

In the linear stage of the MI, when perturbations to the condensate are small, wave-action spectrum $I_{k}$ almost coincides with the spectrum of the condensate (\ref{spectra_condensate}), spatial correlation function is almost indistinguishable from unity $g(x)\approx 1$, and the PDF represents a very high and thin peak around $|\Psi|^{2}=1$. All of the moments (\ref{Mn}) are unity $M^{(n)}\approx 1$, kinetic energy is zeroth $\langle H_{d}\rangle\approx 0$, and potential energy is equal to $\langle H_{4}\rangle\approx -0.5$. 

With the development of the MI this situation changes; for our initial data the system arrives to the nonlinear stage approximately at $t\sim 10$ (the characteristic time of the instability is $1/\gamma_{0}=1$). In the nonlinear stage we observe that the moments $M^{(n)}(t)$ with exponents $n\neq 2$, and also kinetic and potential energies oscillate with time around their asymptotic values, and the amplitudes of these oscillations decay as time increases. The evolution of wave-action spectrum, spatial correlation function and the PDF is more complex, as in addition to oscillations they simultaneously change their forms. These functions also evolve towards some asymptotic forms that they take at late times. Thus, it is possible to say that during the nonlinear stage of the MI the system evolves from the condensate state towards some asymptotic turbulent state. The asymptotic state is characterized by independent on time wave-action spectrum, spatial correlation function, the PDF, the moments, and also 
kinetic and potential energies, that we below call asymptotic ones. We study these asymptotic characteristics in the current Section, while the temporal evolution towards the asymptotic state - in the next Section.

We find asymptotic characteristics by averaging the corresponding functions over time $t\in [950, 1000]$, where the deviations of these functions with time are sufficiently small. Nevertheless, their temporal evolution is still visible even at $t=1000$. However, this evolution is oscillatory-like, and averaging over sufficiently long time interval should provide us very good approximation to the asymptotic characteristics. We would like to stress that such time-averaging procedure gives very similar results already starting from time interval $t\in [150, 200]$. Therefore, we believe that integration beyond $t=1000$ on larger computational boxes $L$ should not provide us very different behavior.

\begin{figure}[t] \centering
\includegraphics[width=8.0cm]{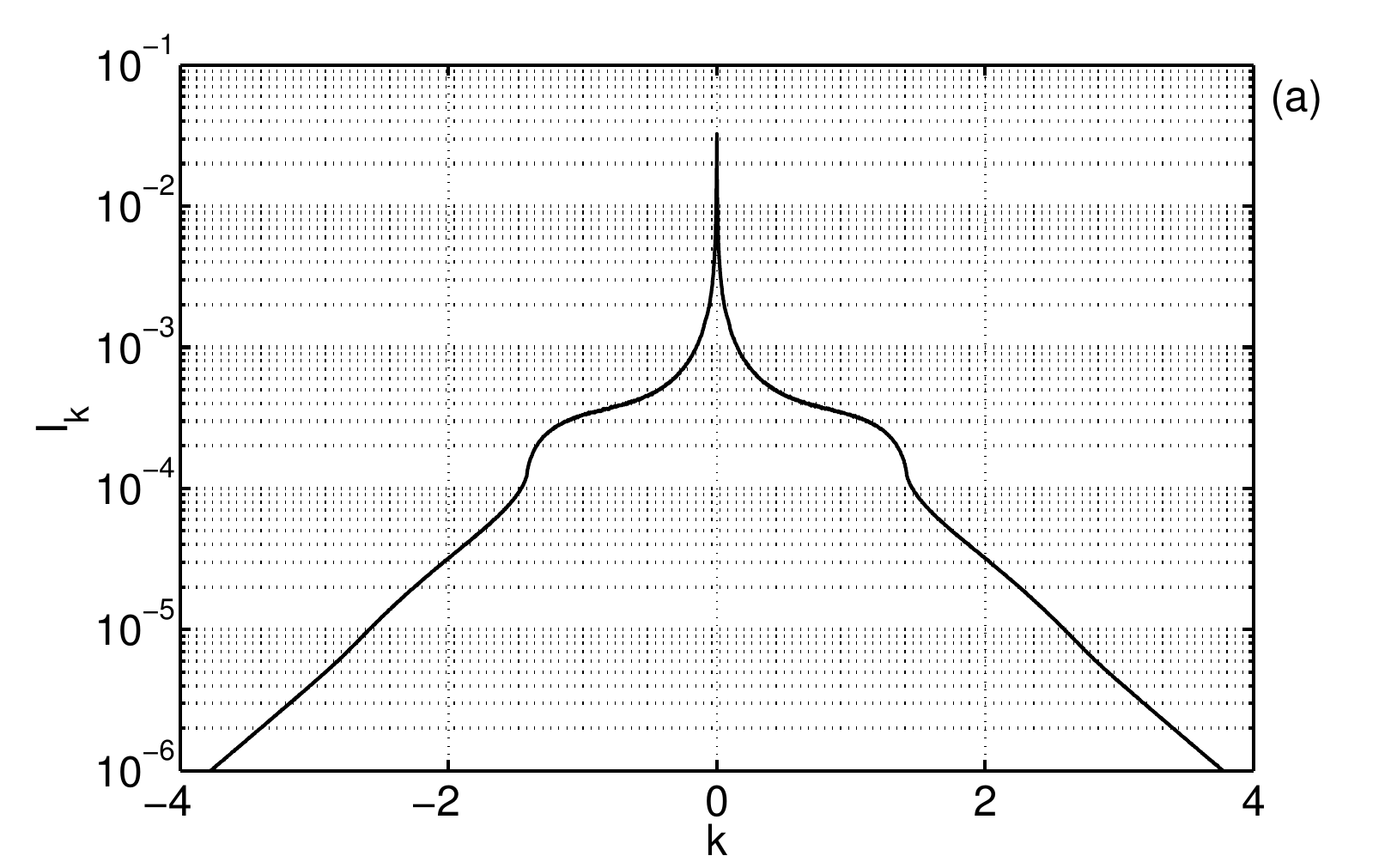}
\includegraphics[width=8.0cm]{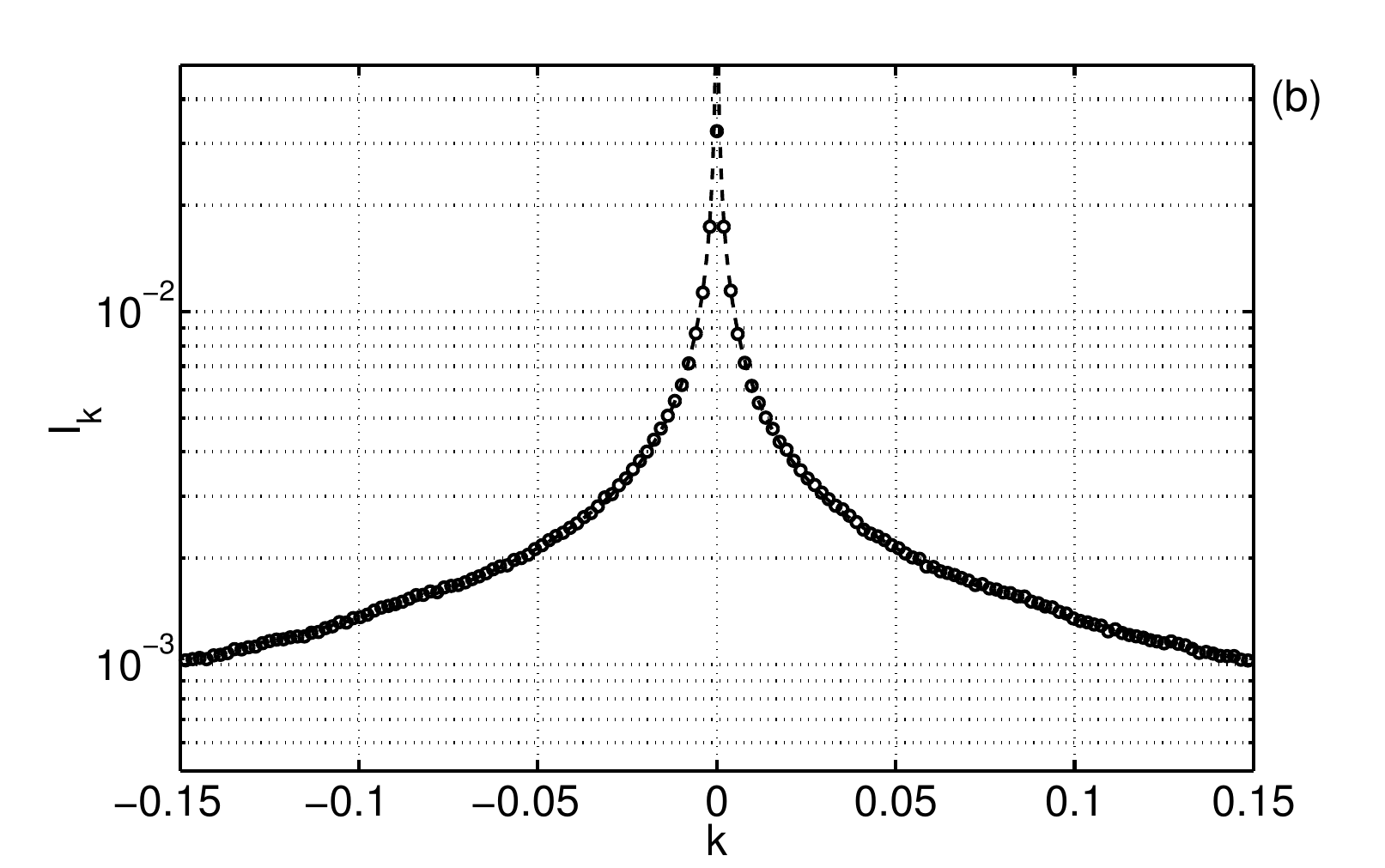}

\caption{\small {\it Graph (a): asymptotic wave-action spectrum $I_{k}$. Graph (b): asymptotic wave-action spectrum $I_{k}$ in the vicinity of $k=0$ (circles) and it's fit by function $f(k)=b|k|^{-\alpha}$, $\alpha\approx 0.659$, $b\approx 2.97\times 10^{-4}$ (dashed line). Graph (a) contains about 6000 harmonics and graph (b) contains about 150 harmonics.}}
\label{fig:spectra_asymptotic}
\end{figure}

FIG.~\ref{fig:spectra_asymptotic}a,b shows asymptotic wave-action spectrum $I_{k}$. The spectrum decays monotonically as $|k|\to +\infty$. This decay is slower at $0.4\lesssim |k|\lesssim 1$, very fast near $|k|=0$ and $|k|=\sqrt{2}$, and close to exponential from $|k|>1.5$. The remarkable property of the spectrum is that it has singularity at zeroth harmonic $k=0$. As will be shown in the next Section, in the beginning of the nonlinear stage of the MI this singularity represents a high peak occupying the zeroth harmonic only. The peak decays in oscillatory way, remaining detectable up to $t\sim 150$. At later times we observe the formation of power-law dependence $\sim\,|k|^{-\alpha}$ for $|k|\le 0.15$ with exponent $\alpha$ close to 2/3 (see FIG.~\ref{fig:spectra_asymptotic}b). The asymptotic wave-action spectrum for these wavenumbers is very well approximated by the function
\begin{equation}\label{spectra_assymptotic_015}
I_{k}\approx b|k|^{-\alpha},
\end{equation}
where $\alpha\approx 0.659$ and $b\approx 2.97\times 10^{-4}$. At $k=0$ the asymptotic spectrum has finite value $I_{0}\approx 0.032$.

\begin{figure}[t] \centering
\includegraphics[width=8.0cm]{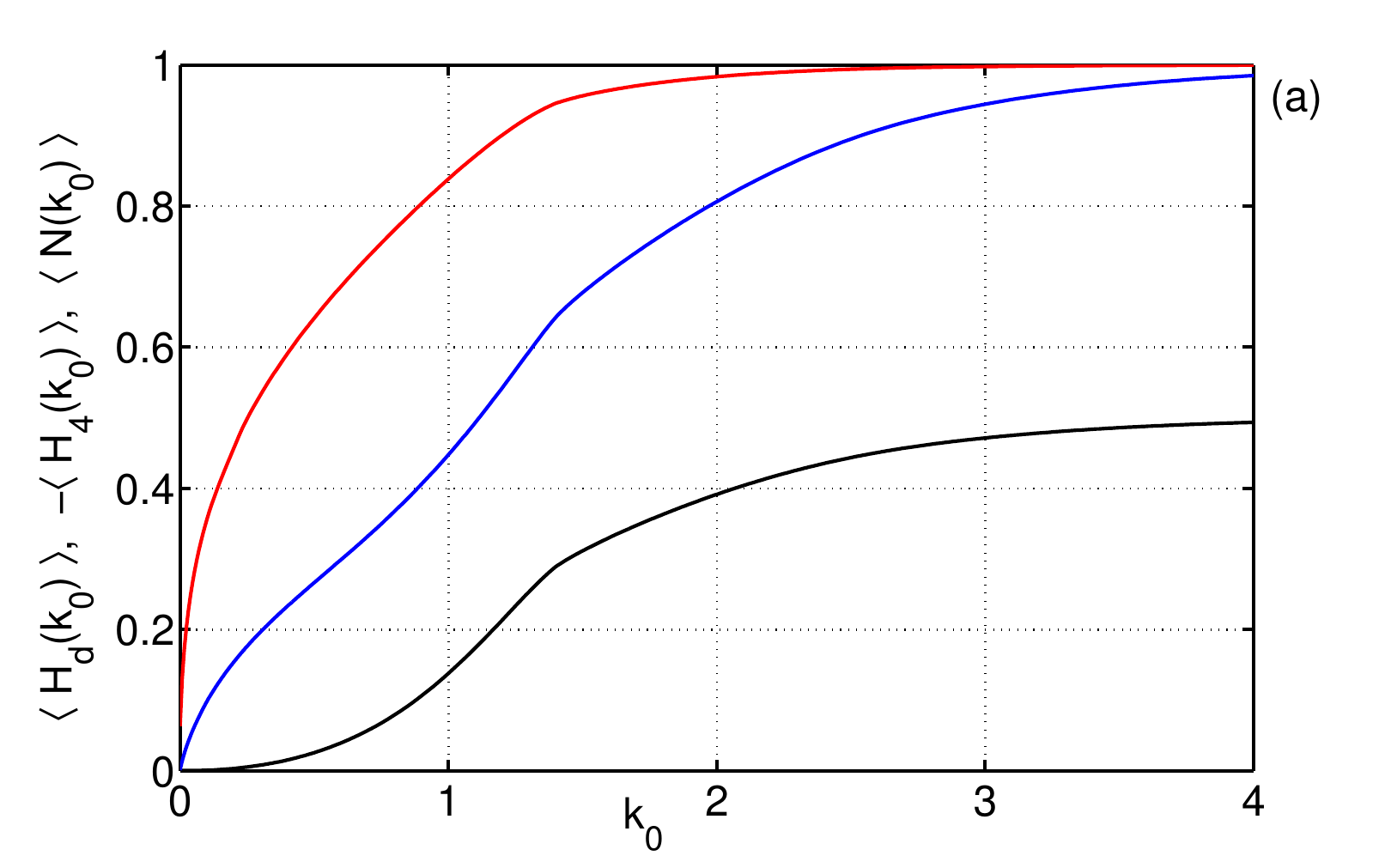}
\includegraphics[width=8.0cm]{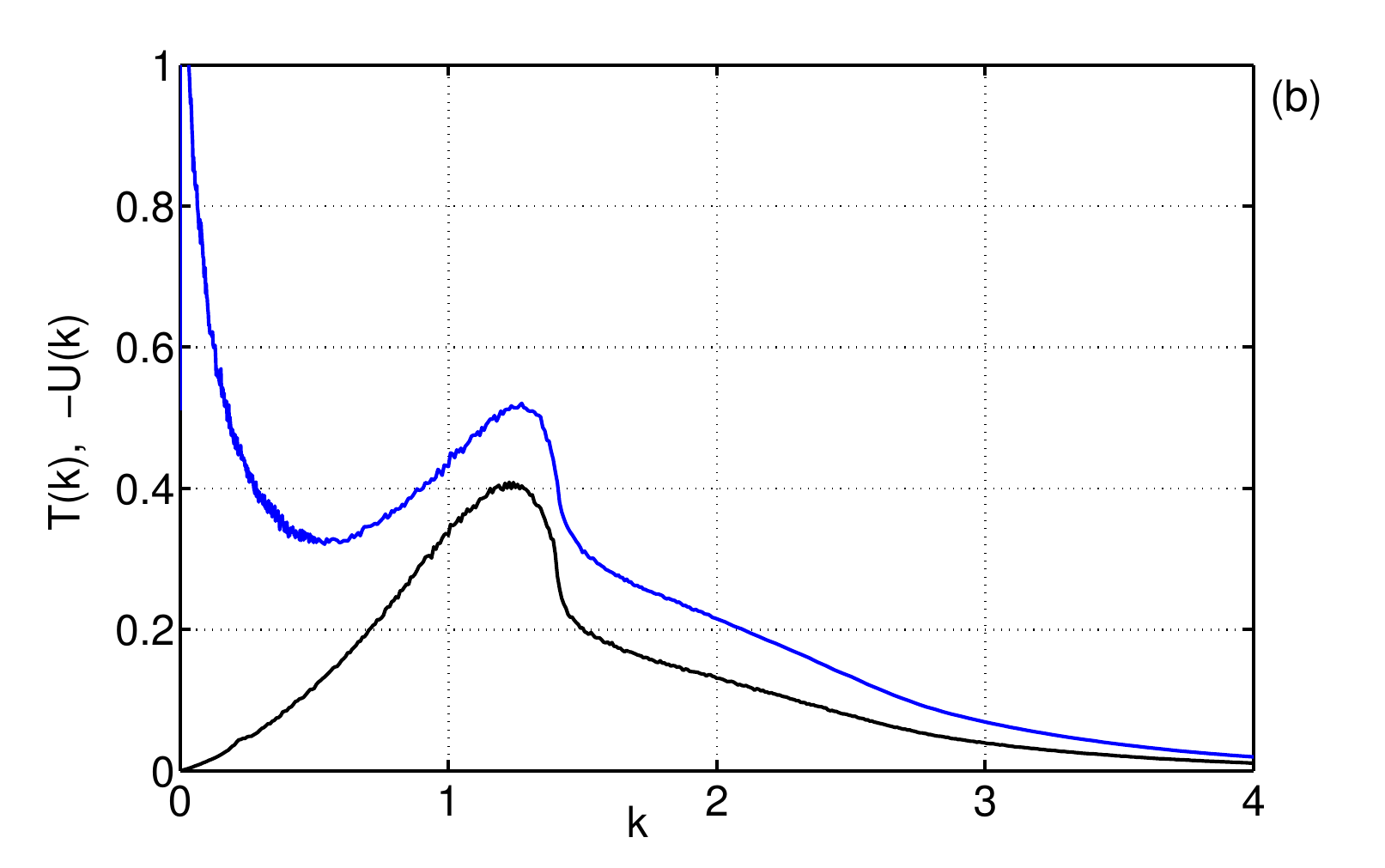}

\caption{\small {\it  (Color on-line) Graph (a): kinetic energy $\langle H_{d}(k_{0})\rangle$ (\ref{Hdk}) (black), potential energy $(-1)\times\langle H_{4}(k_{0})\rangle$ (\ref{H4_3}) (blue) and wave-action $\langle N(k_{0})\rangle$ (\ref{Ik_in_modes}) (red) concentrated in modes $|k|\le k_{0}$ in the asymptotic turbulent state. Graph (b): asymptotic spectral density of kinetic $T(k)$ (\ref{Hdkd}) (black) and potential $(-1)\times U(k)$ (\ref{H4_4}) (blue) energies.}}
\label{fig:H_asymptotic}
\end{figure}

The power-law behavior of the asymptotic spectrum means that the corresponding modes near $k=0$ are relatively large. Wave action concentrated in modes $|k|\le k_{0}$ can be calculated as
\begin{equation}\label{Ik_in_modes}
\langle N(k_{0})\rangle=\sum_{|k|\le k_{0}}I_{k}.
\end{equation}
It coincides with the ensemble average wave action (\ref{wave_action}) in the limit $k_{0}\to +\infty$. It turns out that approximately 41\% of wave action is concentrated in modes from the power-law region $|k|\le k_{0}$, $k_{0}=0.15$, since $\langle N(k_{0})\rangle\approx 0.41$, and about 7\% of wave action is concentrated in just 3 modes $k=0$ and $k=\pm 2\pi/L$ (see also FIG.~\ref{fig:H_asymptotic}a). Note that integration of the asymptotic (\ref{spectra_assymptotic_015}) over modes $|k|\le k_{0}$, $k_{0}=0.15$, gives very similar result:
$$
N(k_{0})=\sum_{|k|\le k_{0}}I_{k}\approx \frac{2}{\Delta k}\int_{0}^{k_{0}} I_{k}dk \approx \frac{2 b}{(1-\alpha)\Delta k}k_{0}^{1-\alpha}\approx 0.47,
$$
where $\Delta k = 2\pi/L=2^{-9}\approx 0.002$ is the distance between the subsequent wavenumbers. The modes $|k|\le 0.15$ have scales in the physical space comparable with the length of the integration box $L$, and thus can be called "quasi-condensate".

The macroscopic wave action concentration into quasi-condensate from one hand, and the FPU phenomenon from the other are the reasons why the numerical simulations that we perform must be implemented on computational boxes with very large lengths. For smaller computational boxes the distance between wavenumbers $2\pi/L$ is too large, and the region of wavenumbers $|k|\le 0.15$ containing about 40\% of wave action cannot be carefully resolved.

\begin{figure}[t] \centering
\includegraphics[width=8.0cm]{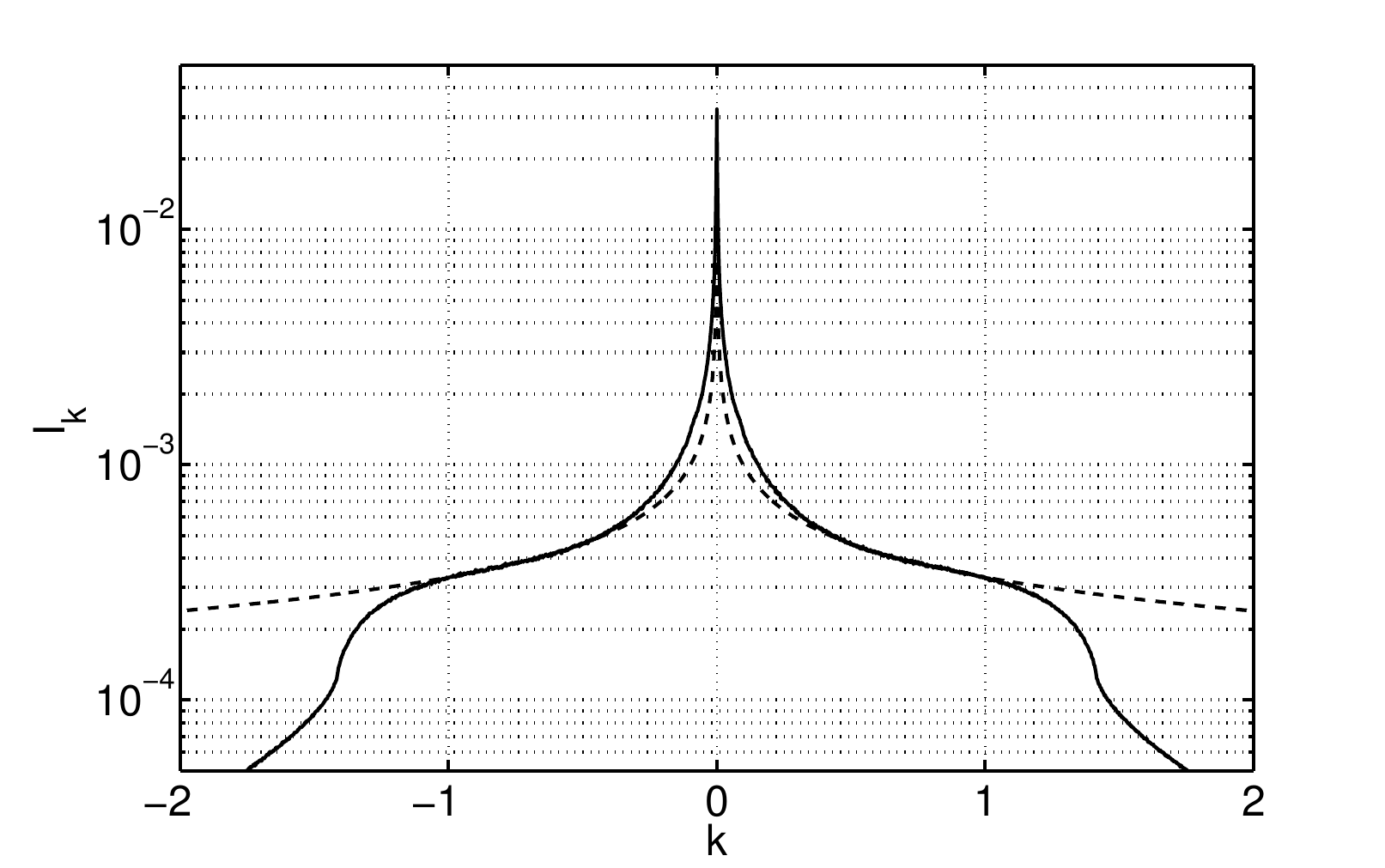}

\caption{\small {\it Asymptotic wave-action spectrum $I_{k}$ (solid line) and it's fit in the region $k\in[0.4, 1]$ by function $f(k)=b|k|^{-\alpha}$, $\alpha\approx 0.474$, $b\approx 3.32\times 10^{-4}$ (dashed line). }}
\label{fig:spectra_asymptotic2}
\end{figure}

We also detect another region of power-law dependence of the asymptotic wave-action spectrum at wavenumbers $|k|\in[0.4, 1]$, where the spectrum decays close to $|k|^{-1/2}$ (see FIG.~\ref{fig:spectra_asymptotic2}b). The maximum growth rate (\ref{max_growth_rate}) of the MI is realized at $|k|=1$; the corresponding modulations have characteristic scale $\ell=2\pi$ in the physical space. Thus, the second power-law region corresponds to modes with scales $\,\sim 2\pi-5\pi$, or 1--2.5 characteristic scales of the MI. These modes acquire about 25\% of wave-action. We think that after a very long evolution computed on a very large computational box the two power-law regions in the spectrum might merge in one $I_{k}\sim |k|^{-\alpha}$, $|k|\le 1$, with some shared exponent $\alpha$. However, the computational resources that we have at our disposal are insufficient to check this hypothesis. 

Modes with wavenumbers $|k|>1.5$, corresponding to scales smaller than $4\pi/3$ in the physical space, decay in the asymptotic spectrum close to exponential law $\sim\,e^{-\beta |k|}$, $\beta\approx 0.9$. These modes have about 5\% of wave-action. 

\begin{figure}[t] \centering
\includegraphics[width=8.0cm]{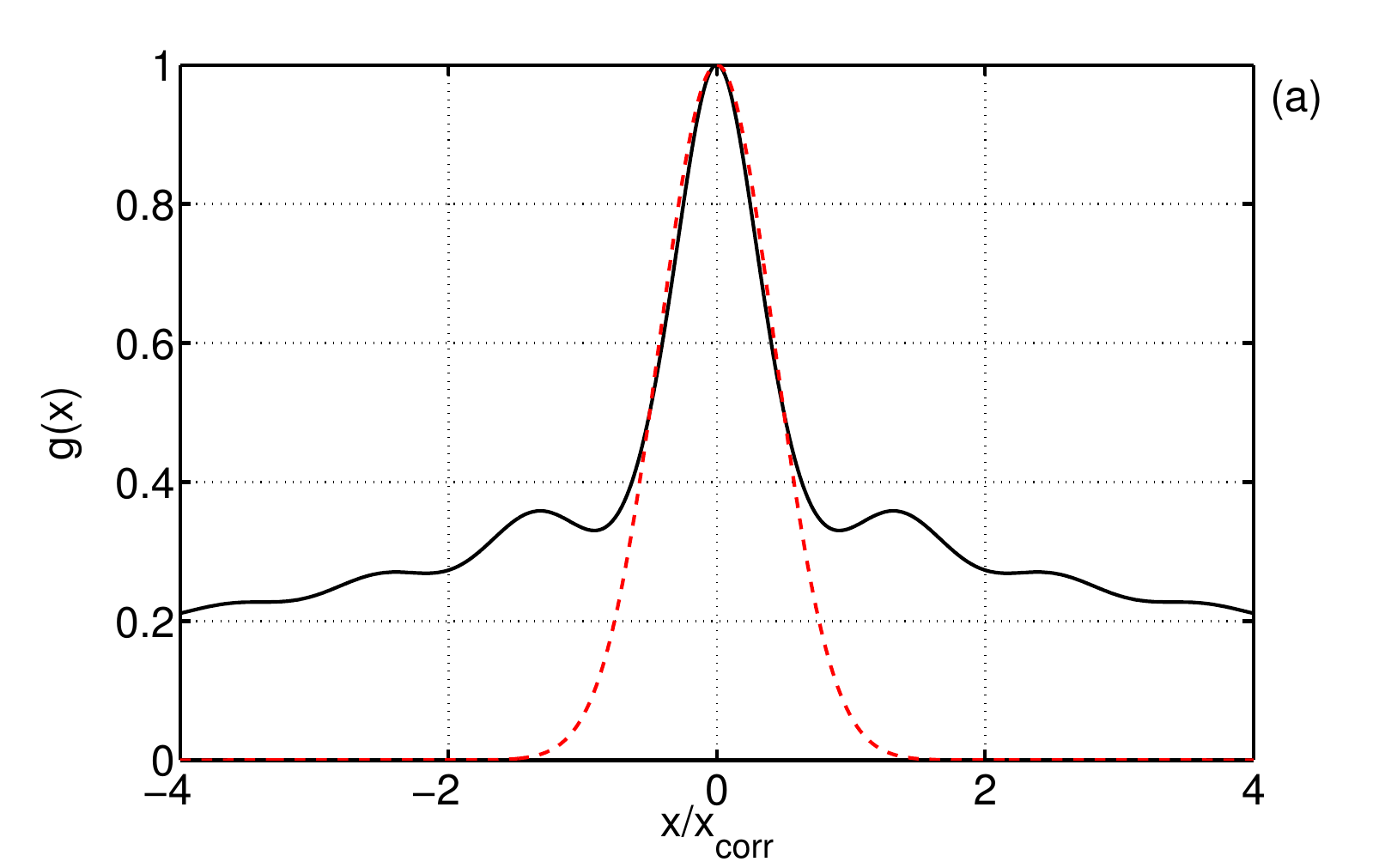}
\includegraphics[width=8.0cm]{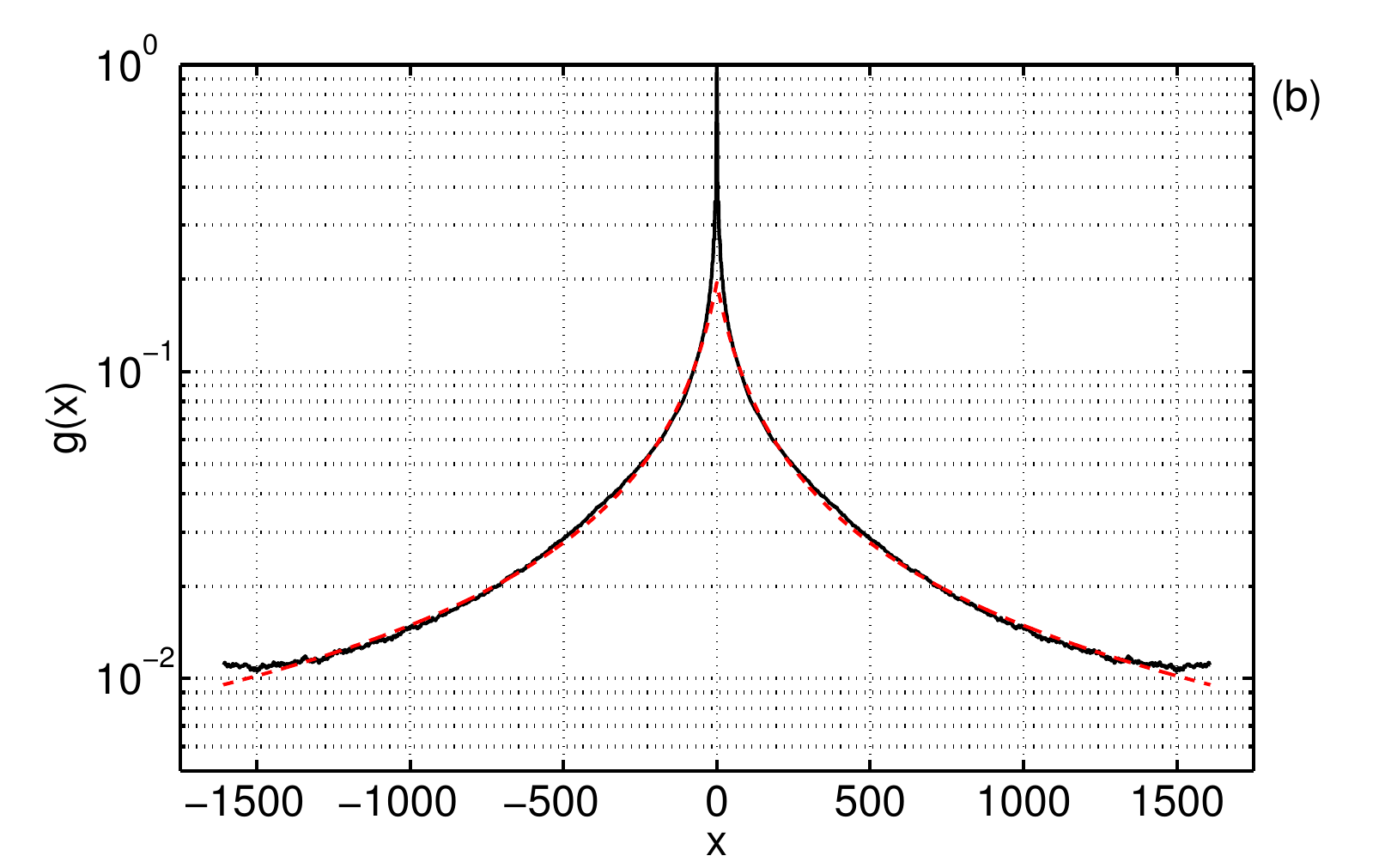}

\caption{\small {\it (Color on-line) Graph (a): asymptotic spatial correlation function $g(x)$ (solid black line) and Gaussian distribution (\ref{correlation_universal}) (dashed red line) versus $x/x_{corr}$, where $x_{corr}\approx 4.016$ is full width at half maximum for $g(x)$. Graph (b): asymptotic spatial correlation function $g(x)$ (solid black line) and it's fit by function $f(x)=b_{1}/(|x|+b_{2})$, $b_{1}\approx 16.1$, $b_{2}\approx 82.7$ (dashed red line).}}
\label{fig:corr_asymptotic}
\end{figure}

The asymptotic spatial correlation function is shown on FIG.~\ref{fig:corr_asymptotic}a,b. It's characteristic scale, defined as full width at half maximum, is $x_{corr}\approx 4.016$. At lengths $|x|<x_{corr}/2$ it is close to Gaussian (see Eq. (\ref{corr_x_0}): $g(0,t)\approx 1$),
\begin{equation}\label{correlation_universal}
g(x)\approx \exp\bigg[-4\ln 2\bigg(\frac{x}{x_{corr}}\bigg)^{2}\bigg].
\end{equation}
At $|x|>x_{corr}/2$ the asymptotic correlation function decays very slowly to about $g(L/2)\approx 0.01$ as $|x|\to L/2$. In the region $|x|\in[100,1500]$ this decay is very well approximated as
\begin{equation}\label{correlation_condensate_decay}
g(x)\approx \frac{b_{1}}{|x|+b_{2}},
\end{equation}
with the coefficients $b_{1}\approx 16.1$ and $b_{2}\approx 82.7$ (see FIG.~\ref{fig:corr_asymptotic}b). In the region $|x|\in[1500,L/2]=[1500,512\pi]$ the decay is even slower, but we believe that this is the effect of the finiteness of the computational box. Indeed, by construction spatial correlation function is periodic $g(L/2)=g(-L/2)$, and should be even $g(x)=g(-x)$, therefore it's derivatives at the borders of the computational box should be zeroth $g_{x}(\pm L/2)=0$. Thus, near the borders the behavior of the correlation function should deviate from (\ref{correlation_condensate_decay}).

\begin{figure}[t] \centering
\includegraphics[width=8.0cm]{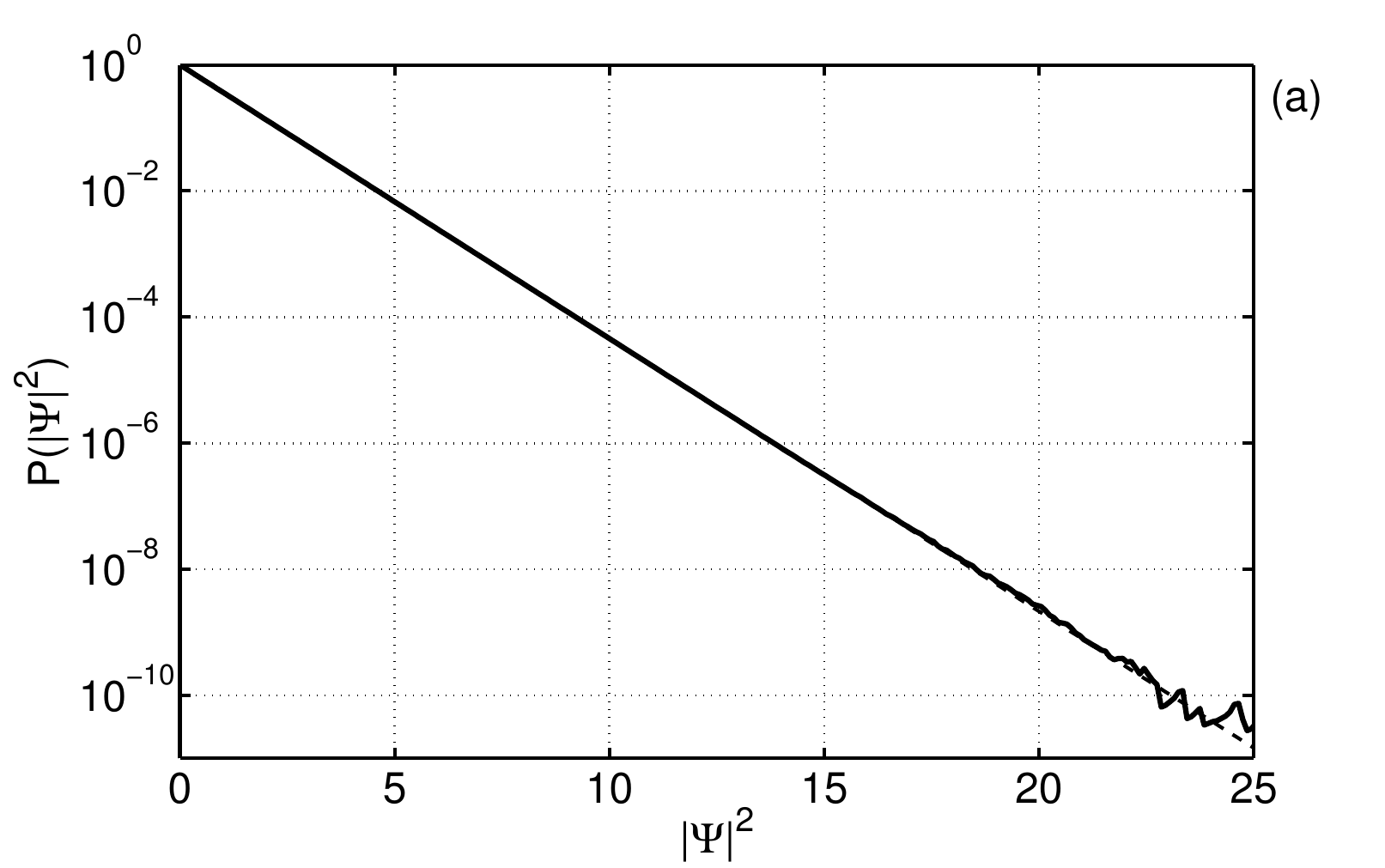}
\includegraphics[width=8.0cm]{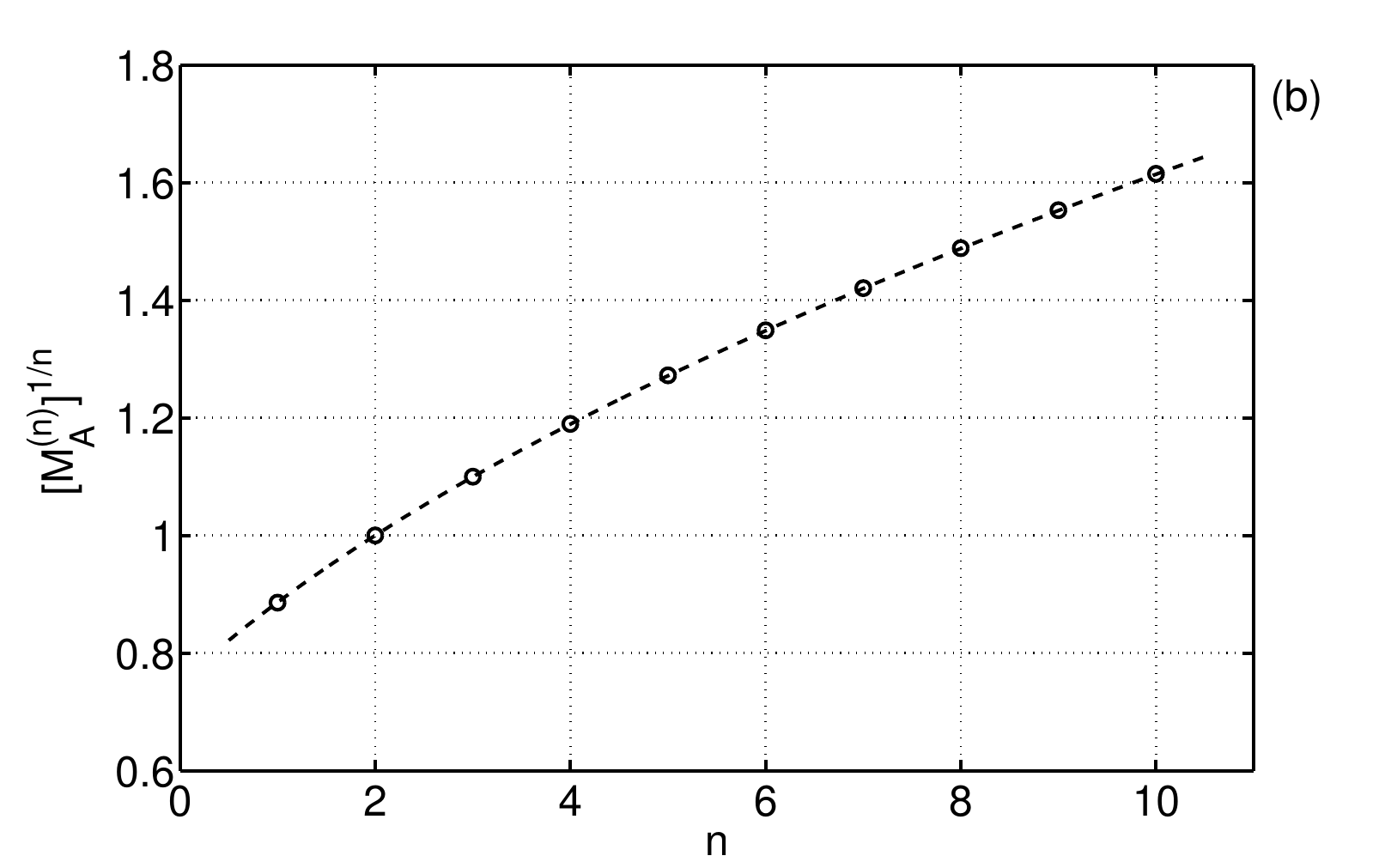}

\caption{\small {\it Graph (a): asymptotic squared amplitude PDF $P(|\Psi|^{2})$ (solid line) and Rayleigh PDF (\ref{Rayleigh_3}) (dashed line). Graph (b): asymptotic values of the moments $[M^{(n)}(t)]^{1/n}$ (circles), $n=1,...,10$, and their Rayleigh prediction $[\Gamma(n/2+1)]^{1/n}$ (\ref{MnR1}) (dashed line).}}
\label{fig:PDF}
\end{figure}

The squared amplitude asymptotic PDF $P(|\Psi|^{2})$ coincides with Rayleigh PDF (\ref{Rayleigh_3}), as shown on FIG.~\ref{fig:PDF}a. Thus, we come to surprising conclusion, that despite the nonlinearity of the NLS equation it's asymptotic PDF is the same that would be for a wave field described by linear equations. We additionally checked this conclusion by calculating the asymptotic values of the moments (\ref{Mn}) $M_{A}^{(n)}$, and found that these values coincide with their Rayleigh predictions (\ref{MnR1}), at least for exponents $n=1,...,10$ (see FIG.~\ref{fig:PDF}b). The asymptotic moment $M_{A}^{(4)}\approx 2$ allows us to calculate the ensemble average potential energy $\langle H_{4}\rangle$ in the asymptotic state,
\begin{equation}\label{asymptotic_H4}
\langle H_{4}\rangle = -\frac{1}{2}M^{(4)}_{A}\approx -1.
\end{equation}
Combined with the conservation of total energy $\langle H_{d}+H_{4}\rangle\approx -0.5$, this yields the ensemble average kinetic energy $\langle H_{d}\rangle$,
\begin{equation}\label{asymptotic_Hd}
\langle H_{d}\rangle \approx 0.5.
\end{equation}
We also calculated $\langle H_{d}\rangle$ independently with the same result. Thus, in the asymptotic state we have "moderately strong" turbulence with $Q=|\langle H_{4}\rangle|/|\langle H_{d}\rangle|\approx 2$. This makes the conclusion of Rayleigh statistics for amplitudes $|\Psi(x,t)|$ even more surprising.

In fact, the relation (\ref{asymptotic_H4}) is itself truly remarkable. According to (\ref{Fourier2}),
\begin{equation}\label{H4_Fourier}
-\langle H_{4}\rangle = \frac{1}{2}\sum_{k_{1},k_{2},k_{3},k_{4}}\langle\Psi_{k_{1}}\Psi_{k_{2}}\Psi_{k_{3}}^{*}\Psi_{k_{4}}^{*}\rangle\delta(k_{1}+k_{2}-k_{3}-k_{4}),
\end{equation}
where $\delta(k)$ is Kronecker delta
$$
\delta(k)
=\left\{ \begin{array}{ll} 
1, & k=0, \\ 
0, & k\neq 0.
\end{array}\right.
$$
The four-wave momentum in (\ref{H4_Fourier}) can be represented as
\begin{equation}\label{cumulant}
\langle\Psi_{k_{1}}\Psi_{k_{2}}\Psi_{k_{3}}^{*}\Psi_{k_{4}}^{*}\rangle = I_{k_{1}}I_{k_{2}}\bigg(\delta(k_{1}-k_{3})\delta(k_{2}-k_{4}) + \delta(k_{1}-k_{4})\delta(k_{2}-k_{3})\bigg) + J_{k_{1},k_{2},k_{3},k_{4}},
\end{equation}
where $J_{k_{1},k_{2},k_{3},k_{4}}$ is the cumulant. Since $\sum I_{k}=\langle N\rangle\approx 1$, we obtain
\begin{equation}\label{cumulant2}
-\langle H_{4}\rangle \approx 1 + \frac{1}{2}\sum_{k_{1},k_{2},k_{3},k_{4}}J_{k_{1},k_{2},k_{3},k_{4}}\delta(k_{1}+k_{2}-k_{3}-k_{4}).
\end{equation}
Together with the relation (\ref{asymptotic_H4}) this yields
\begin{equation}\label{cumulant3}
\bigg|\sum_{k_{1},k_{2},k_{3},k_{4}}J_{k_{1},k_{2},k_{3},k_{4}}\delta(k_{1}+k_{2}-k_{3}-k_{4})\bigg| \ll 1.
\end{equation}
The latter result might mean that the cumulant in the asymptotic turbulent state is zeroth. Thus, the relation $\langle H_{4}\rangle\approx -1$ can be considered as an indication that the stationary integrable turbulence is purely Gaussian. The calculation of the cumulant is a cumbersome problem, and we will study it in a separate publication. 

It is interesting to examine the spectral distribution of the ensemble average kinetic and potential energies. The spectral density of kinetic energy $k^{2}I_{k}$ can be obtained from Eq. (\ref{energy}) and Eq. (\ref{Fourier1})-(\ref{Fourier2}). For convenience we divide it by the distance between the subsequent wavenumbers $\Delta k$, and use it in the following form:
\begin{equation}\label{Hdkd}
T(k) = \frac{k^{2}(I_{k}+I_{-k})}{\Delta k}.
\end{equation}
Kinetic energy concentrated within modes $|k|\le k_{0}$ can be calculated as
\begin{equation}\label{Hdk}
\langle H_{d}(k_{0})\rangle = \sum_{|k|\le k_{0}}k^{2}I_{k} = \Delta k \sum_{0\le k\le k_{0}}T(k).
\end{equation}

Calculation of the spectral distribution of potential energy is more complex. For this purpose we introduce the new function $\tilde{\Psi}(x,t)$,
\begin{equation}\label{H4_1}
\tilde{\Psi}(x,t) = \mathscr{F}^{-1}[\tilde{\Psi}_{k}(t)],
\end{equation}
that contains only modes with $|k|\le k_{0}$ from the original solution of the NLS equation $\Psi(x,t)$,
\begin{equation}\label{H4_2}
\tilde{\Psi}_{k}(t)
=\left\{ \begin{array}{ll} 
\Psi_{k}(t), & |k| \le k_{0}, \\ 
0, & |k| > k_{0},
\end{array}\right.
\end{equation}
and $\Psi_{k}(t)=\mathscr{F}[\Psi(x,t)]$ is Fourier components of original solution. Then we find potential energy concentrated in these modes and average it across the realizations of initial data,
\begin{equation}\label{H4_3}
\langle H_{4}(k_{0})\rangle=-\bigg\langle\frac{1}{2L}\int_{-L/2}^{L/2}|\tilde{\Psi}(x,t)|^{4}\,dx\bigg\rangle,
\end{equation}
Hence, the spectral density of potential energy can be calculated as
\begin{equation}\label{H4_4}
U(k)=\frac{\langle H_{4}(k+\Delta k)\rangle-\langle H_{4}(k)\rangle}{\Delta k}.
\end{equation}
In the limit $k_{0}\to +\infty$ the quantities (\ref{Hdk}) and (\ref{H4_3}) coincide with the ensemble average kinetic and potential energies (\ref{energy}) respectively.

The distribution across wavenumbers of kinetic $\langle H_{d}(k_{0})\rangle$ and potential $\langle H_{4}(k_{0})\rangle$ energies, as well as their spectral densities $T(k)$ and $U(k)$, are shown on FIG.~\ref{fig:H_asymptotic}a,b. The spectral density of kinetic energy $T(k)$ monotonically increases from zero at zeroth harmonic $k=0$ to $T(k)\approx 0.4$ at $k\approx 1.2$, where it achieves maximum, sharply decreases at $k\approx \sqrt{2}$, and then decays to zero starting from $k>1.5$. The spectral density of potential energy modulus $|U(k)|$ has sharp maximum at $k=0$, monotonically decreases to $|U(k)|\approx 0.3$ at $k\approx 0.5$, where it achieves local minimum, and then behaves similar to $T(k)$, acquiring another maximum $|U(k)|\approx 0.5$ at $k\approx 1.2$. 

It is interesting that $|U(k)|$ is always larger than $T(k)$, so that in the asymptotic turbulent state all modes are essentially nonlinear. The "most nonlinear modes" are those of the quasi-condensate $|k|\le 0.15$, that contain less than 1\% of kinetic energy, about 10\% of potential energy, and about 40\% of wave action. Modes with $0.15\le |k|\le 1.5$ contain about 60\% of kinetic and potential energies and about 55\% of wave action, while the exponentially decaying modes $|k|>1.5$ have about 5\% of wave action, about 40\% of kinetic energy and 30\% of potential energy.


\section{The nonlinear stage of the modulation instability: evolution towards the asymptotic state.}

\begin{figure}[t] \centering
\includegraphics[width=8.0cm]{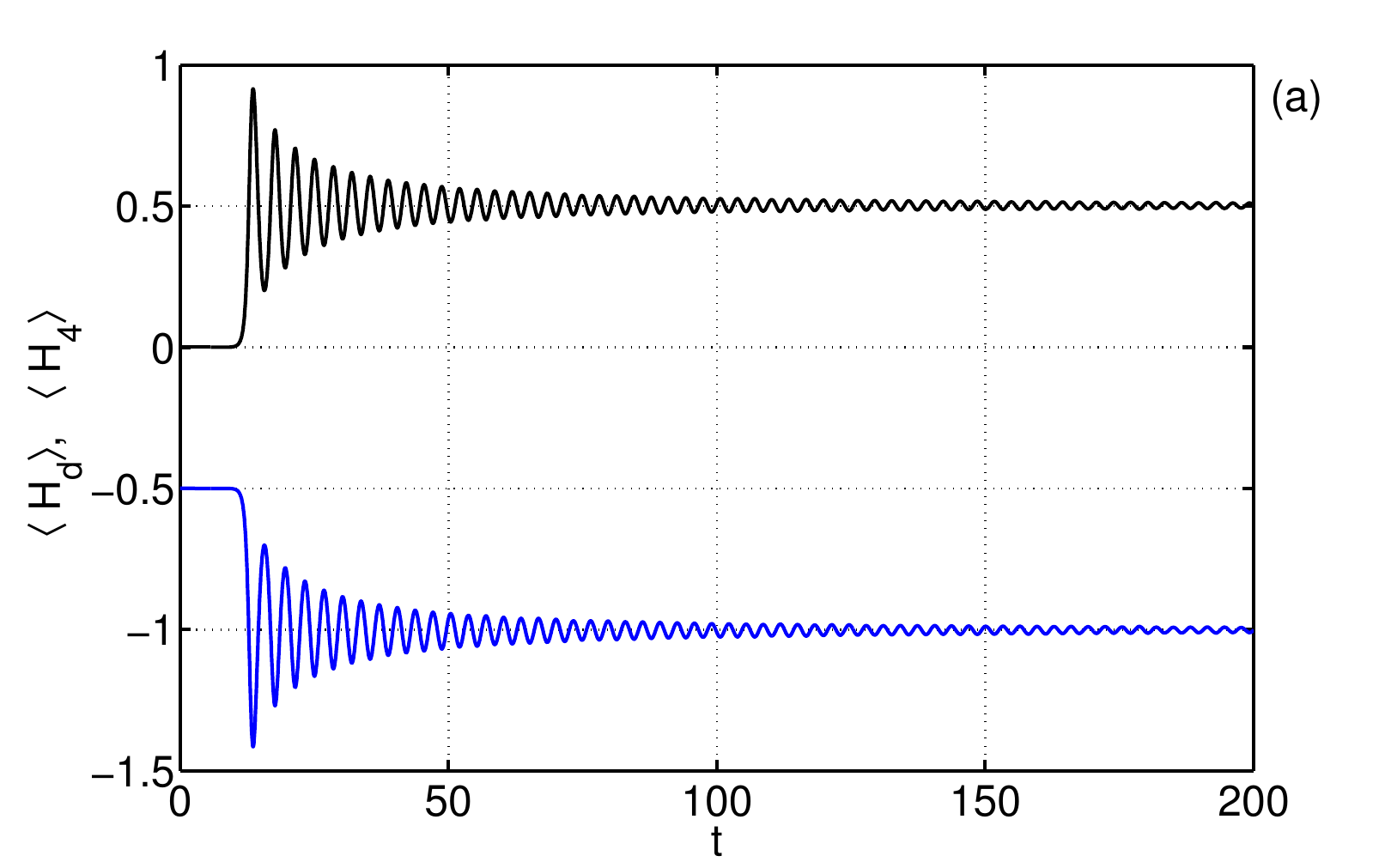}
\includegraphics[width=8.0cm]{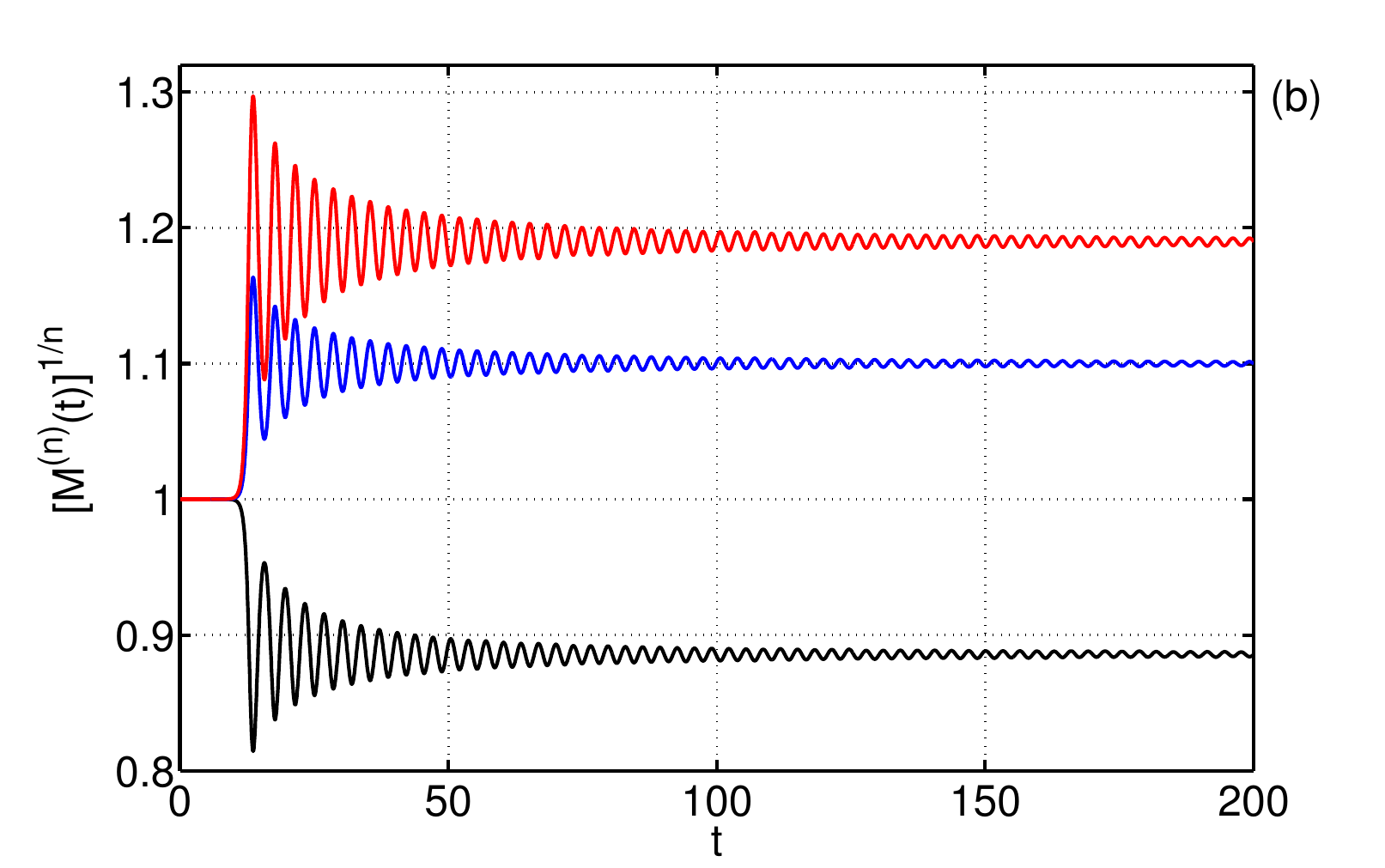}

\caption{\small {\it  (Color on-line) Evolution of ensemble average (a) kinetic $\langle H_{d}\rangle$ (black) and potential $\langle H_{4}\rangle$ (blue) energies and (b) moments $M^{(1)}(t)$ (black), $[M^{(3)}(t)]^{1/3}$ (blue) and $[M^{(4)}(t)]^{1/4}$ (red).}}
\label{fig:HamiltonianMoments}
\end{figure}

It will be convenient for us to study the evolution towards the asymptotic turbulent state on the example of ensemble average kinetic $\langle H_{d}\rangle$ and potential $\langle H_{4}\rangle$ energies, and also the moments $M^{(n)}(t)$ (see FIG.~\ref{fig:HamiltonianMoments}a,b). Up to $t\sim 10$ the perturbations to the condensate are small, so that the moments and the energies do not change substantially from their initial values $M^{(n)}\approx 1$, $\langle H_{d}\rangle\approx 0$ and $\langle H_{4}\rangle\approx -0.5$ respectively. At $t\sim 10$ the MI arrives to it's nonlinear stage; the moments start to oscillate around their asymptotic Rayleigh values (\ref{MnR1}), kinetic energy - around 0.5, and potential energy -- around -1. The moment $M^{(1)}(t)$ oscillates in-phase with potential energy $\langle H_{4}\rangle$, and antiphase with the moments $M^{(n)}(t)$, $n\geq 3$, and kinetic energy $\langle H_{d}\rangle$, so that the positions in 
time of local maximums and minimums of $M^{(1)}(t)$ and $\langle H_{4}\rangle$ coincide with the positions of local minimums and maximums of $M^{(n)}(t)$, $n\geq 3$, and $\langle H_{d}\rangle$ respectively.

\begin{figure}[t] \centering
\includegraphics[width=8.0cm]{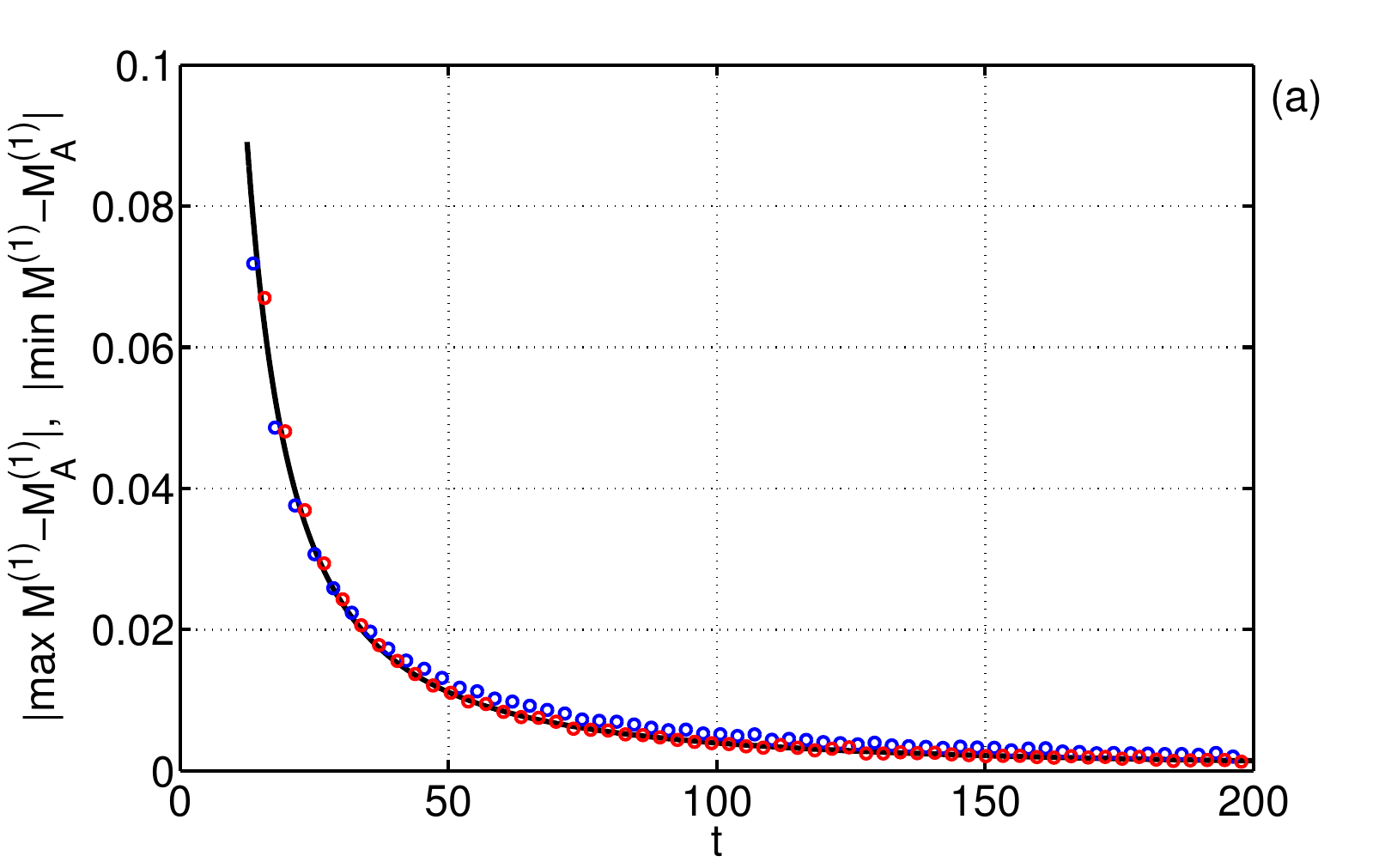}
\includegraphics[width=8.0cm]{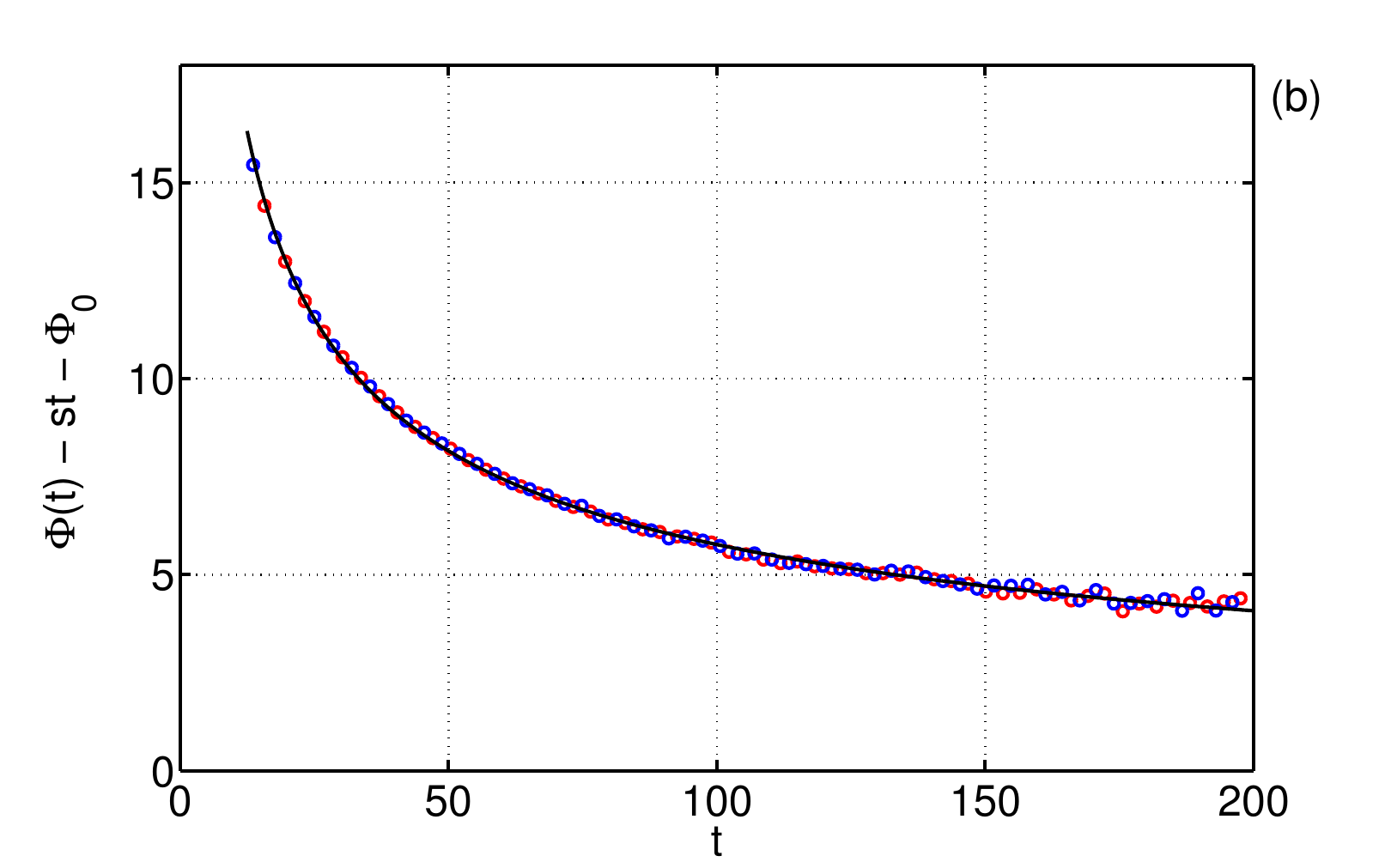}

\caption{\small {\it  (Color on-line) Graph (a): amplitude of the oscillations of the moment $M^{(1)}(t)$ (circles), calculated as the modulus of the deviations of the extremums of $M^{(1)}(t)$ from it's asymptotic value $M^{(1)}_{A}$, depending on time $t$. Graph (b): nonlinear phase shift (\ref{nps}) calculated at the extremums of $M^{(1)}(t)$ (circles), depending on time $t$. Red circles mark local maximums, blue circles - local minimums of $M^{(1)}(t)$. Black line on graph (a) is fit by function $p/t^{3/2}$, $p\approx 3.94$, on graph (b) -- is fit by function $q/\sqrt{t}$, $q\approx 57.7$.}}
\label{fig:oscillations_laws}
\end{figure}

We study the time dependence of the oscillations on the example of moment $M^{(1)}(t)$. FIG.~\ref{fig:oscillations_laws}a shows that the amplitude of the oscillations of $M^{(1)}(t)$, that we measure as the modulus of deviations of local maximums and minimums of $M^{(1)}(t)$ from it's asymptotic value $M^{(1)}_{A}$, is very well approximated by the function $p/t^{3/2}$ with the prefactor $p=(3.94\pm 0.03)$. The period of the oscillations changes from $\Delta T\sim 4$ at $t\sim 20$ to $\Delta T\sim 3$ at $t\sim 200$. We think that this is the effect analogous to the nonlinear phase shift. Indeed, one can search for the approximation of $M^{(1)}(t)$ in the form
\begin{equation}\label{f1}
M^{(1)}(t)\approx M^{(1)}_{A} + \frac{p}{t^{3/2}}\sin(\Phi(t)),\quad \Phi(t)=st+\phi_{nl}(t)+\Phi_{0},
\end{equation}
where $\Phi(t)$ is phase, $s$ is constant frequency, $\Phi_{0}$ is constant phase, and the nonlinear phase shift $\phi_{nl}(t)$ should be proportional to the amplitude of the oscillations $p/t^{3/2}$ multiplied by time $t$, or $\phi_{nl}(t)=q/\sqrt{t}$ with constant $q$. Then, the phases $\Phi$ at the local maximums $t_{max}$ of $M^{(1)}(t)$ should be equal to 
$$
\Phi(t_{max}) = st_{max}+\frac{q}{\sqrt{t_{max}}}+\Phi_{0} = \frac{\pi}{2}+2\pi m,
$$
and at the local minimums $t_{min}$ -- to
$$
\Phi(t_{min}) = st_{min}+\frac{q}{\sqrt{t_{min}}}+\Phi_{0} = \frac{3\pi}{2}+2\pi m,
$$
where $m$ is integer number. We find all the subsequent extremums $t_{max}$ and $t_{min}$ of $M^{(1)}(t)$ from one hand, and their phases $\Phi$ from the other hand by setting $m=0$ for the first maximum, $m=1$ for the second maximum, and so on. Then, with the help of the least squares method we determine the coefficients $s\approx 1.99$, $q\approx 57.7$ and $\Phi_{0}\approx -44.1$. After that we check that the nonlinear phase shift
\begin{equation}\label{nps}
\Phi(t)-st-\Phi_{0},
\end{equation}
calculated at the extremums of $M^{(1)}(t)$, indeed is very well approximated by the function $q/\sqrt{t}$, as shown on FIG.~\ref{fig:oscillations_laws}b.

\begin{figure}[t] \centering
\includegraphics[width=16cm]{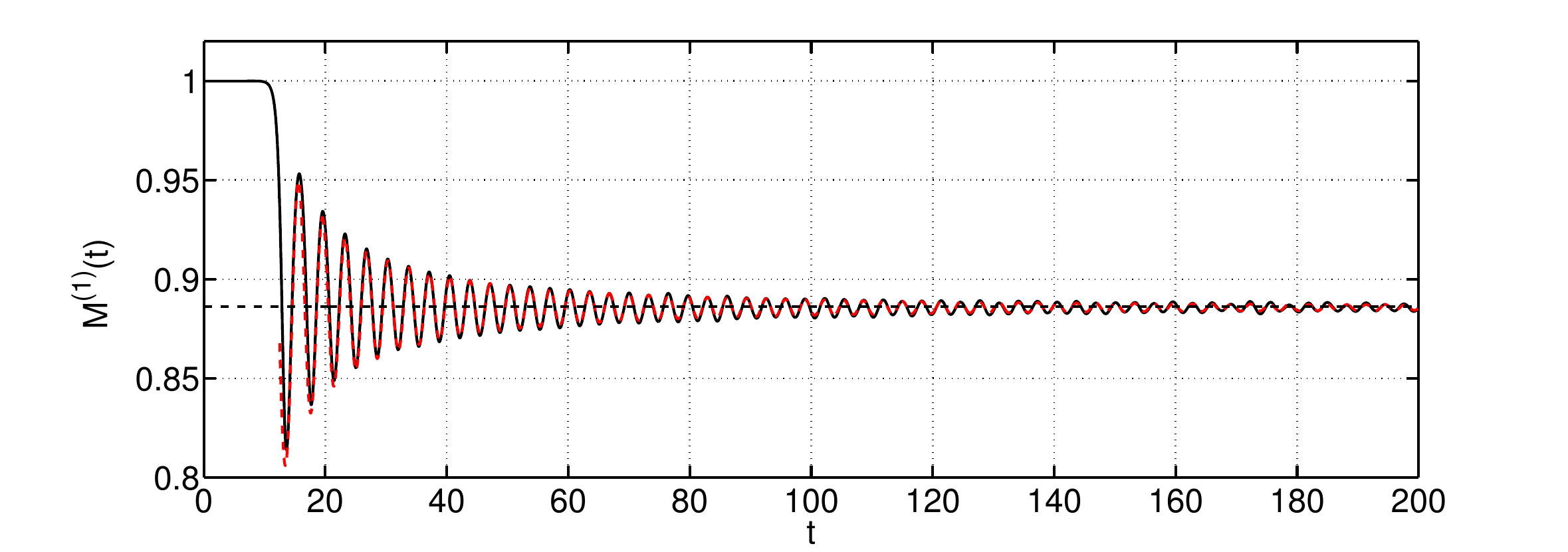}

\caption{\small {\it  (Color on-line) Evolution of the moment $M^{(1)}(t)$ (solid black line), it's fit by function $f(t)=M^{(1)}_{A} + [p/t^{3/2}]\sin(st + q/\sqrt{t} + \Phi_{0})$ with parameters $M^{(1)}_{A}\approx 0.886$, $p\approx 3.94$, $s\approx 1.99$, $q\approx 57.7$, $\Phi_{0}\approx -44.1$ (dashed red line), and the Rayleigh value of the moment $M^{(1)}_{R}\approx 0.886$ (\ref{MnR1}) (dashed black line).}}
\label{fig:oscillations_fit}
\end{figure}

We observe that anzats (\ref{f1}) fits very well to the experimental time dependence of the moments, and also kinetic and potential energies as well. The example of such fit for $M^{(1)}(t)$ is shown on FIG.~\ref{fig:oscillations_fit}. The phases $\Phi_{0}$ for the moment $M^{(1)}(t)$ and potential energy $\langle H_{4}\rangle$ coincide, and differ by $\pi$ from the phases $\Phi_{0}$ for the moments $M^{(n)}(t)$, $n\geq 3$, and kinetic energy $\langle H_{d}\rangle$. We checked that anzats (\ref{f1}) without the nonlinear phase shift, or with the exponent of the nonlinear phase shift significantly different from --0.5, fits significantly worse to the experimental data. It is interesting to note that the period of the oscillations $2\pi/s\approx 3.16$ is almost equal to $\pi$, and their frequency $s$ is almost equal to the double maximum growth rate of the MI (\ref{max_growth_rate}), $s\approx 2\gamma_{0}$. We think that the frequency $s$ should coincide with 2, and the 
period should accordingly coincide with $\pi$, and we measure the frequency $s\approx 2$ almost the same for all of our experiments irrespective of the statistics of initial noise. However, the nature of such correspondence is unclear for us yet.

Wave-action spectrum, spatial correlation function and the PDF of squared amplitudes also evolve in oscillatory way with time, approaching to their asymptotic forms at late times. The "turning points" for the evolution of these functions -- the points in time where the motion of $I_{k}(t)$, $g(x,t)$ and $P(|\Psi|^{2},t)$ at fixed $k$, $x$ and $|\Psi|^{2}$ respectively changes to roughly the opposite -- approximately coincide with the local maximums and minimums of the moments, and also kinetic and potential energies. For definiteness, below we will refer to such points in time on the example of extremums of potential energy modulus $|\langle H_{4}\rangle|$. This choice has also straightforward physical sense: at the local maximums of $|\langle H_{4}\rangle|$ the effect of nonlinearity is the largest, and at the local minimums -- the smallest. Note, that the spectrum, the correlation function and the PDF 
do not evolve exactly as $|\langle H_{4}\rangle|$, since these functions simultaneously change their forms with time.

\begin{figure}[t] \centering
\includegraphics[width=8.0cm]{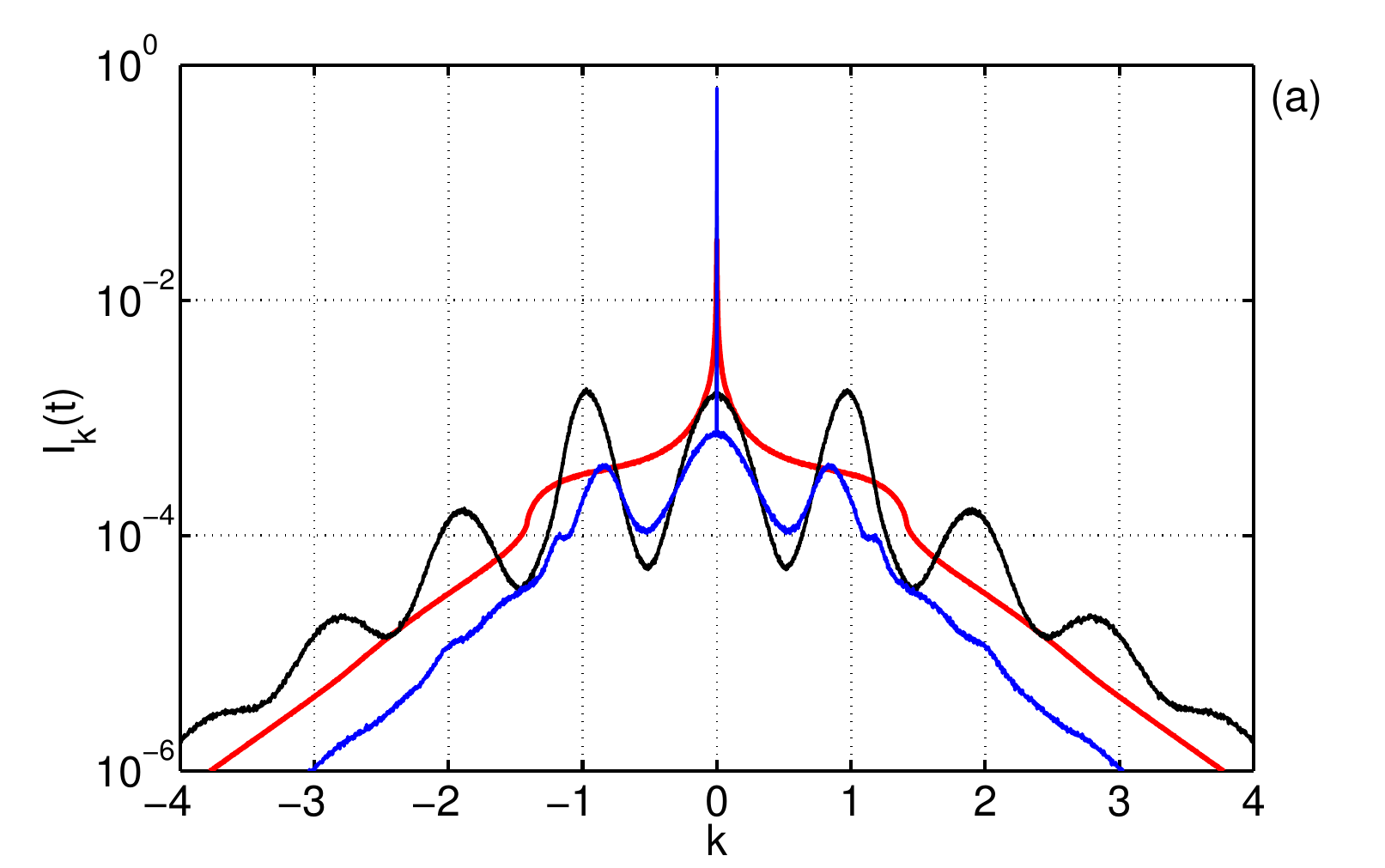}
\includegraphics[width=8.0cm]{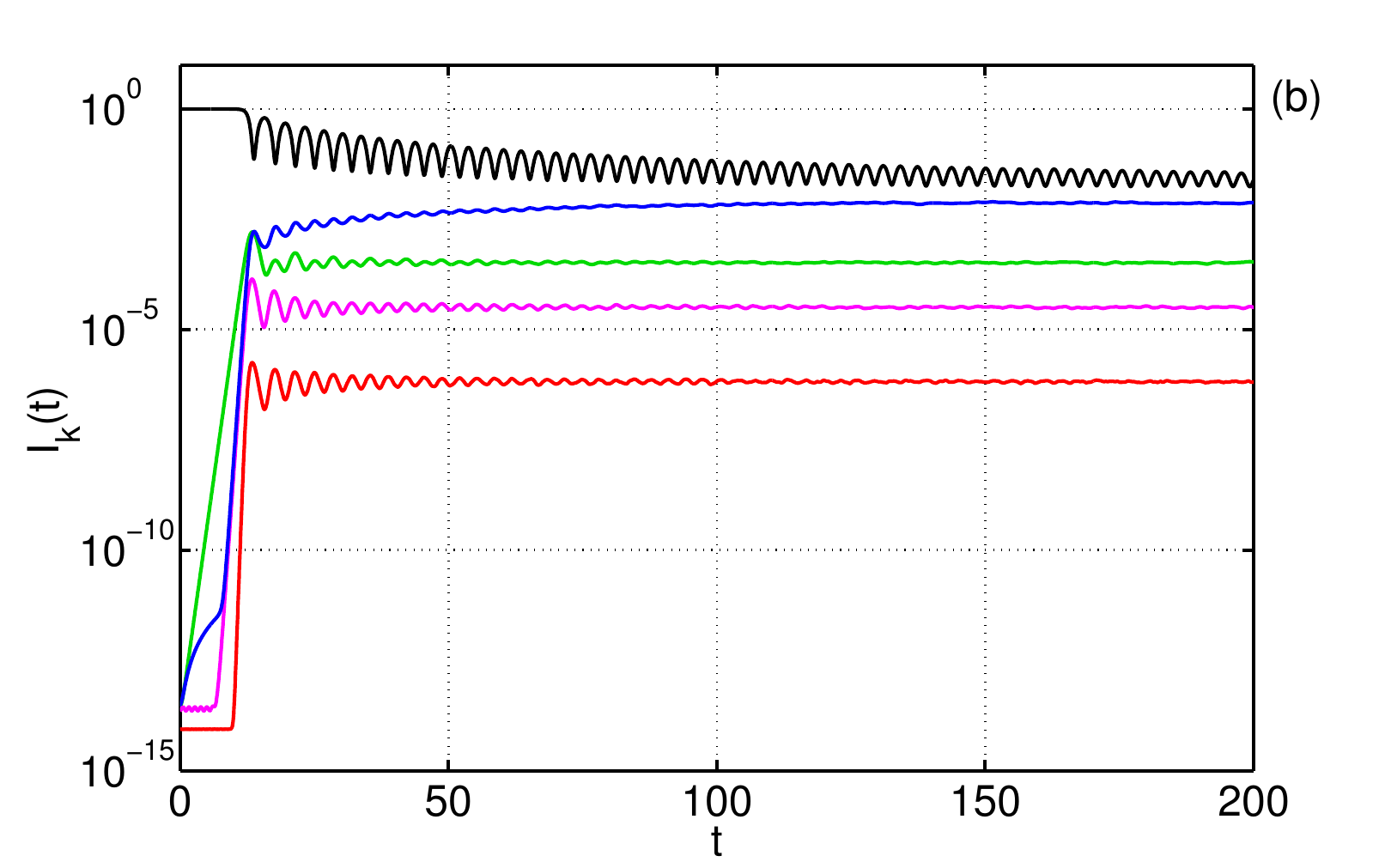}

\caption{\small {\it  (Color on-line) Graph (a): wave-action spectrum $I_{k}(t)$ at the points of time corresponding to the first local maximum of the ensemble average potential energy modulus $|\langle H_{4}\rangle|$ at $t=13.7$ (black) and the first local minimum of 
$|\langle H_{4}\rangle|$ at $t=15.8$ (blue), and also the asymptotic wave-action spectrum (red). Graph (b): evolution of wave-action spectrum $I_{k}(t)$ at $k=0$ (black), $k=0.01$ (blue), $k=1$ (green), $k=2$ (pink) and $k=4$ (red).}}
\label{fig:oscillations_spectra}
\end{figure}

The evolution of wave-action spectrum is shown on FIG.~\ref{fig:oscillations_spectra}a,b (see also more detailed graphs in the Appendix A). In full correspondence with (\ref{MI_condensate}), we observe that in the linear stage of the MI the modes with wavenumbers from the instability band $|k|<\sqrt{2}$ grow exponentially, and the fastest growth rate is achieved at $|k|=1$, while modes outside the instability band $|k|>\sqrt{2}$ do not change with time substantially. Nonlinear interaction produces multiple harmonics, so that starting from some time when the instability band is sufficiently exited, the entire spectral band -- except for the zeroth harmonic -- should rise with the highest growth for modes with integer wavenumbers. We observe such effect starting from $t\sim 5$ (see the fast growth of modes $k=2$ and $k=4$ on FIG.~\ref{fig:oscillations_spectra}b from $t\sim 5$ up to $t\sim 10$, and also wave-action spectrum at $t=13.7$ shown on FIG.~\ref{fig:oscillations_spectra}a).

\begin{figure}[t] \centering
\includegraphics[width=8.0cm]{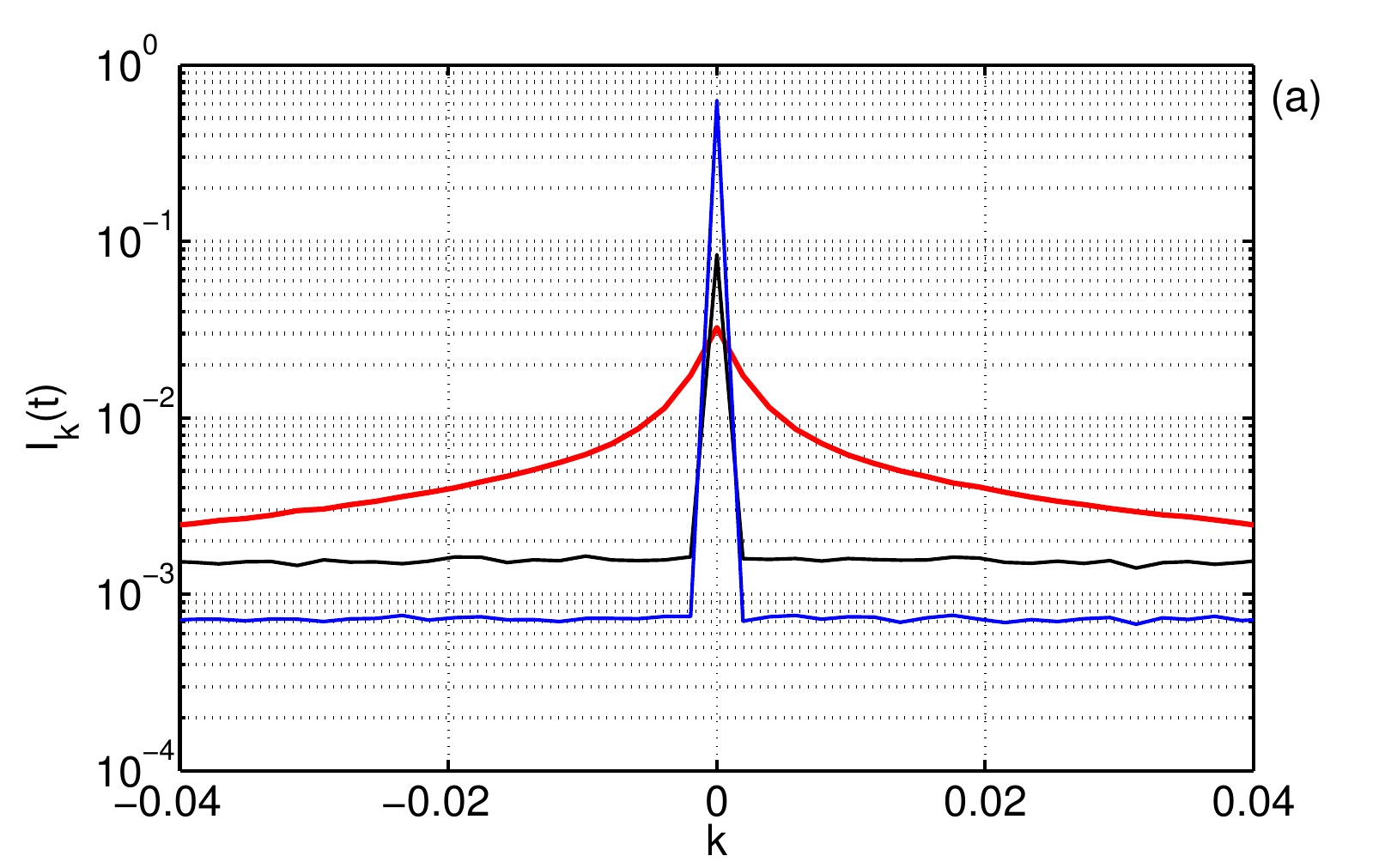}
\includegraphics[width=8.0cm]{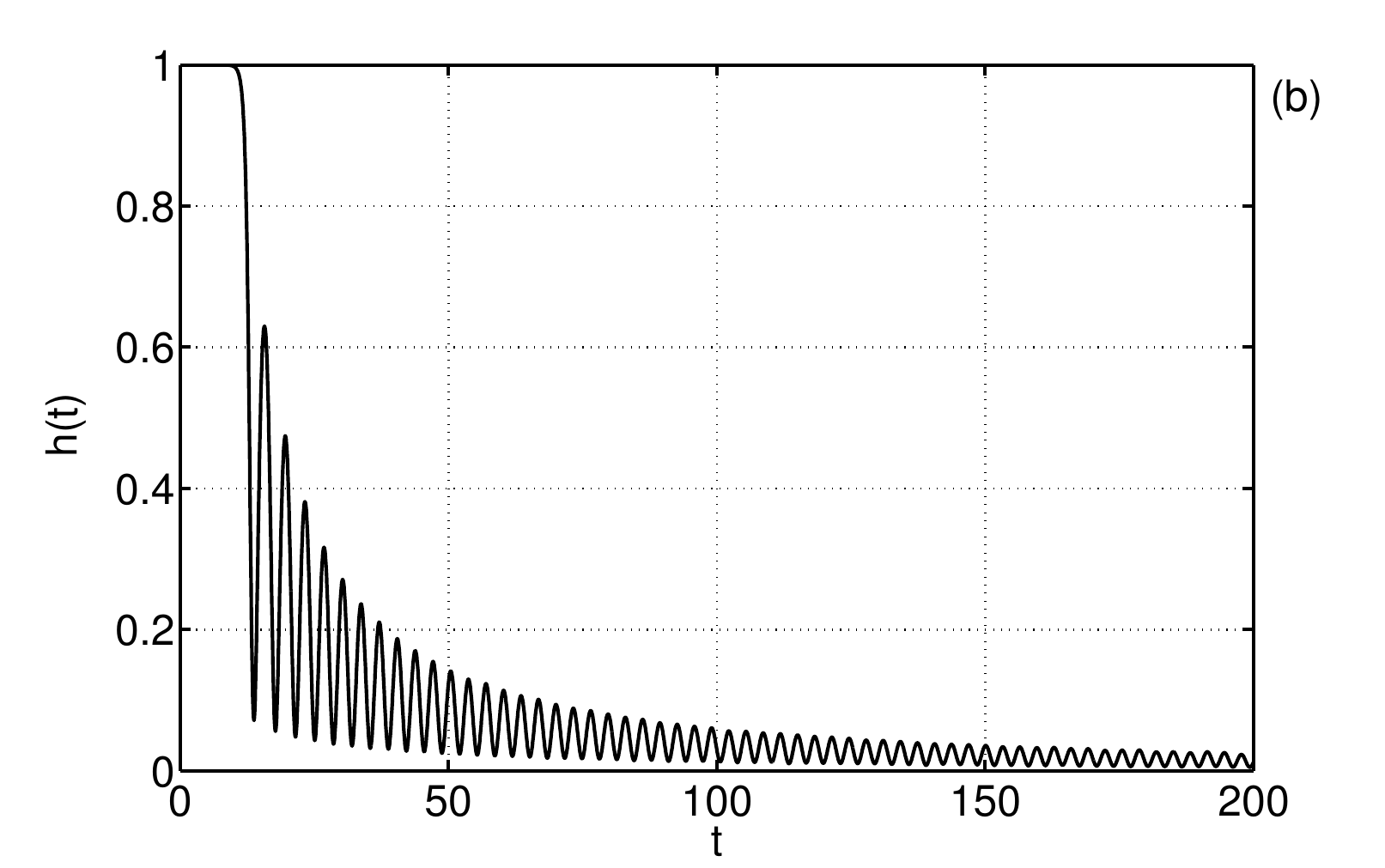}

\caption{\small {\it  (Color on-line) Graph (a): wave-action spectrum $I_{k}(t)$ at the points of time corresponding to the first local maximum of the ensemble average potential energy modulus $|\langle H_{4}\rangle|$ at $t=13.7$ (black) and the first local minimum of $|\langle H_{4}\rangle|$ at $t=15.8$ (blue), and also the asymptotic wave-action spectrum (red). Graph (b): evolution of the peak at zeroth harmonic $h(t)$ (\ref{spectra_peak}).}}
\label{fig:oscillations_spectra_peak}
\end{figure}

In the linear stage of the MI, and also for a long time in the nonlinear stage, wave-action spectrum has discontinuity at $k=0$ in the form of a high peak occupying the zeroth harmonic only (see FIG.~\ref{fig:oscillations_spectra_peak}a). This peak appears from the initial data (\ref{ensemble}), when we add the singular spectrum of the condensate (\ref{spectra_condensate}) to the continuous spectrum of noise (\ref{noise}). We observe that this peak does not fully disappear in the nonlinear stage, but decays in oscillatory way, as shown on FIG.~\ref{fig:oscillations_spectra_peak}b, where we measure the peak as the difference between the zeroth harmonic $k=0$ and the arithmetic average of the two neighbor harmonics $k=\pm 2\pi/L$,
\begin{equation}\label{spectra_peak}
h(t) = I_{0}(t) - \frac{1}{2}[I_{2\pi/L}(t) + I_{-2\pi/L}(t)].
\end{equation}

Both the peak in the spectrum $h(t)$ and the zeroth harmonic $I_{0}(t)$ take (locally) minimal values at the local maximums of potential energy modulus $|\langle H_{4}\rangle|$, and (locally) maximal values at the local minimums of $|\langle H_{4}\rangle|$ (see also FIG.~\ref{fig:spectra_0_lin} in the Appendix A). The rest of the spectrum $I_{k}(t)$, $|k|>0$, evolves similar to antiphase with $I_{0}(t)$, so that we observe decaying with time oscillatory exchange of wave action between the zeroth harmonic from one hand, and all other harmonics from the other. At late times $t\sim 150$ the peak in the spectrum disappears and the discontinuity at $k=0$ transforms into singularity of $\sim\,|k|^{-a}$ type with exponent $\alpha$ close to 2/3.

\begin{figure}[t] \centering
\includegraphics[width=8.0cm]{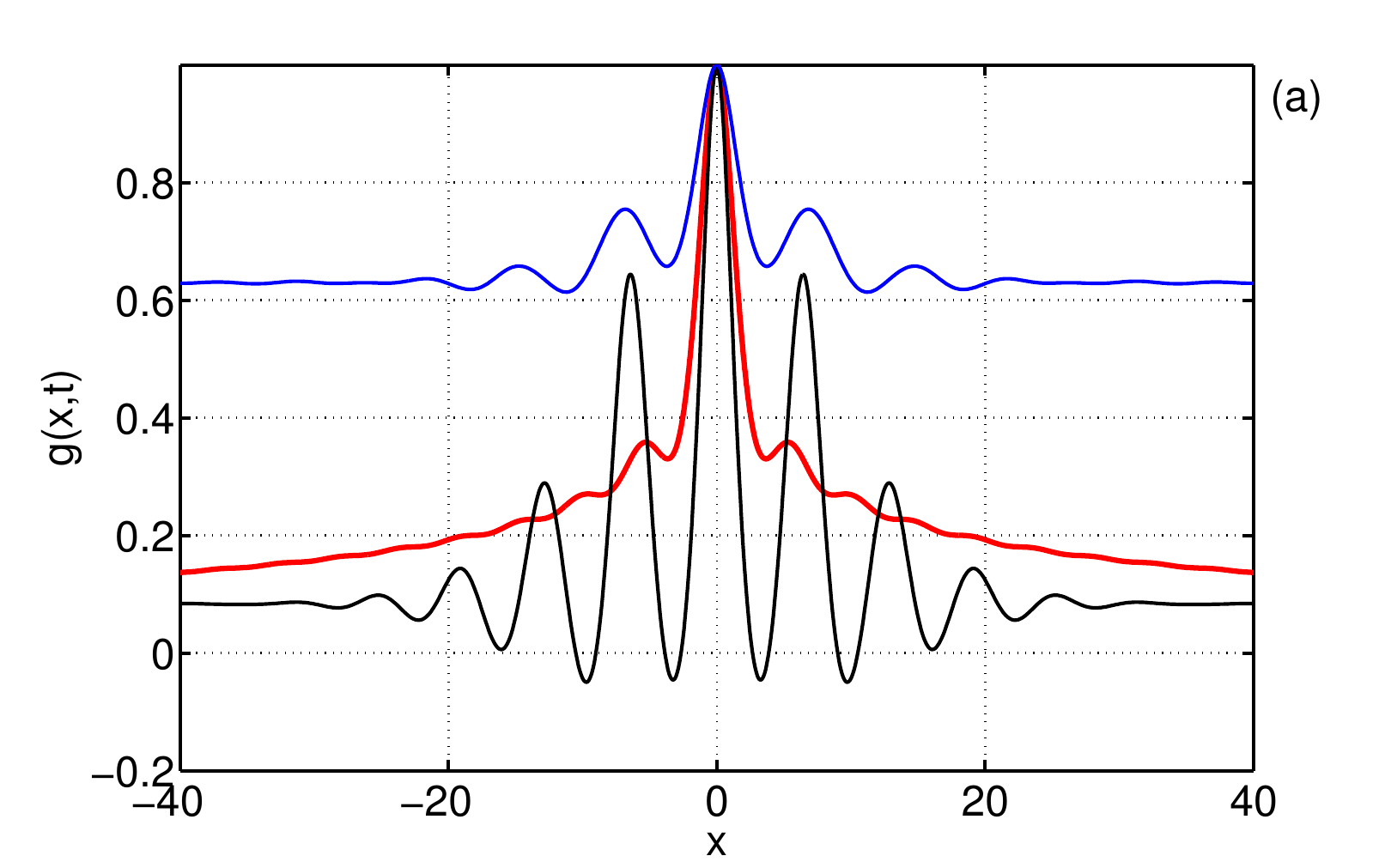}
\includegraphics[width=8.0cm]{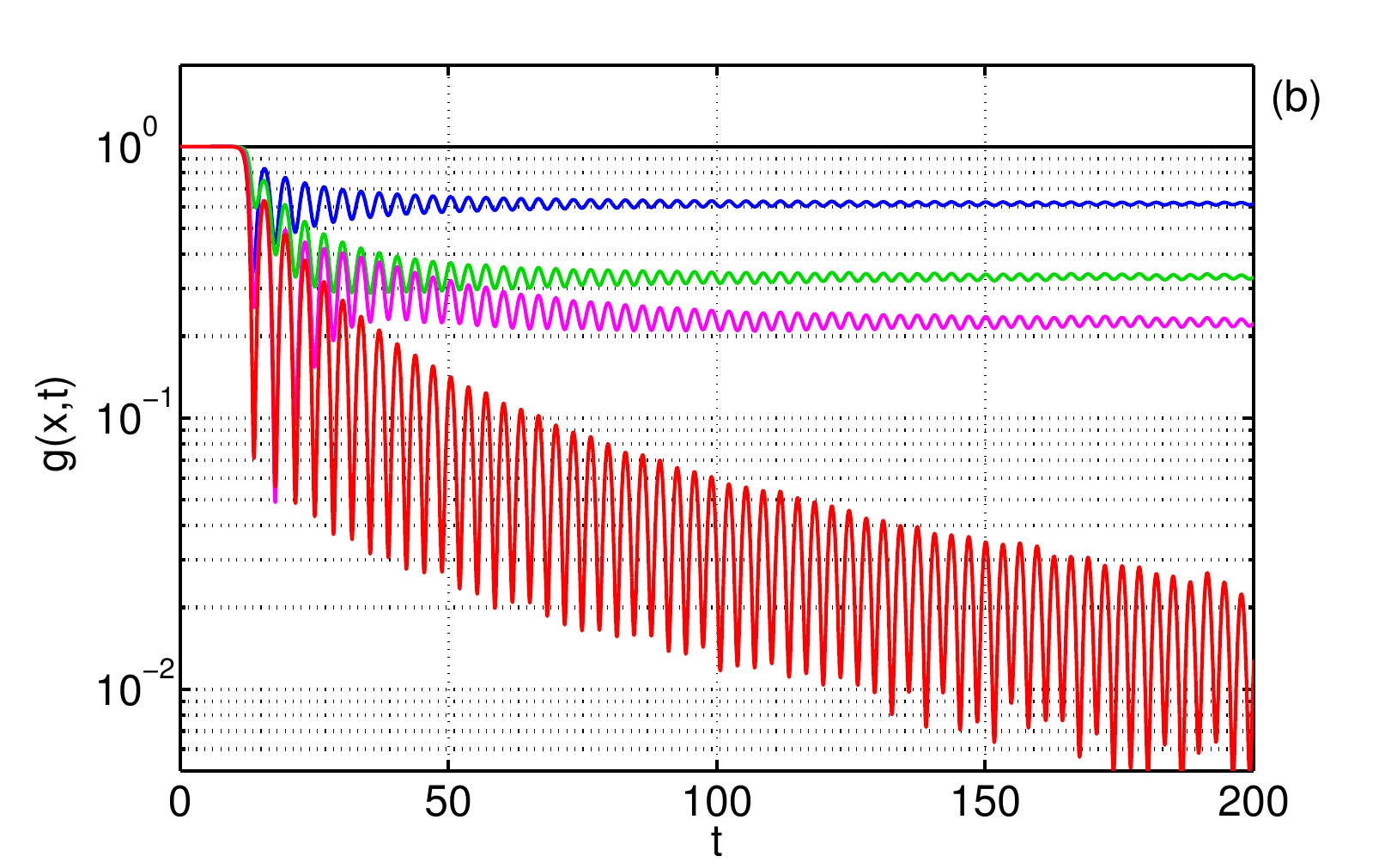}

\caption{\small {\it  (Color on-line) Graph (a): spatial correlation function $g(x,t)$ at the points of time corresponding to the first local maximum of the ensemble average potential energy modulus $|\langle H_{4}\rangle|$ at $t=13.7$ (black) and the first local minimum of $|\langle H_{4}\rangle|$ at $t=15.8$ (blue), and also the asymptotic spatial correlation function (red). Graph (b): evolution of spatial correlation function $g(x,t)$ at $x=0$ (black), $x=\pi/2$ (blue), $x=2\pi$ (green), $x=4\pi$ (pink) and at the border of the computational box $x=L/2$ (red).}}
\label{fig:oscillations_corr_x}
\end{figure}

The evolution of spatial correlation function is shown on FIG.~\ref{fig:oscillations_corr_x}a,b (see also more detailed graphs in the Appendix A). In the linear stage of the MI $t\lesssim 10$ the correlation function is close to unity $g(x,t)\approx 1$ since at this time $\Psi(x,t)=1+\zeta(x,t)$, $|\zeta(x,t)|\ll 1$. In the nonlinear stage $g(x,t)$ evolves in oscillatory way, approaching at late times to it's asymptotic form. Due to conservation of wave action, the maximum value of the correlation function at $x=0$ is fixed to unity $g(0,t)\approx 1$ (see Eq. (\ref{corr_x_0}) and FIG.~\ref{fig:oscillations_corr_x}b). At fixed $|x|>0$ the correlation function evolves similar to antiphase with potential energy modulus $|\langle H_{4}\rangle|$.

The remarkable property of spatial correlation function is that it decays at $|x|\to L/2$ to some clearly nonzero level, and this situation takes place not only in the linear stage of the MI, but also for a long time during the nonlinear stage. This behavior is the result of the presence of the peak at zeroth harmonic in wave-action spectrum. We checked this fact by subtracting from the correlation function it's level at the edges of the computational box $|x|=L/2$, 
$$
\tilde{g}(x,t) = g(x,t) - \frac{1}{2}[\lim_{x\to L/2}g(x,t) + \lim_{x\to -L/2}g(x,t)],
$$
and calculating the new wave-action spectrum as
$$
\tilde{I}_{k}(t)=\mathscr{F}[\tilde{g}(x,t)].
$$
As follows from (\ref{Fourier1})-(\ref{Fourier2}), such transformation changes only the zeroth harmonic in wave-action spectrum, so that $\tilde{I}_{k}(t)=I_{k}(t)$ for all $|k|>0$. It turns out that the peak at the zeroth harmonic in the new spectrum $\tilde{I}_{k}(t)$ is very small and changes it's sign with time. Therefore, the peak in the original wave-action spectrum directly corresponds to the nonzero level to which spatial correlation function decays at large lengths $|x|$, and vice versa.

\begin{figure}[t] \centering
\includegraphics[width=8.0cm]{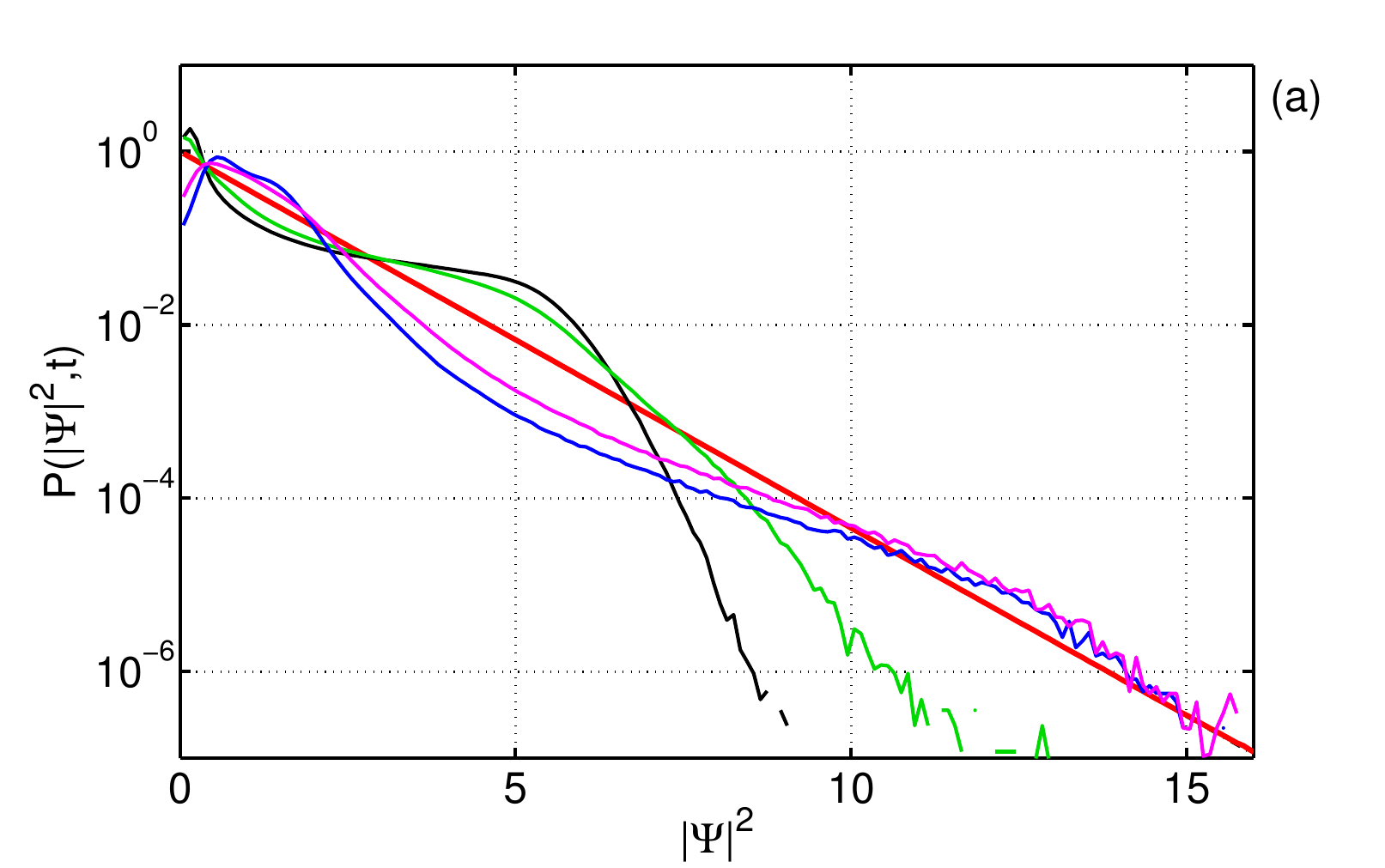}
\includegraphics[width=8.0cm]{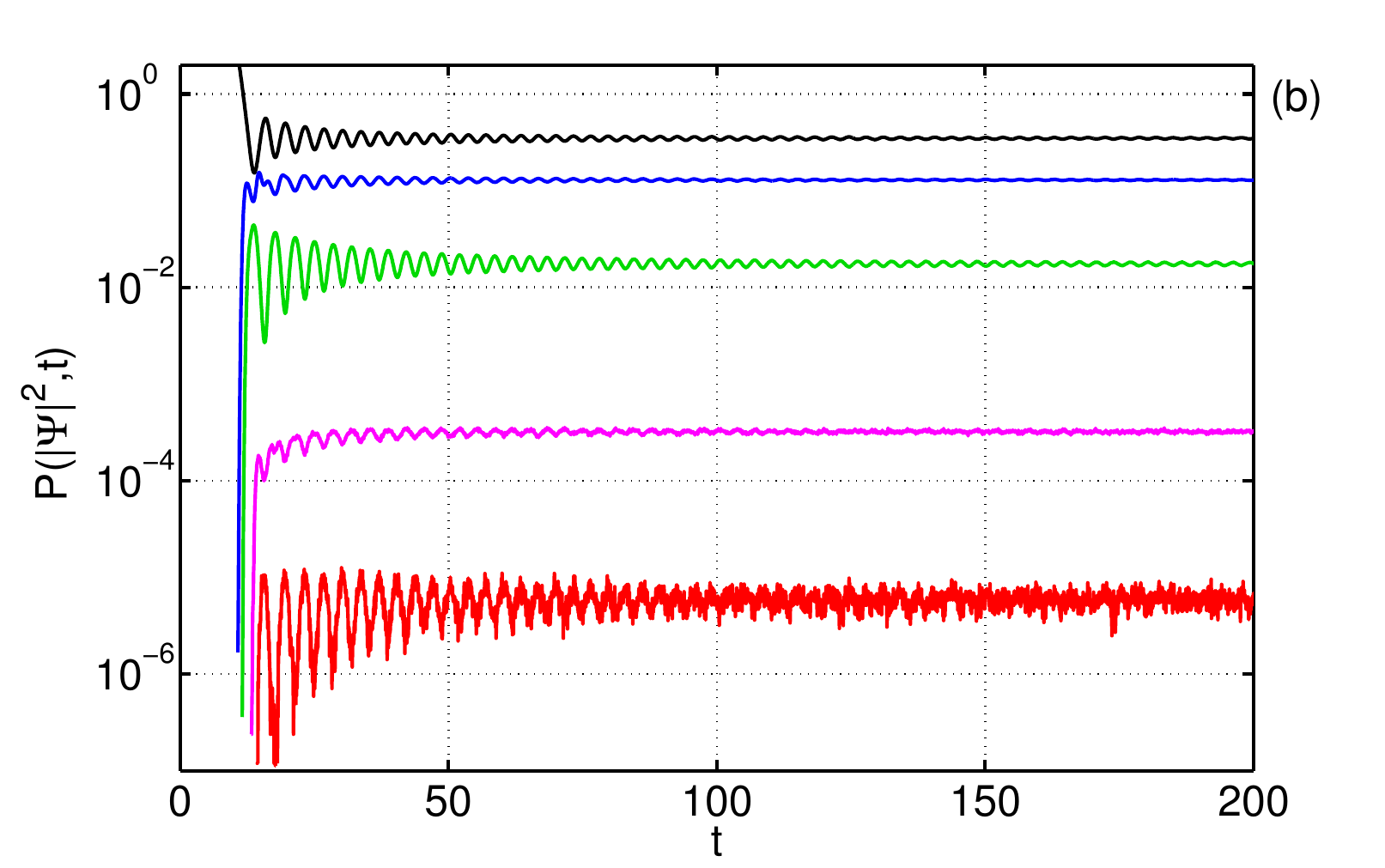}

\caption{\small {\it  (Color on-line) Graph (a): squared amplitude PDF $P(|\Psi|^{2},t)$ at the points of time corresponding to extremums of the ensemble average potential energy modulus $|\langle H_{4}\rangle|$ -- at $t=13.7$ (black line, local maximum of $|\langle H_{4}\rangle|$), $t=15.8$ (blue, minimum), $t=17.7$ (green, maximum), $t=19.6$ (pink, minimum), and the asymptotic squared amplitude PDF (thick red line). Graph (b): time dependence of the squared amplitude PDF $P(|\Psi|^{2},t)$ at $|\Psi|^{2}=1$ (black), $|\Psi|^{2}=2$ (blue), $|\Psi|^{2}=4$ (green), $|\Psi|^{2}=8$ (pink) and $|\Psi|^{2}=12$ (red).}}
\label{fig:oscillations_PDF}
\end{figure}

\begin{figure}[t] \centering
\includegraphics[width=8.0cm]{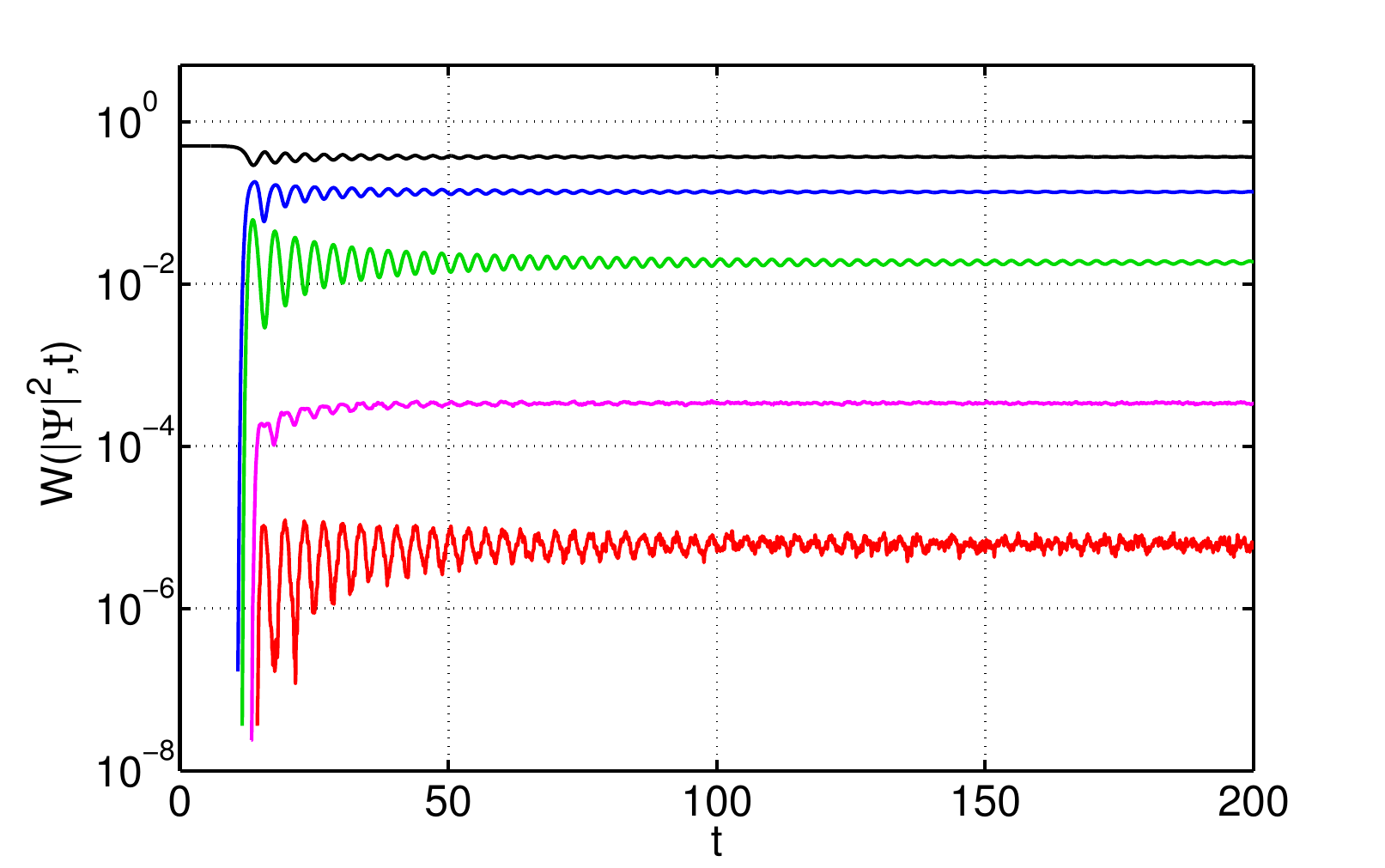}

\caption{\small {\it  (Color on-line) Probability of occurrence of waves $W(|\Psi|^{2},t)$ (\ref{WY}) with amplitudes $|\Psi|^{2}>1$ (black), $|\Psi|^{2}>2$ (blue), $|\Psi|^{2}>4$ (green), $|\Psi|^{2}>8$ (pink) and $|\Psi|^{2}>12$ (red), versus time.}}
\label{fig:oscillations_WY}
\end{figure}

The evolution of the squared amplitude PDF $P(|\Psi|^{2},t)$ is shown on FIG.~\ref{fig:oscillations_PDF}a,b. In the linear stage of the MI $t\lesssim 10$ the PDF represents a very thin peak at $|\Psi|^{2}=1$, gradually widening with time. In the nonlinear stage the PDF evolves in oscillatory way and becomes almost indistinguishable from Rayleigh PDF (\ref{Rayleigh_3}) already at $t\sim 100$. The time dependence of the PDF is quite similar to that of potential energy modulus $|\langle H_{4}\rangle|$, especially at $|\Psi|^{2}\in(0.5, 1.5)$ and $|\Psi|^{2}\in(4, 6)$, where at fixed $|\Psi|^{2}$ the PDF oscillates according to anzats (\ref{f1}) antiphase and in-phase with $|\langle H_{4}\rangle|$ respectively. FIG.~\ref{fig:oscillations_WY} shows that the probability of occurrence of waves $W(|\Psi|^{2},t)$ (\ref{WY}) evolves very similar to the PDF $P(|\Psi|^{2},t)$, oscillating around the probabilities determined by Rayleigh PDF (\ref{WYR}). More detailed graphs for the probability of rogue waves occurrence 
are shown in the Appendix 
B.

The evolution of the PDF demonstrates that in the beginning of the nonlinear stage of the MI there are two regions of squared amplitudes $3\lesssim |\Psi|^{2}\lesssim 7$ and $10\lesssim |\Psi|^{2}\lesssim 15$, where the PDF $P(|\Psi|^{2},t)$ significantly exceeds Rayleigh PDF (\ref{Rayleigh_3}). The maximum increase in comparison with Rayleigh PDF takes place for these regions of waves at the points of time corresponding to local maximums and local minimums of potential energy modulus $|\langle H_{4}\rangle|$ respectively. Thus, at the local maximums of $|\langle H_{4}\rangle|$ the probability of occurrence of waves with $|\Psi|^{2}>4$, that is approximately the center of the first region, is by about 3 times larger than Rayleigh one (\ref{WYR}) (see FIG.~\ref{fig:oscillations_WY}). Below we will refer to these waves as to "imperfect" rogue waves since they do not match the standard criterion $|\Psi|^{2}>8$.

\begin{figure}[t] \centering
\includegraphics[width=8.0cm]{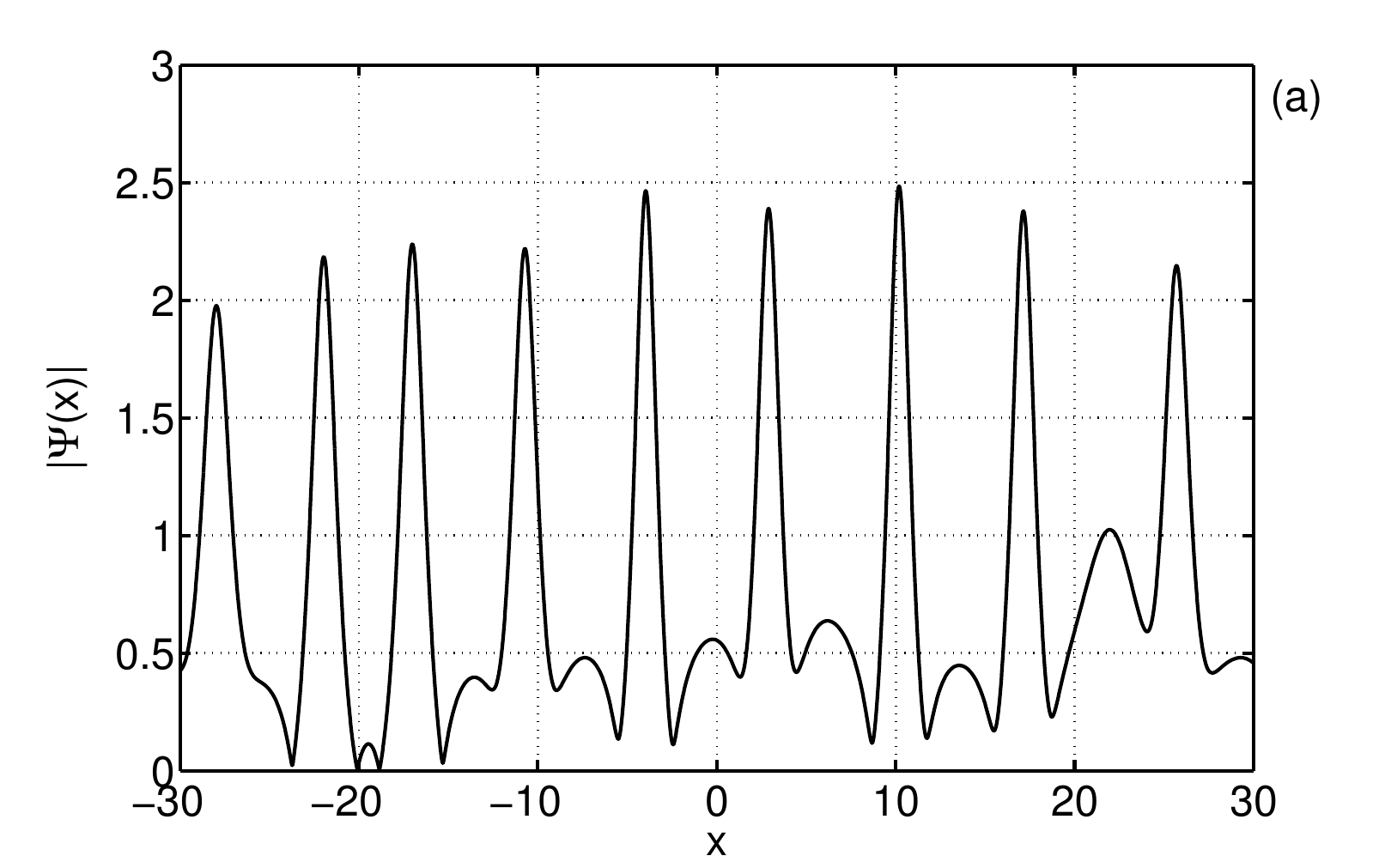}
\includegraphics[width=8.0cm]{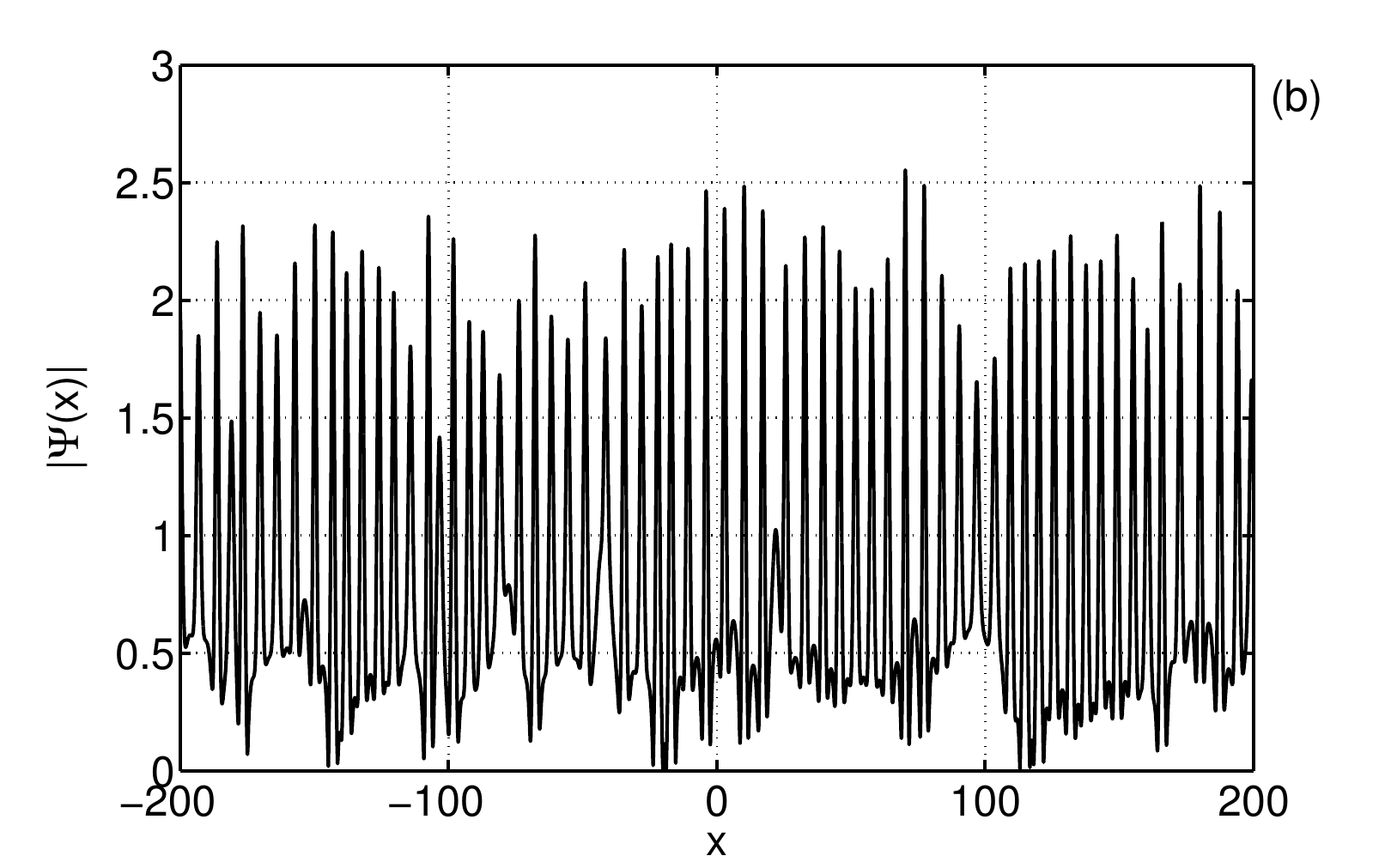}

\caption{\small {\it Spatial distribution of the amplitude $|\Psi(x)|$ at the first local maximum of the ensemble average potential energy modulus $|\langle H_{4}\rangle|$ at $t=13.7$ for one of the realizations of initial data at $x\in[-30, 30]$ (a) and $x\in[-200, 200]$ (b).}}
\label{fig:oscillations_field}
\end{figure}

The typical amplitude distribution $|\Psi(x)|$ at the first local maximum of $|\langle H_{4}\rangle|$ at $t=13.7$, shown on FIG.~\ref{fig:oscillations_field}a,b, indeed contains significant fraction of the "imperfect" rogue waves that exceed about two times the initial condensate amplitude. The "imperfect" rogue waves are the typical outcome of the MI, and we observe such waves at the first several local maximums of $|\langle H_{4}\rangle|$. In space these waves form a modulated lattice of large waves with distance between them close to characteristic length $\ell=2\pi$ of the MI. The crests of the "imperfect" rogue waves are mostly composed of imaginary part of wave field $\Psi(x)$, $|\mathrm{Re}\,\Psi|\ll |\mathrm{Im}\,\Psi|$. At the first, third, and so on, local maximums of $|\langle H_{4}\rangle|$ it is positive $\mathrm{Im}\,\Psi > 0$, and at the second, fourth, and so on, local maximums -- negative  $\mathrm{Im}\,\Psi < 0$. We observe such behavior for sufficiently long time, at least up to $t\sim 
50$. We checked these facts directly and also by measuring the evolution of the PDF for real $\mathrm{Re}\,\Psi$ and imaginary $\mathrm{Im}\,\Psi$ parts of wave field. We will report our results for the evolution of such PDFs in the next publication.

The similar scenario is realized for the Akhmediev breather that corresponds to the maximum growth rate of the MI. The Akhmediev breathers \cite{akhmediev1987exact, akhmediev2009extreme, akhmediev2009waves} are the solutions of the NLS equation that are periodic in space and localized in time,
\begin{equation}\label{Akhmediev}
\Psi_{AB}(x,t)=e^{-2i\phi} \frac{\cosh(\omega t-2i\phi)-\cos(\phi)\cos(bx)}{\cosh(\omega t)-\cos(\phi)\cos(bx)},
\end{equation}
where $0<\phi<\pi/2$ is free parameter and 
\begin{equation}\label{Akhmediev2}
\omega=\sin(2\phi),\quad b=\sqrt{2}\sin{\phi}.
\end{equation}
These solutions appear at $t\to -\infty$ on the background of the condensate $\Psi=1$, develop with the growth rate $\omega$, become maximal at $t=0$, and then decay into the condensate $e^{-4i\phi}$ as $t\to +\infty$. At $t=0$ these solutions have phase $(-1)\times e^{-2i\phi}$ at the points of their maximal amplitude. 

Thus, for $\phi=\pi/4$ both the growth rate $\omega$ and the period of the Akhmediev breather $2\pi/b$ are equal to the maximum growth rate $\gamma_{0}=1$ and the characteristic length $\ell=2\pi$ of the MI respectively. At $t=0$ this solution is purely imaginary, and at it's maximums $x=2\pi n$ where $n$ is integer, the imaginary part is positive $\mathrm{Im}\,\Psi > 0$. After the decay this solution changes the phase of the condensate to $e^{i\pi}=-1$. Thus, the subsequent Akhmediev breather -- if it appears -- should have negative imaginary part at it's maximums $\mathrm{Im}\,\Psi < 0$, the third Akhmediev breather -- positive imaginary part, and so on. However, it is unclear how these solutions may appear one after another with short interval between them from random statistically homogeneous is space noise. The interval between the maximal elevation of these solutions must be equal then to the period of the oscillations of potential energy modulus $|\langle H_{4}\rangle|$, which 
is close to 4 in the beginning of the nonlinear stage and approaches to $\pi$ with time.

There are also significant distinctions between the Akhmediev breather solution with $\phi=\pi/4$ and the "imperfect" rogue waves. Thus, spatial correlation function of the Akhmediev breather is periodic with period $2\pi$, is maximal at $x=2\pi n$ where $n$ is integer, and is equal to unity at these maximums. For the Akhmediev breather with small perturbations it is natural to expect that the corresponding spatial correlation function should have pronounced peaks at the same points, and the magnitude of these peaks should sufficiently slowly decay with distance $|x|$. In our experiments we observe that at the first local maximum of $|\langle H_{4}\rangle|$ spatial correlation function have just 5 pronounced peaks at $x=2\pi n$ for $n=0,\pm 1,\pm 2$, and at the second local maximum - just 3 peaks for $n=0,\pm 1$ (see FIG.~\ref{fig:oscillations_corr_x} and also FIG.~\ref{fig:oscillations_sp_corr01},~\ref{fig:oscillations_sp_corr03} in the Appendix A). The magnitude of these peaks quickly decays with distance 
$|x|$, and then spatial correlation function soon becomes almost constant and equal to $\sim\,0.1$. It is interesting that at the local maximums of $|\langle H_{4}\rangle|$ spatial correlation function takes (locally in time) minimal values.

Wave-action spectrum of the Akhmediev breather with $\phi=\pi/4$ is composed of harmonics with only integer wavenumbers. Thus, the spectrum of this solution immersed in a field of small perturbations should have very pronounced peaks at integer wavenumbers. However, in our experiments we observe in the spectrum only 5 peaks $k=0,\pm 1, \pm 2$ at the first local maximum of $|\langle H_{4}\rangle|$, and just 3 peaks $k=0,\pm 2$ at the second local maximum of $|\langle H_{4}\rangle|$ (in the latter case at $k=\pm 1$ the spectrum takes local minimums; see FIG.~\ref{fig:oscillations_spectra} and also FIG.~\ref{fig:spectra_0_lin},~\ref{fig:oscillations_sp_corr01},~\ref{fig:oscillations_sp_corr03} in the Appendix A). It is interesting that at the local maximums of $|\langle H_{4}\rangle|$ wave-action spectrum has (locally in time) minimal presence of the zeroth harmonic with the rest of the spectrum maximally excited.

In the beginning of the nonlinear stage of the MI and at the local minimums of potential energy modulus $|\langle H_{4}\rangle|$ we observe that waves with $|\Psi|^{2}>12$, that is approximately the center of the second region of squared amplitudes $10\lesssim |\Psi|^{2}\lesssim 15$, appear about two times more frequently than predicted by Rayleigh PDF (\ref{WYR}). These waves are rogue waves. It is noteworthy that the "standard" rogue waves $|\Psi|^{2}>8$ appear in the beginning of the nonlinear stage even less frequently than Rayleigh prediction (\ref{WYR}) (see FIG.~\ref{fig:oscillations_WY_lin01},~\ref{fig:oscillations_WY_lin05} in the Appendix B). 

\begin{figure}[t] \centering
\includegraphics[width=8.0cm]{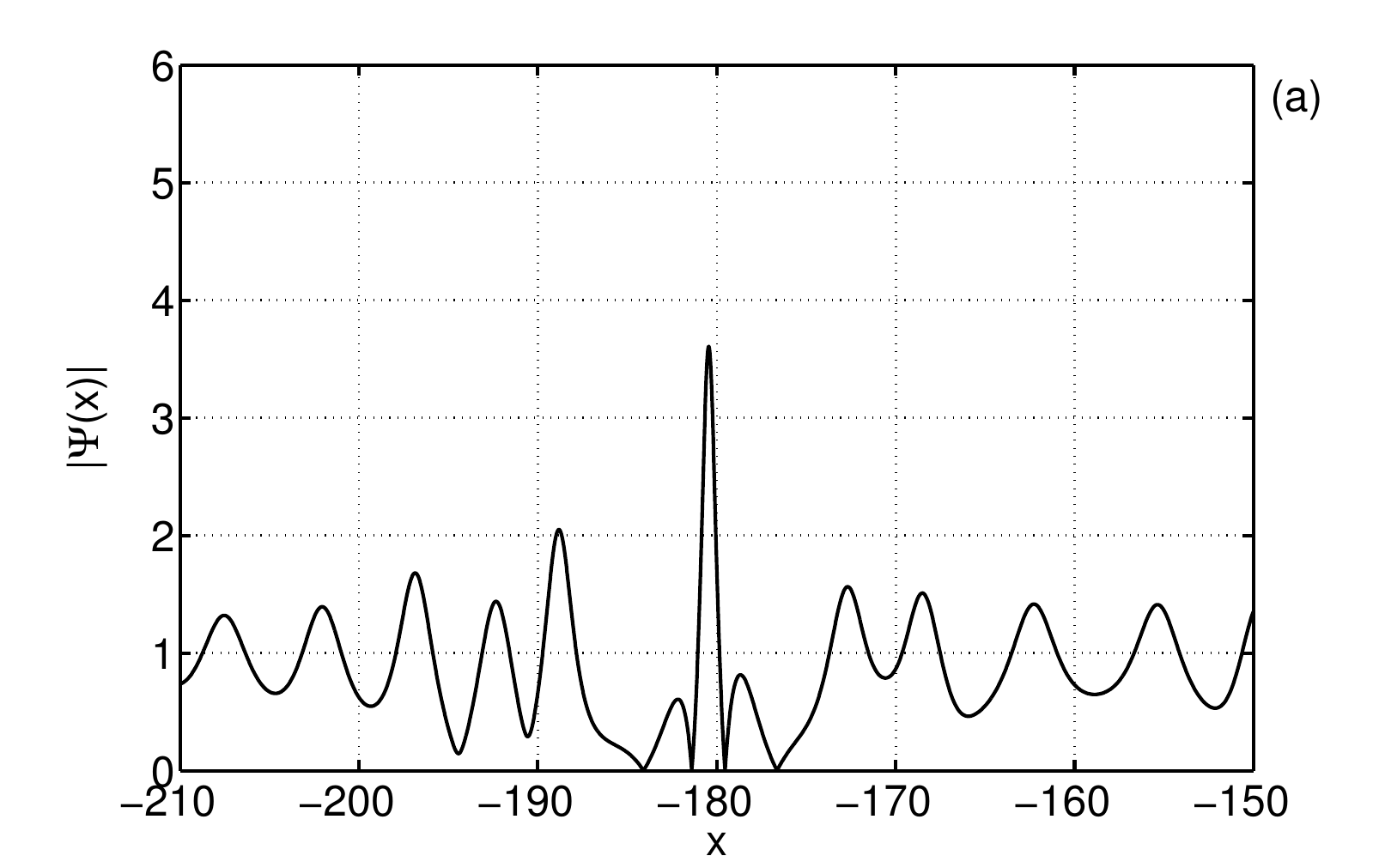}
\includegraphics[width=8.0cm]{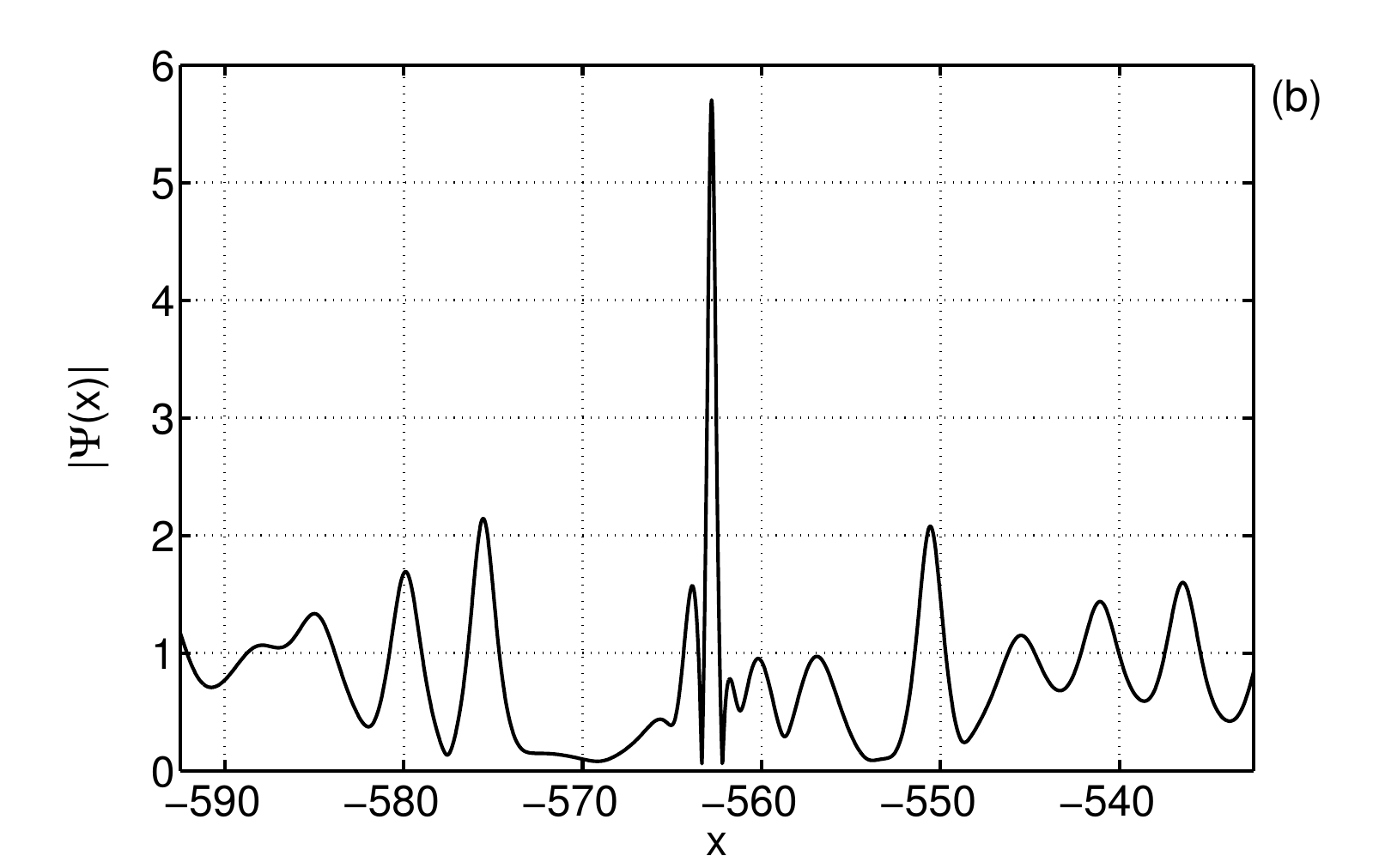}

\caption{\small {\it Spatial distribution of the amplitude $|\Psi(x)|$ for rogue wave events. Graph (a): a typical rogue wave occurred near the local minimum of $|\langle H_{4}\rangle|$. The maximum amplitude $\max|\Psi|\approx 3.6$ is achieved at $t\approx 19.8$, that is near the second local minimum of $|\langle H_{4}\rangle|$ at $t\approx 19.6$; the duration of this event was $\Delta T\sim 1$. Graph (b): extremely large rogue wave occurred $t\approx 715.1$ with maximum amplitude $\max|\Psi|\approx 5.7$. The duration of this event was $\Delta T\sim 0.5$.}}
\label{fig:large_wave_event}
\end{figure}

These "large" rogue waves $10\lesssim |\Psi|^{2}\lesssim 15$ represent in space a singular high peak with full width at half maximum of about $x_{FW}\sim 1$ and duration in time of about $\Delta T\sim 1$ (see FIG.~\ref{fig:large_wave_event}a). These peaks are very rare events and appear on the background of perturbed wave field that is usually less than $|\Psi|<1.5$ in amplitude (see FIG.~\ref{fig:oscillations_PDF}a). Statistically at this time wave field is strongly correlated, spatial correlation function takes (locally in time) maximal values and wave-action spectrum has (locally in time) maximal zeroth harmonic with the rest of the spectrum minimally excited (see FIG.~\ref{fig:oscillations_spectra},~\ref{fig:oscillations_corr_x}, and also FIG.~\ref{fig:spectra_0_lin},~\ref{fig:oscillations_sp_corr02},~\ref{fig:oscillations_sp_corr04} in the Appendix A). 

The crests of the "large" rogue waves are mostly composed of real part of wave field $\Psi(x)$, $|\mathrm{Im}\,\Psi|\ll |\mathrm{Re}\,\Psi|$. At the first, third, and so on, local minimum of $|\langle H_{4}\rangle|$ it is negative $\mathrm{Re}\,\Psi<0$, and at the second, fourth, and so on, local minimum -- positive $\mathrm{Re}\,\Psi>0$. We observe such behavior for sufficiently long time, at least up to $t\sim 50$. It is interesting that the Peregrine solution \cite{peregrine1983water} of the NLS equation has similar property. This localized in space and time algebraic solution
\begin{equation}\label{Peregrine}
\Psi_{P}(x,t)=1-\frac{4(1+2 i t)}{1+2x^{2}+4t^{2}},
\end{equation}
appears at $t\to -\infty$ on the background of the condensate $\Psi=1$, at the time of its maximum elevation $t=0$ is purely real and equal to $\Psi_{P}(0,0)=-3$ at it's maximum amplitude, and then decay back to the condensate $\Psi=1$. However, the amplitudes of the "large" rogue waves $10\lesssim |\Psi|^{2}\lesssim 15$ are slightly larger than the maximum amplitude of the Peregrine solution $\max|\Psi_{P}|^{2}=9$.

We also observe the occurrence of extremely large waves with amplitudes up to $|\Psi|\sim 6$. These waves represent in space a singular high peak with full width at half maximum of about $x_{FW}\sim 1$ and duration in time of about $\Delta T\sim 0.5$ (see FIG.~\ref{fig:large_wave_event}b). It is interesting that we often observe such waves very similar to each other, despite the fact that they are generated at different times from different realizations of initial data. These very large waves are extremely rare events, and the accuracy of our simulations does not allow us to determine the evolution of the PDF and the probability of occurrence for such waves with time. The only way we can observe the influence of these waves on the PDF is the averaging the PDF over sufficiently long time interval. The latter allows us to conclude that in the asymptotic turbulent state these waves are distributed according to Rayleigh PDF (\ref{Rayleigh_3}) (see FIG.~\ref{fig:PDF}).


\section{Conclusions and acknowledgements.}

In the current publication we performed the systematic study of the statistics of the MI developing from the condensate solution in the framework of the focusing NLS equation. Our goal was two-fold: first, to study the asymptotic stationary turbulent state to which the system evolves at late times. The investigation of this state has fundamental importance as the example of stationary integrable turbulence, that can be considered as thermodynamically equilibrium state defined by infinite number of conserved quantities. Second, to examine the beginning of the nonlinear stage of the MI and the subsequent evolution towards the asymptotic state. This study is important in relation to rogue waves phenomenon \cite{kharif2003physical, dysthe2008oceanic, onorato2013rogue}.

We found that the asymptotic integrable turbulence is "moderately strong", with kinetic energy $\langle H_{d}\rangle \approx 0.5$ and potential energy $\langle H_{4}\rangle \approx -1$. The PDF of wave amplitudes and their momenta in this state are Rayleigh ones (\ref{Rayleigh_3}) and (\ref{MnR1}) with a very good accuracy. These results would be natural for a random wave field governed by linear equations, that has Gaussian statistics. The result $\langle H_{4}\rangle \approx -1$ is itself truly remarkable, since it indicates that the cumulant in the asymptotic turbulent state might be zeroth (\ref{cumulant3}). The calculation of the cumulant is a cumbersome problem, but we hope to publish these results in the near future.

Note that in the recent publication \cite{walczak2015optical} the authors also studied the asymptotic state of the integrable turbulence in the framework of the focusing NLS equation, but with incoherent wave field initial conditions. In this case the PDF significantly deviates from Rayleigh one (\ref{Rayleigh_3}) and has ``fat tails'', while the probability of occurrence of large waves exceeds the corresponding Rayleigh distribution (\ref{WYR}) by orders of magnitude. As we study the integrable system, it is not surprising that it's long-time evolution depends on the initial conditions. However, further study is necessary to characterize this dependence.

At small wavenumbers $|k|\le 0.15$ the asymptotic wave-action spectrum has power-law dependence $I_{k}\sim |k|^{-\alpha}$ with exponent $\alpha$ close to 2/3. At $k=0$ the spectrum has finite value. The modes with $|k|\le 0.15$ have very large scales in the physical space, contain about 40\% of wave action, less than 1\% of kinetic energy and about 10\% of potential energy, and can be called "quasi-condensate". In the region $0.15\le |k|\le 1.5$ the spectrum decays monotonically, with another area of power-law dependence $I_{k}\sim |k|^{-\alpha}$ at $0.4 \le |k|\le 1$ with exponent $\alpha$ close to 1/2. In the small vicinity of $|k|=\sqrt{2}$ the spectrum decays sharply, and starting from $|k|> 1.5$ the spectrum decays close to exponential law $I_{k}\sim e^{-\beta|k|}$, $\beta\approx 0.9$. Modes with $0.15\le |k|\le 1.5$ contain about 55\% of wave action, and about 60\% of kinetic and potential energies, while the exponentially decaying modes $|k|>1.5$ have about 5\% of wave action, about 40\% of kinetic 
energy and about 30\% of potential energy. The 
asymptotic spatial correlation function has full width at half maximum $x_{corr}\approx 4$, is close to Gaussian at $|x|<x_{corr}/2$, and slowly decays with length $|x|\to +\infty$ as $1/|x|$. 

We think that after a very long evolution computed on a very large computational box $L$ the two power-law regions in wave-action spectrum may merge in one $I_{k}\sim |k|^{-\alpha}$, $|k|\le 1$, with some shared exponent $\alpha$. However, so far we are unable to check this hypothesis.

Approaching to the asymptotic turbulent state is a long oscillatory process. During this process the moments $M^{(n)}(t)$ (\ref{Mn}) with exponents $n\neq 2$, and also kinetic and potential energies, oscillate with time according to anzats (\ref{f1}) around their asymptotic (Rayleigh) values. The amplitudes of these oscillations decay with time $t$ as $t^{-3/2}$, the phases contain the nonlinear phase shift that decays as $t^{-1/2}$, and the period of the oscillations is equal to $\pi$. Thus, the frequency of the oscillations is equal to the double maximum growth rate of the MI, $s\approx 2\gamma_{0}$. So far we do not have analytical model to describe this beautiful phenomenon. We hope that some study can be done in the approximation of "quasi-kinetic" equation, developed in the publications of S.Y. Annenkov and V.I. Shrira \cite{annenkov2006role, annenkov2009evolution}.

During the evolution towards the asymptotic state, wave-action spectrum, spatial correlation function and the PDF also evolve with time in oscillatory way, approaching to their asymptotic forms at late times. The "turning points" for the evolution of these functions, where their motion with time changes to roughly the opposite, approximately coincide with the extremums of the ensemble average potential energy modulus $|\langle H_{4}\rangle|$. 

The zeroth harmonic in wave-action spectrum $I_{0}(t)$ evolves similar to antiphase, and the rest of the spectrum - similar to in-phase with $|\langle H_{4}\rangle|$. Thus, we observe decaying with time oscillatory exchange of wave action between the zeroth harmonic and the rest of the spectrum. During this exchange the zeroth harmonic decays from $I_{0}\approx 1$ at $t=0$ to it's asymptotic value $I_{0}\approx 0.032$, and at the same time the quasi-condensate modes increase. In the beginning of the MI the spectrum has discontinuity at $k=0$ in the form of a high peak occupying the zeroth harmonic only. This peak appears from the initial data (\ref{ensemble}), (\ref{noise}). The peak does not disappear with the arrival to the nonlinear stage of the MI, but decays in oscillatory way remaining detectable for a long time in the nonlinear stage. After the peak finally disappears, the discontinuity in the spectrum transforms to power-law dependence $\,\sim |k|^{-\alpha}$ at $|k|\le 0.15$ with exponent $\alpha$ 
close to 2/3. While the peak in the spectrum is present, spatial correlation function decays with length $|x|\to +\infty$ to some nonzero level determined by the magnitude of the peak. After the discontinuity at $k=0$ transforms to quasi-condensate, spatial correlation function decays with length $|x|\to +\infty$ as $1/|x|$. At fixed $|x|>0$ spatial correlation function $g(x,t)$ evolves similar to antiphase with $|\langle H_{4}\rangle|$.

We now come to the impact of our results on the theory of rogue waves. Our study was inspired in part by the idea that the MI of a narrow-banded spectrum could be an effective mechanism for rogue waves formation \cite{kharif2003physical, dysthe2008oceanic, onorato2013rogue}. We tried to check this notion in the framework of the most popular model for the description of rogue waves, which is the focusing NLS equation. Our results turned out to be dubious. We found that after a very long development of the MI we arrive to the asymptotic stationary turbulent state that has Rayleigh PDF of wave amplitudes (\ref{Rayleigh_3}). Probability of occurrence of rogue waves in this state is the same as in a system of non-interacting waves with random phases (\ref{WYR}). 

However, this asymptotic state is reached after a very long oscillatory evolution. In the beginning of this process the PDF 
$P(|\Psi|^{2},t)$ evolves significantly and in oscillatory way, and becomes significantly larger than Rayleigh PDF (\ref{Rayleigh_3}) for two separate regions of squared amplitudes $3\lesssim |\Psi|^{2}\lesssim 7$ and $10\lesssim |\Psi|^{2}\lesssim 15$. The maximal increase of the PDF for these regions of waves takes place at the points of time corresponding to local maximums and local minimums of potential energy modulus $|\langle H_{4}\rangle|$ respectively.

In the beginning of the nonlinear stage of the MI and at the local maximums of $|\langle H_{4}\rangle|$ the waves from the first region appear about three times more frequently for $|\Psi|^{2}>4$ than predicted by Rayleigh PDF (\ref{WYR}). These waves are "imperfect" rogue waves since their amplitudes are smaller than rogue waves criterion $|\Psi|^{2}>8$. The "imperfect" rogue waves are the typical outcome of the MI that we observe at the first several local maximums of $|\langle H_{4}\rangle|$. In space these waves form a modulated lattice of large waves with distance between them close to characteristic length $\ell=2\pi$ of the MI. The crests of the "imperfect" rogue waves are mostly composed of imaginary part of wave field $\Psi(x)$, $|\mathrm{Re}\,\Psi|\ll |\mathrm{Im}\,\Psi|$. This imaginary part is positive $\mathrm{Im}\,\Psi > 0$ at the first, third, and so on, local maximums of $|\langle H_{4}\rangle|$, and negative $\mathrm{Im}\,\Psi < 0$ at the second, fourth, and so on, local maximums. 

The similar scenario is realized for the Akhmediev breather (\ref{Akhmediev}) that corresponds to the maximum growth rate of the MI. At the time of it's maximum elevation this solution is purely imaginary, and at it's maximums the imaginary part is positive. After it's decay, this solution changes the phase of the condensate by $e^{i\pi}=-1$, so that if there appears the following Akhmediev breather, then it should have negative imaginary part at it's crests, the third Akhmediev breather should have positive imaginary part, and so on. 

However, it is unclear how these solutions my appear from random statistically homogeneous in space noise with short interval one after another. This interval then coincides with the period of the oscillations of potential energy, is close to 4 in the beginning of the nonlinear stage of the MI and approaches to $\pi$ with time. Also, spatial correlation function of the "imperfect" rogue waves significantly decreases after a few characteristic lengths $\ell$ of the MI, and takes (locally in time) minimal values. For the Akhmediev breather it remains periodic. Wave-action spectrum of the "imperfect" rogue waves has (locally in time) minimal zeroth harmonic with the rest of the spectrum maximally excited, and it is not very similar to the spectrum of the Akhmediev breather.

In the beginning of the nonlinear stage of the MI and at the local minimums of $|\langle H_{4}\rangle|$ the waves from the second region $10\lesssim |\Psi|^{2}\lesssim 15$ appear by about two times more frequently for $|\Psi|^{2}>12$ than predicted by Rayleigh PDF (\ref{WYR}). These rogue waves are very rare events and represent in space a singular high peak with full width at half maximum of about $x_{FW}\sim 1$ and duration in time of about $\Delta T\sim 1$, and appear on the background of perturbed wave field that is usually less than $|\Psi|<1.5$ in amplitude. Statistically at this time spatial correlation function takes (locally in time) maximal values, and wave-action spectrum has (locally in time) maximal zeroth harmonic with the rest of the spectrum minimally excited. 

The crests of the "large" rogue waves are mostly composed of real part of wave field $\Psi(x)$, $|\mathrm{Im}\,\Psi|\ll |\mathrm{Re}\,\Psi|$. This real part is negative $\mathrm{Re}\,\Psi < 0$ at the first, third, and so on, local minimums of $|\langle H_{4}\rangle|$, and positive $\mathrm{Re}\,\Psi > 0$ at the second, fourth, and so on, local minimums. The Peregrine solution (\ref{Peregrine}) has similar property: at the time of it's maximal elevation this solution is purely real and is negative at it's maximum amplitude $\Psi_{P}(0,0)=-3$. However, it's maximum amplitude is slightly smaller than the amplitudes of the "large" rogue waves.

We also observe extremely large rogue waves with up to six-fold increase in comparison with the initial condensate amplitude. These waves represent in space a singular high peak with full width at half maximum of about $x_{FW}\sim 1$ and duration in time of about $\Delta T\sim 0.5$, and often look quite similar to each other despite the fact that they appear at different times and from different realizations of initial data. However, the accuracy of our simulations is too limited to determine the time dependence of the PDF and the probability of occurrence for these waves.

We think that our results demonstrate that the MI of the condensate in the framework of the focusing NLS equation is not a very promising model for the studies of rogue waves phenomena. From one hand, our study reveals that the maximum increase in the probability of rogue waves occurrence is just about 2 times in comparison with Rayleigh predictions (\ref{WYR}). This is not a very promising value. From the other hand, the NLS equation is too specific model due to it's complete integrability that implies the conservation of infinite number of invariants. Thus, even a small correction to NLS equation that violates it's integrability may significantly change the scenario of rogue waves formation. We hope to illustrate this idea in our next publications.

The authors thank E. Kuznetsov and A. Dyachenko for valuable discussions concerning this publication, M. Fedoruk for access to and V. Kalyuzhny for assistance with Novosibirsk Supercomputer Center. This work was done in the framework of Russian Science Foundation grant No. 14-22-00174 "Wave turbulence: theory, numerical simulation, experiment".


\vspace{1cm}

\appendix
\section{Graphs of the evolution of wave-action spectrum and spatial correlation function.}

In this Appendix we provide more detailed graphs for the evolution of the zeroth harmonic $I_{0}(t)$, and also for wave-action spectrum and spatial correlation function at the first two local maximums and minimums of potential energy modulus $|\langle H_{4}\rangle|$.

\begin{figure}[H] \centering
\includegraphics[width=16cm]{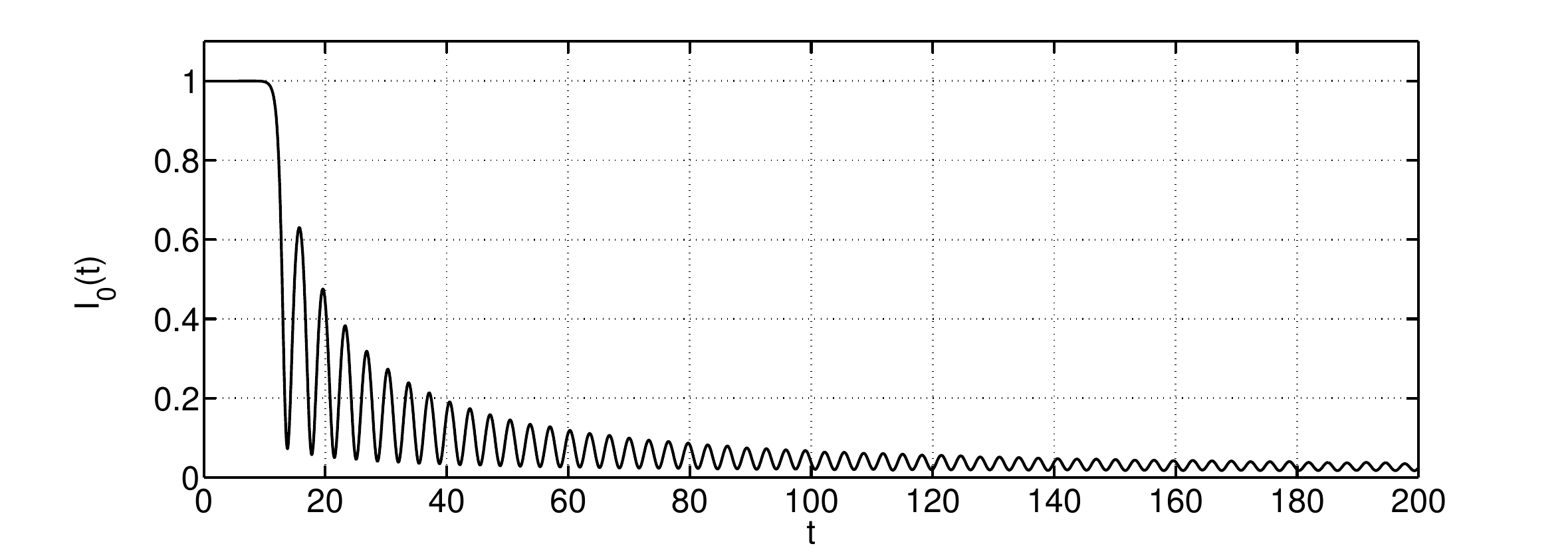}

\caption{\small {\it Evolution of the zeroth harmonic of wave-action spectrum $I_{0}(t)$.}}
\label{fig:spectra_0_lin}
\end{figure}

\begin{figure}[H] \centering
\includegraphics[width=8.0cm]{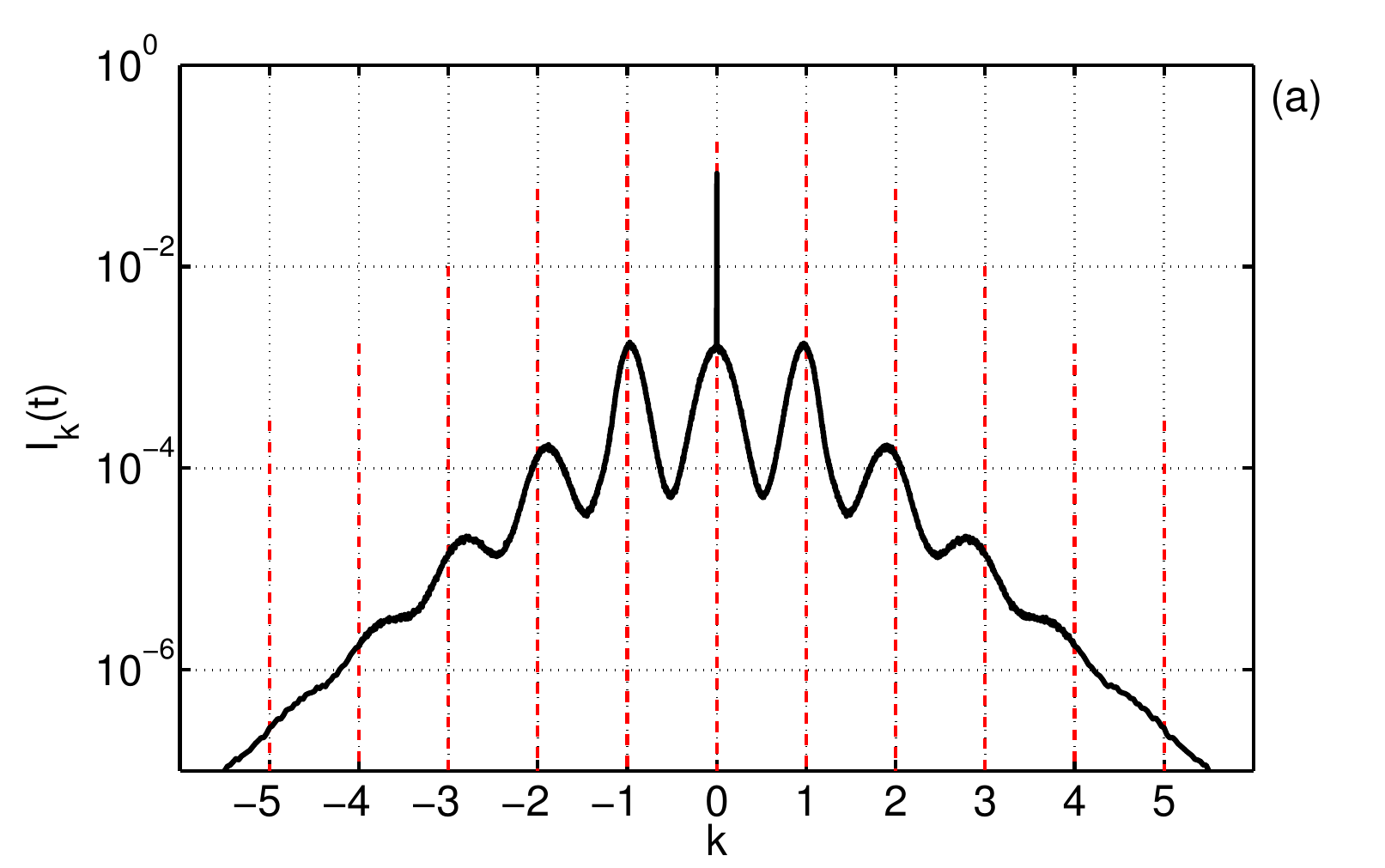}
\includegraphics[width=8.0cm]{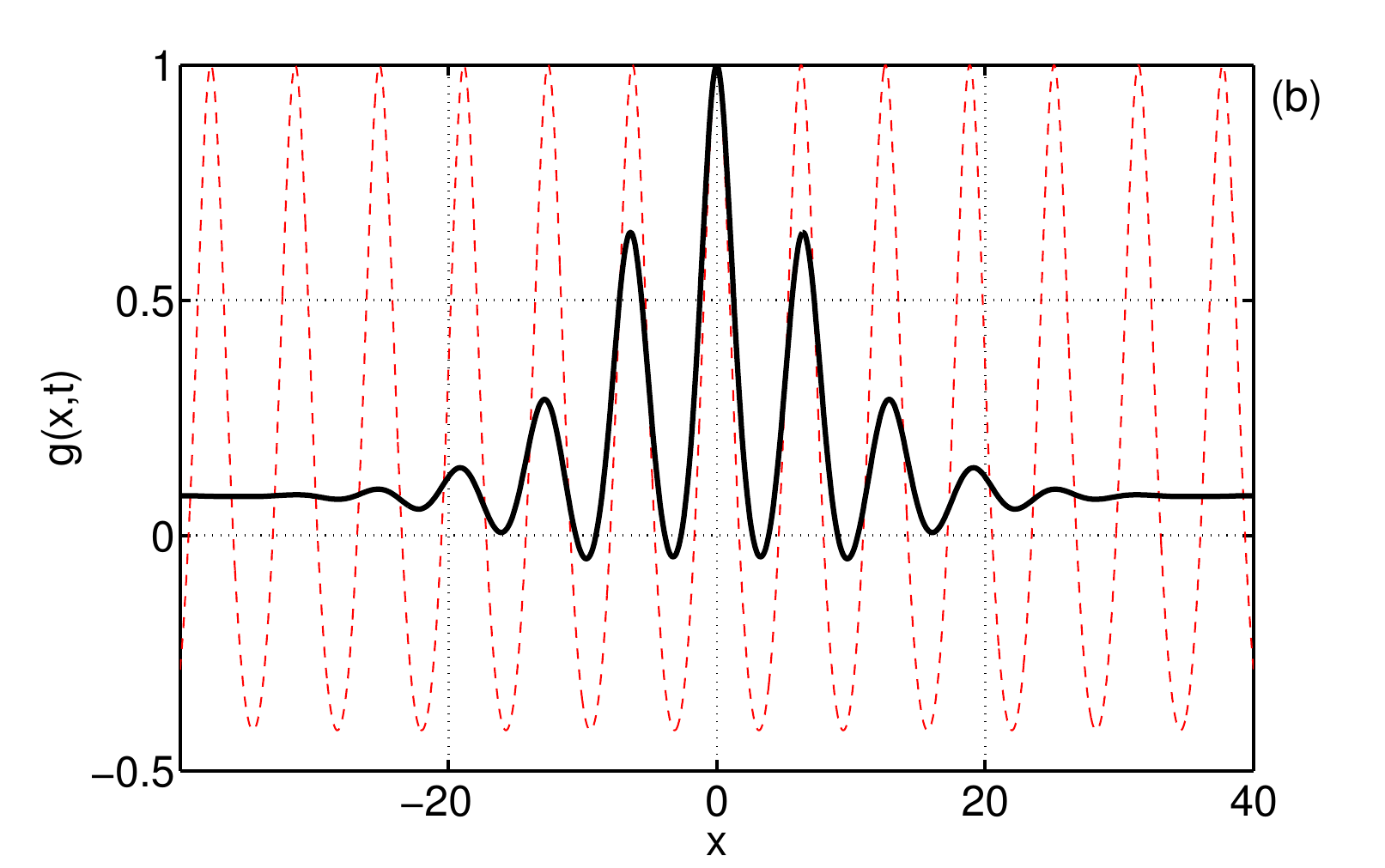}

\caption{\small {\it  (Color on-line) Solid black lines: wave-action spectrum $I_{k}(t)$ (a) and spatial correlation function $g(x,t)$ (b) at the first local maximum $t=13.7$ of the ensemble average potential energy modulus $|\langle H_{4}\rangle|$. Dashed red lines: wave-action spectrum (a) and spatial correlation function (b) of the Akhmediev breather (\ref{Akhmediev}) with $\phi=\pi/4$ at the time of it's maximal elevation.}}
\label{fig:oscillations_sp_corr01}
\end{figure}

\begin{figure}[H] \centering
\includegraphics[width=8.0cm]{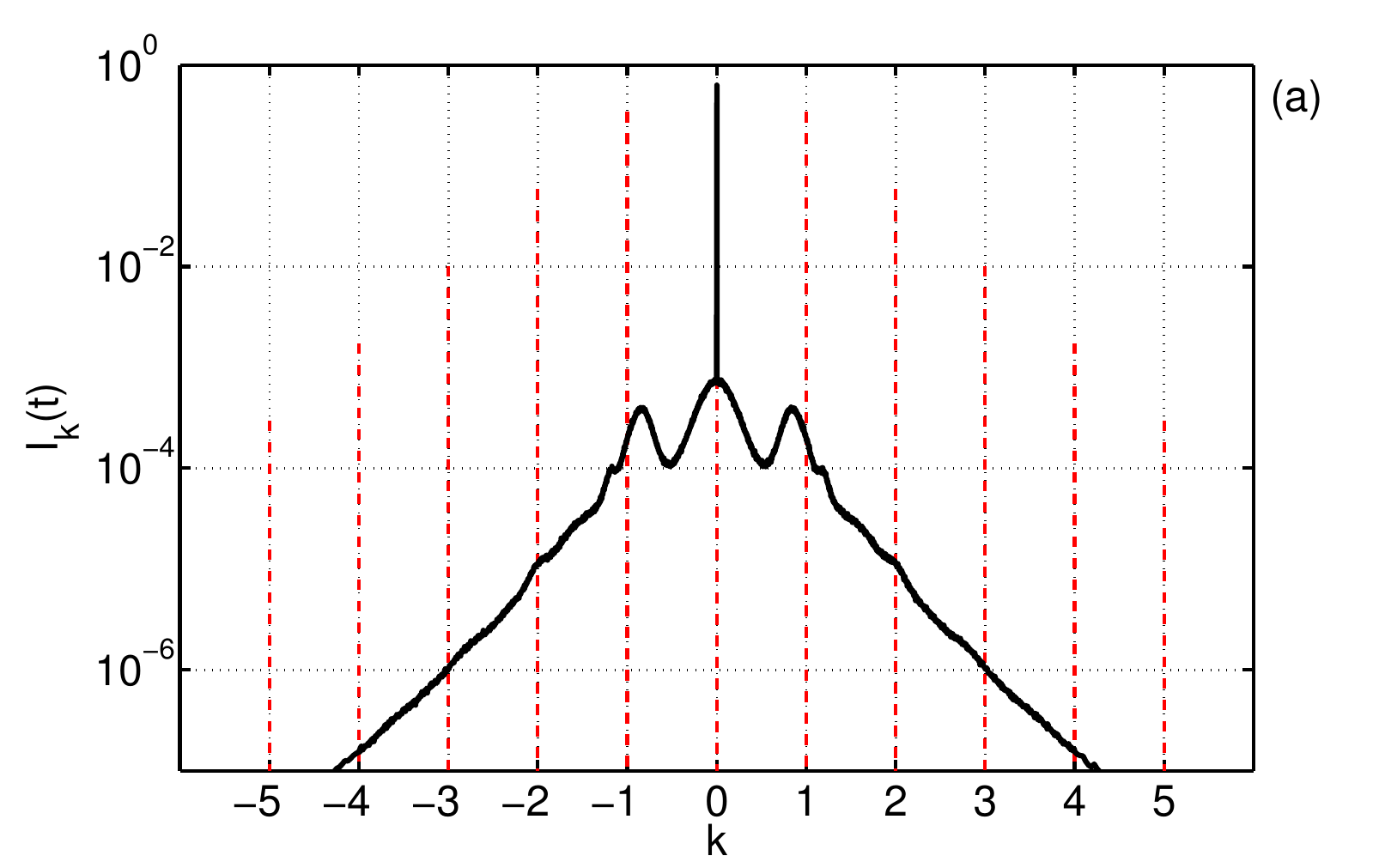}
\includegraphics[width=8.0cm]{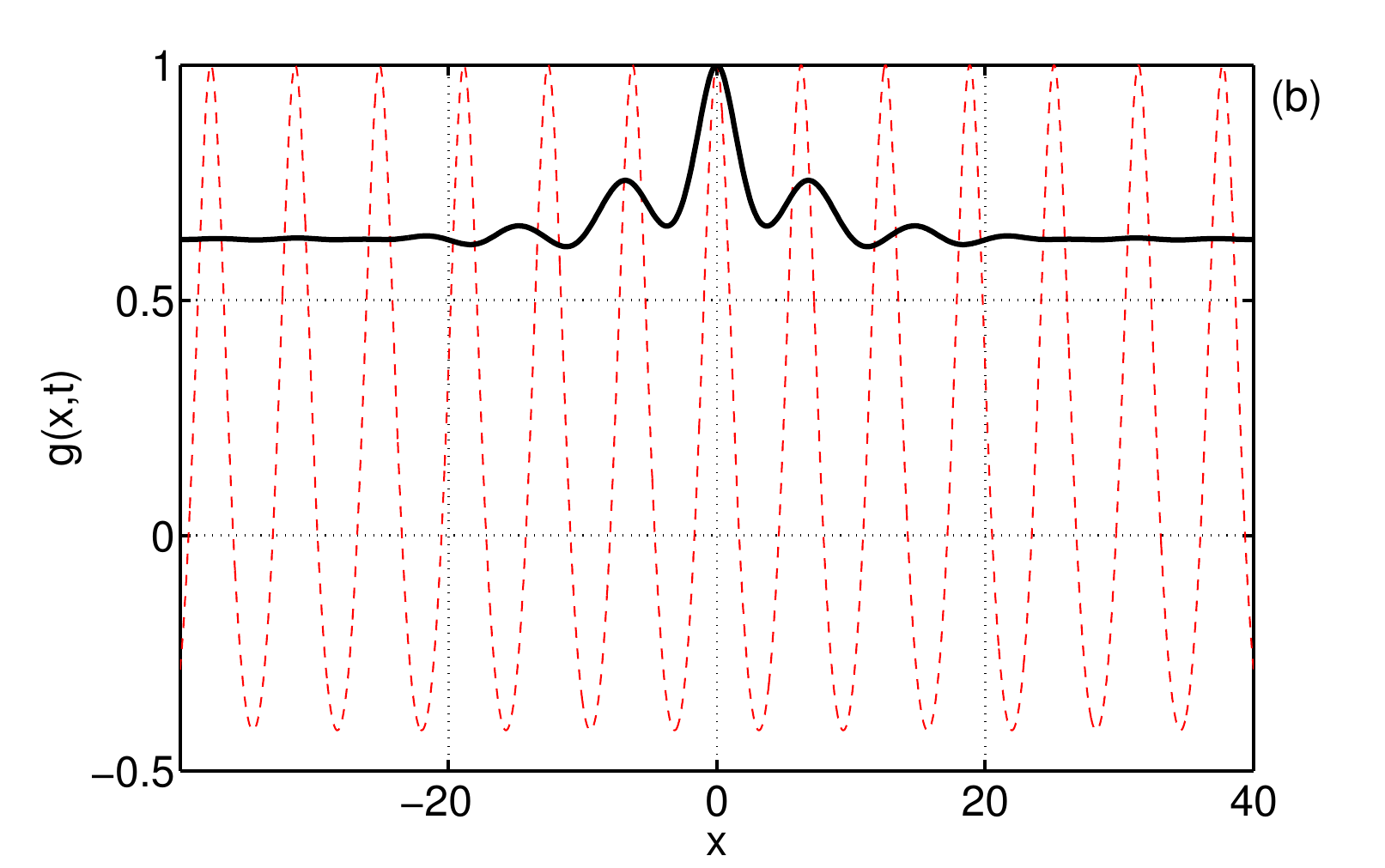}

\caption{\small {\it  (Color on-line) Solid black lines: wave-action spectrum $I_{k}(t)$ (a) and spatial correlation function $g(x,t)$ (b) at the first local minimum $t=15.8$ of the ensemble average potential energy modulus $|\langle H_{4}\rangle|$. Dashed red lines: wave-action spectrum (a) and spatial correlation function (b) of the Akhmediev breather (\ref{Akhmediev}) with $\phi=\pi/4$ at the time of it's maximal elevation.}}
\label{fig:oscillations_sp_corr02}
\end{figure}

\begin{figure}[H] \centering
\includegraphics[width=8.0cm]{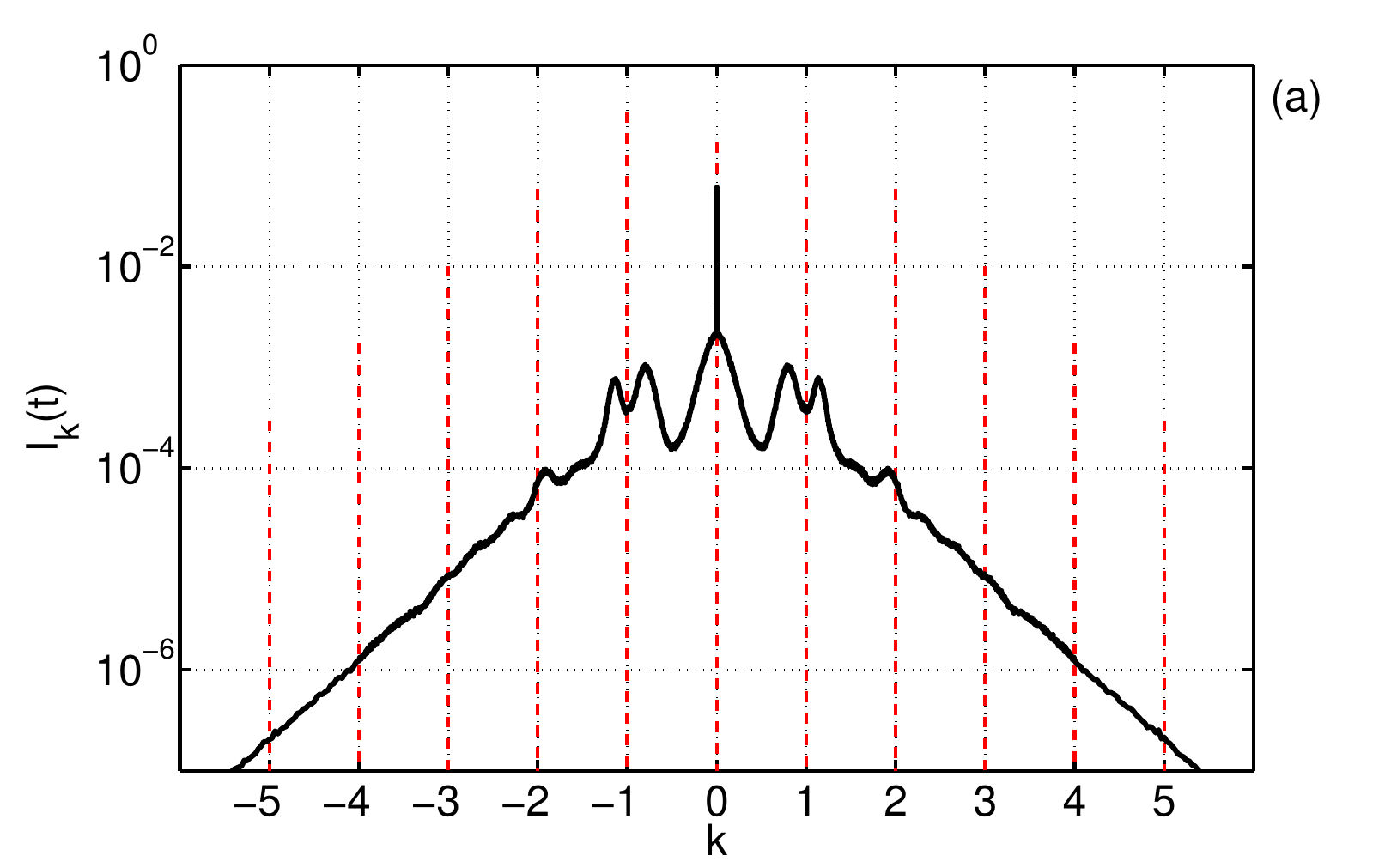}
\includegraphics[width=8.0cm]{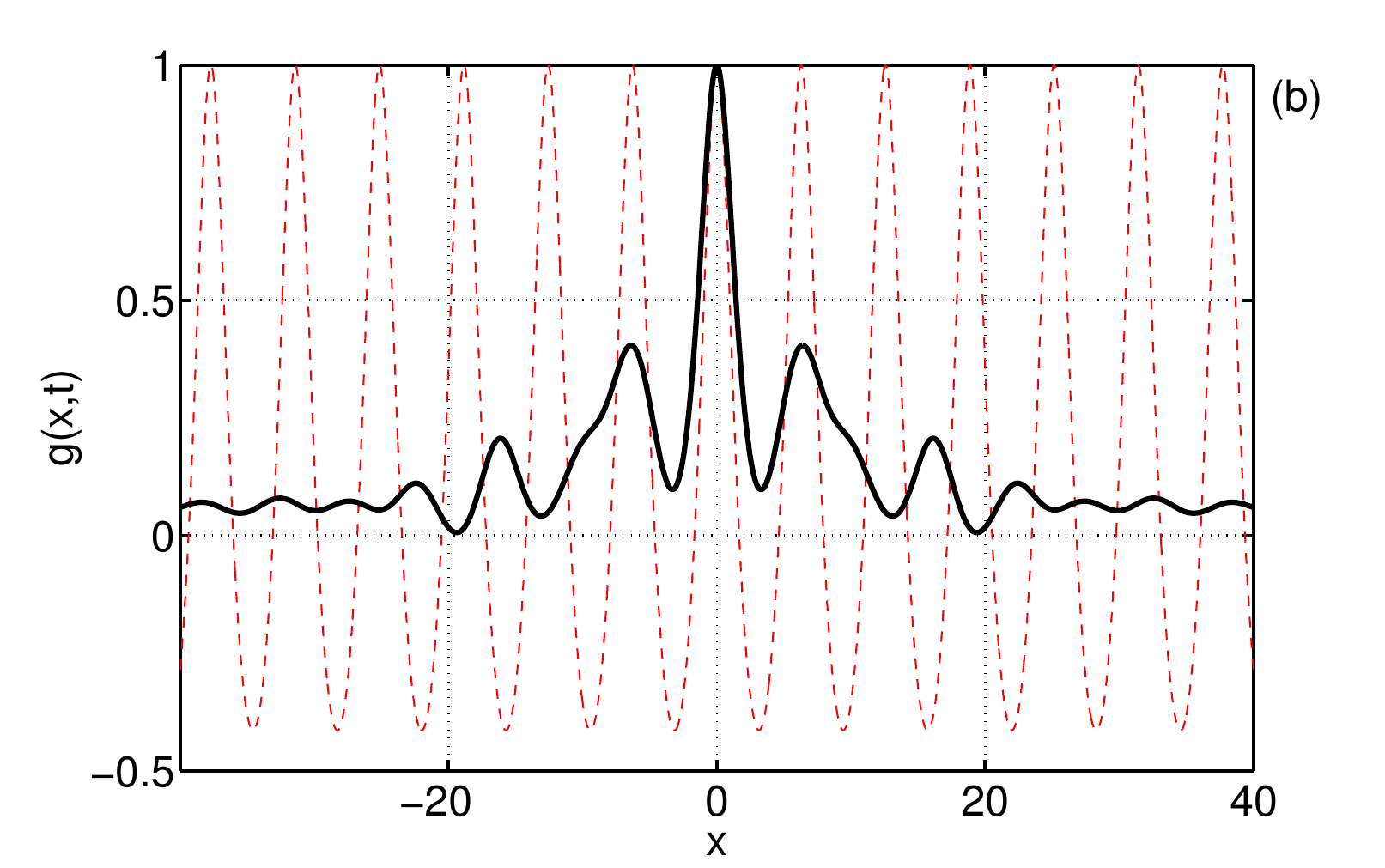}

\caption{\small {\it  (Color on-line) Solid black lines: wave-action spectrum $I_{k}(t)$ (a) and spatial correlation function $g(x,t)$ (b) at the second local maximum $t=17.7$ of the ensemble average potential energy modulus $|\langle H_{4}\rangle|$. Dashed red lines: wave-action spectrum (a) and spatial correlation function (b) of the Akhmediev breather (\ref{Akhmediev}) with $\phi=\pi/4$ at the time of it's maximal elevation.}}
\label{fig:oscillations_sp_corr03}
\end{figure}

\begin{figure}[H] \centering
\includegraphics[width=8.0cm]{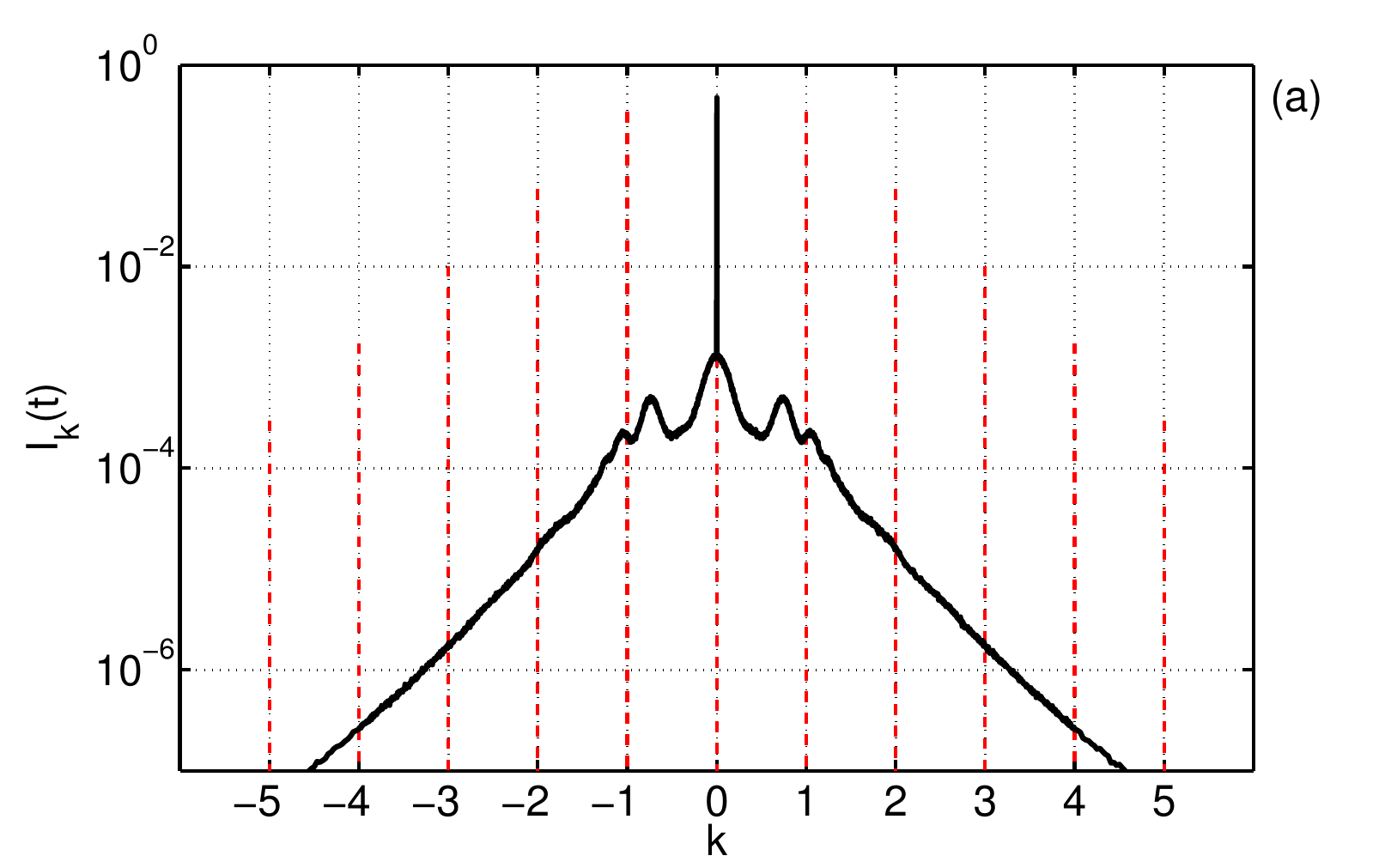}
\includegraphics[width=8.0cm]{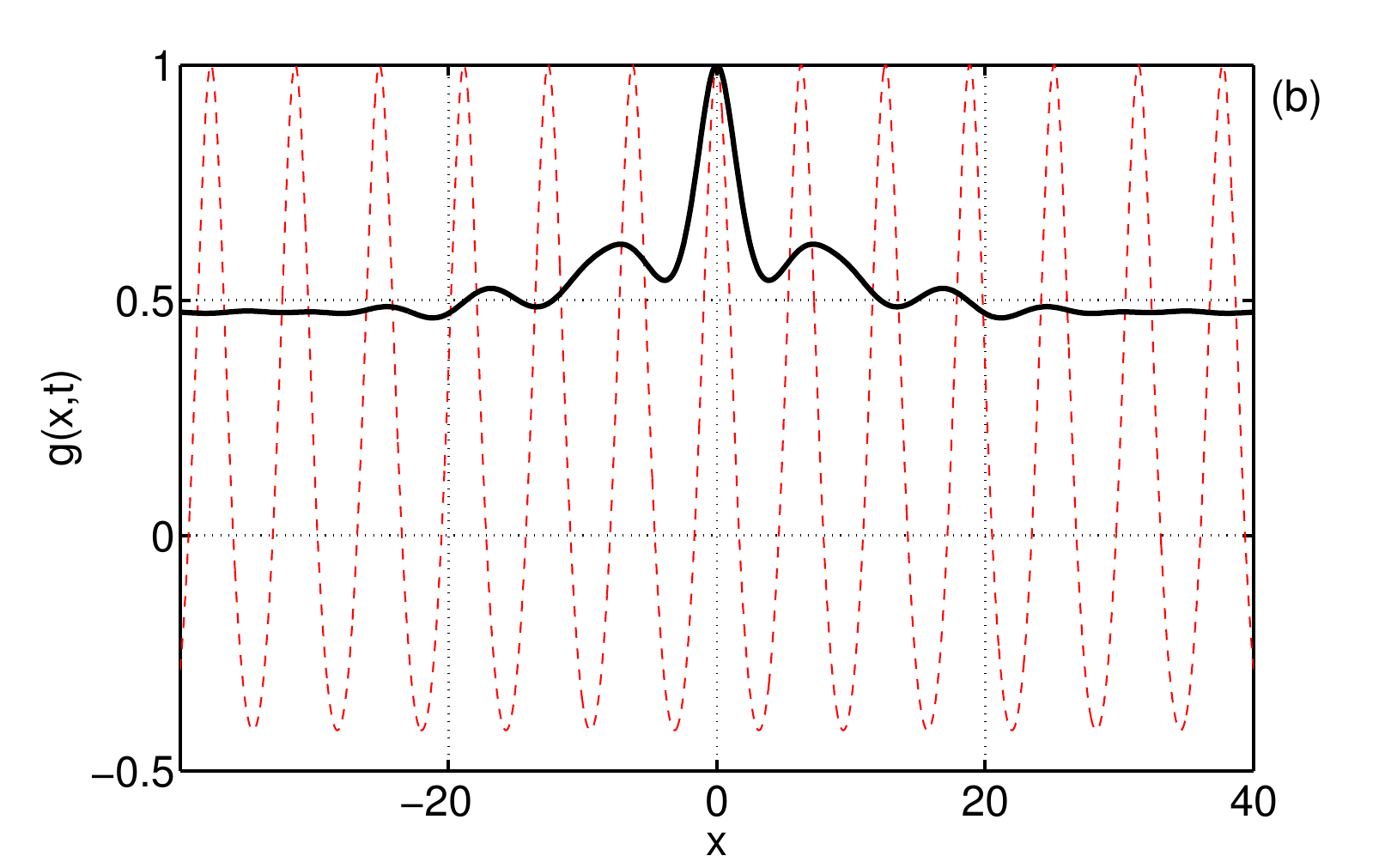}

\caption{\small {\it  (Color on-line) Solid black lines: wave-action spectrum $I_{k}(t)$ (a) and spatial correlation function $g(x,t)$ (b) at the second local minimum $t=19.6$ of the ensemble average potential energy modulus $|\langle H_{4}\rangle|$. Dashed red lines: wave-action spectrum (a) and spatial correlation function (b) of the Akhmediev breather (\ref{Akhmediev}) with $\phi=\pi/4$ at the time of it's maximal elevation.}}
\label{fig:oscillations_sp_corr04}
\end{figure}


\section{Graphs of the probability of rogue waves occurrence.}

In this Appendix we provide more detailed version of FIG.~\ref{fig:oscillations_WY} for the evolution of the probability $W(|\Psi|^{2},t)$ of rogue waves occurrence for waves exceeding $|\Psi|^{2}>8$, $|\Psi|^{2}>10$ and $|\Psi|^{2}>12$ in squared amplitude. Also, we calculate the cumulative probability to meet these waves to time $t$ as
\begin{equation}\label{WYcumul}
R(Y,t) = \int_{0}^{t}W(Y,t)\,dt.
\end{equation}
We calculated these results using computational box $L=256\pi$ and ensemble of about $6\times 10^{4}$ realizations of initial data. As pointed out in the Numerical methods, the results of such simulations coincide with the base experiment with $L=1024\pi$ up to $t\sim 300$. The usage of the smaller computational box allowed us to gather significantly larger statistics, that in turn significantly increased the resolution of the PDF for $|\Psi|^{2}\ge 8$.

\begin{figure}[H] \centering
\includegraphics[width=8.0cm]{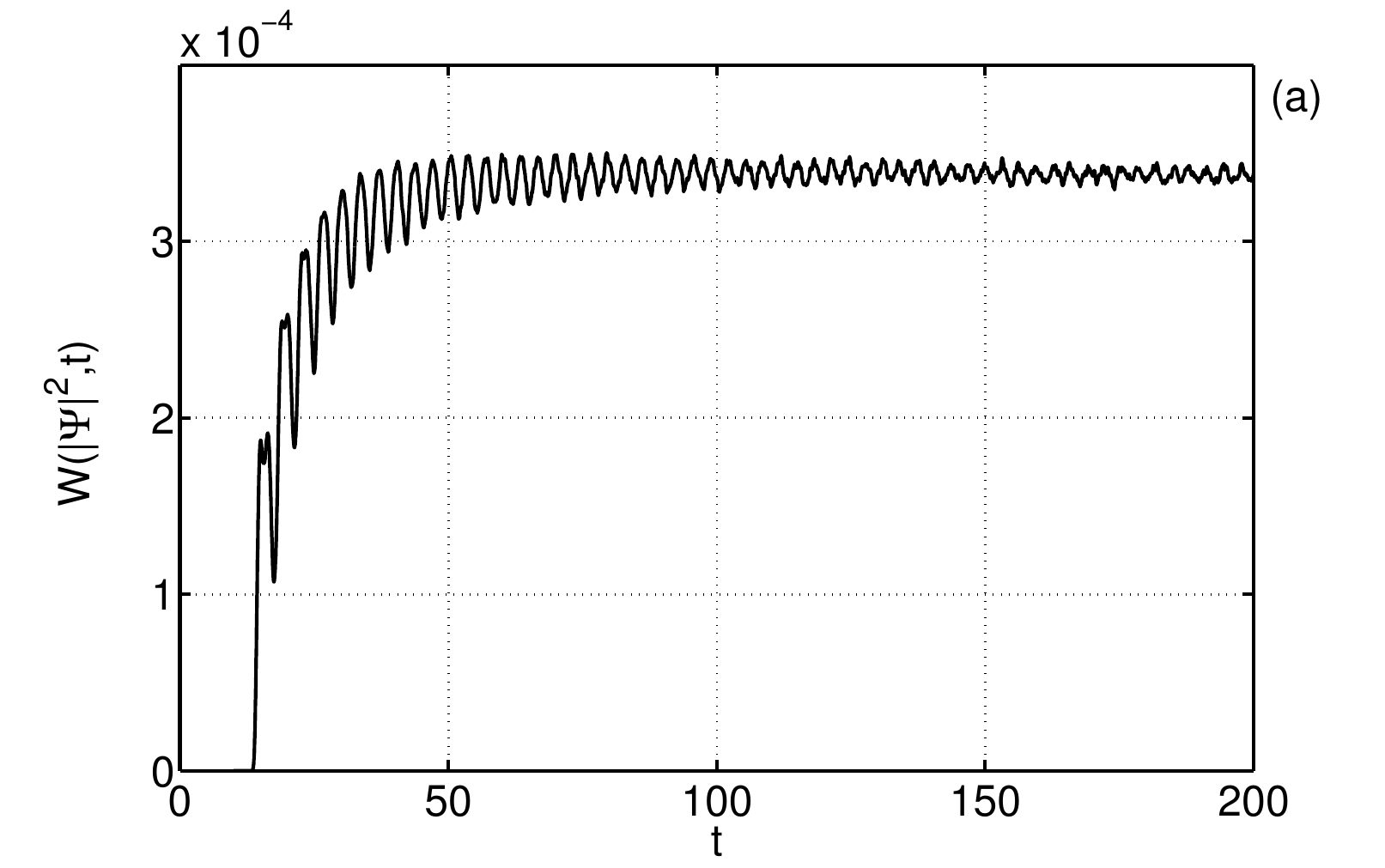}
\includegraphics[width=8.0cm]{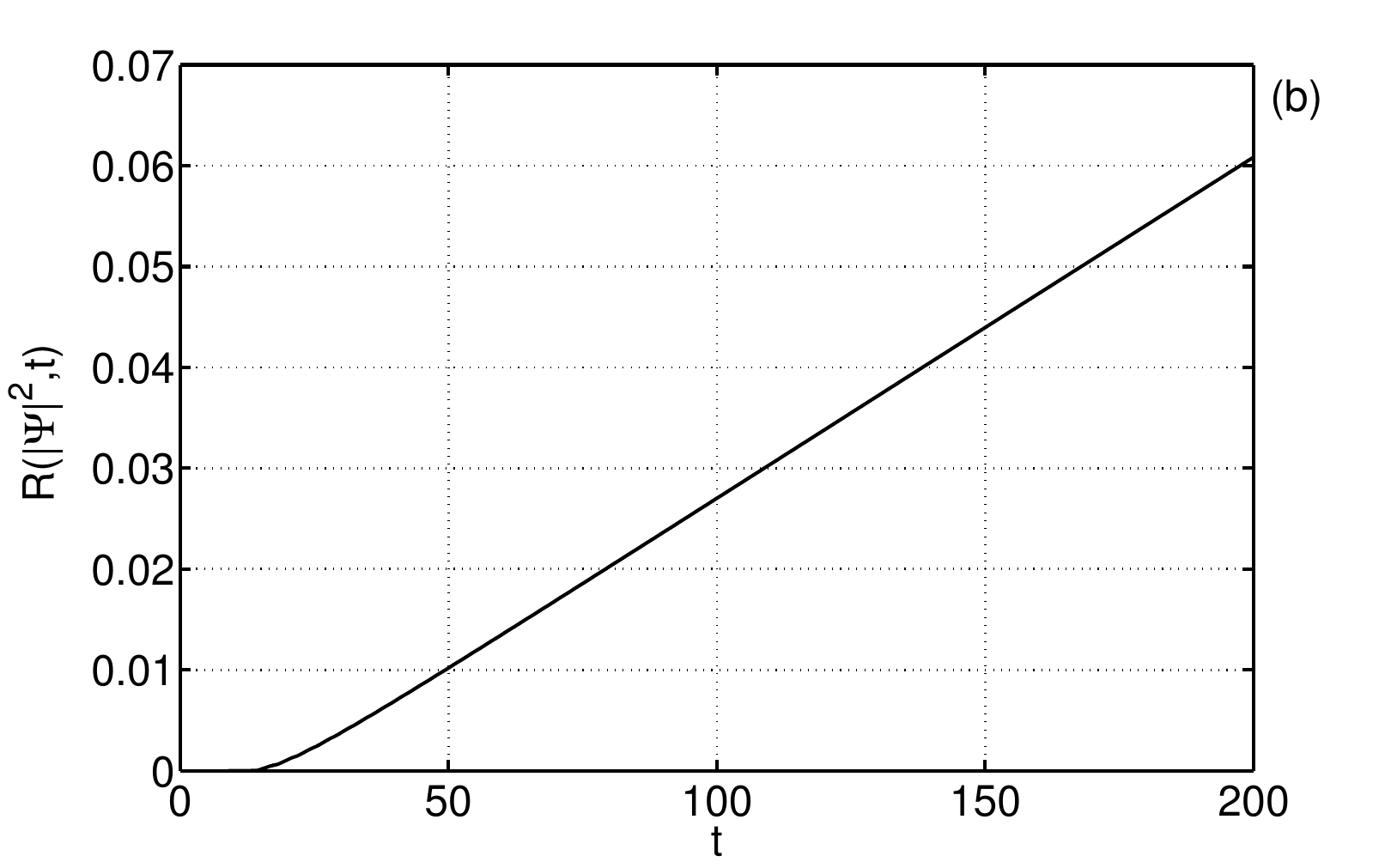}

\caption{\small {\it Time dependence of the probability of occurrence $W(|\Psi|^{2},t)$ (\ref{WY}) (a) and cumulative probability of occurrence $R(|\Psi|^{2},t)$ (\ref{WYcumul}) (b) for waves with amplitudes $|\Psi|^{2}>8$.}}
\label{fig:oscillations_WY_lin01}
\end{figure}

\begin{figure}[H] \centering
\includegraphics[width=8.0cm]{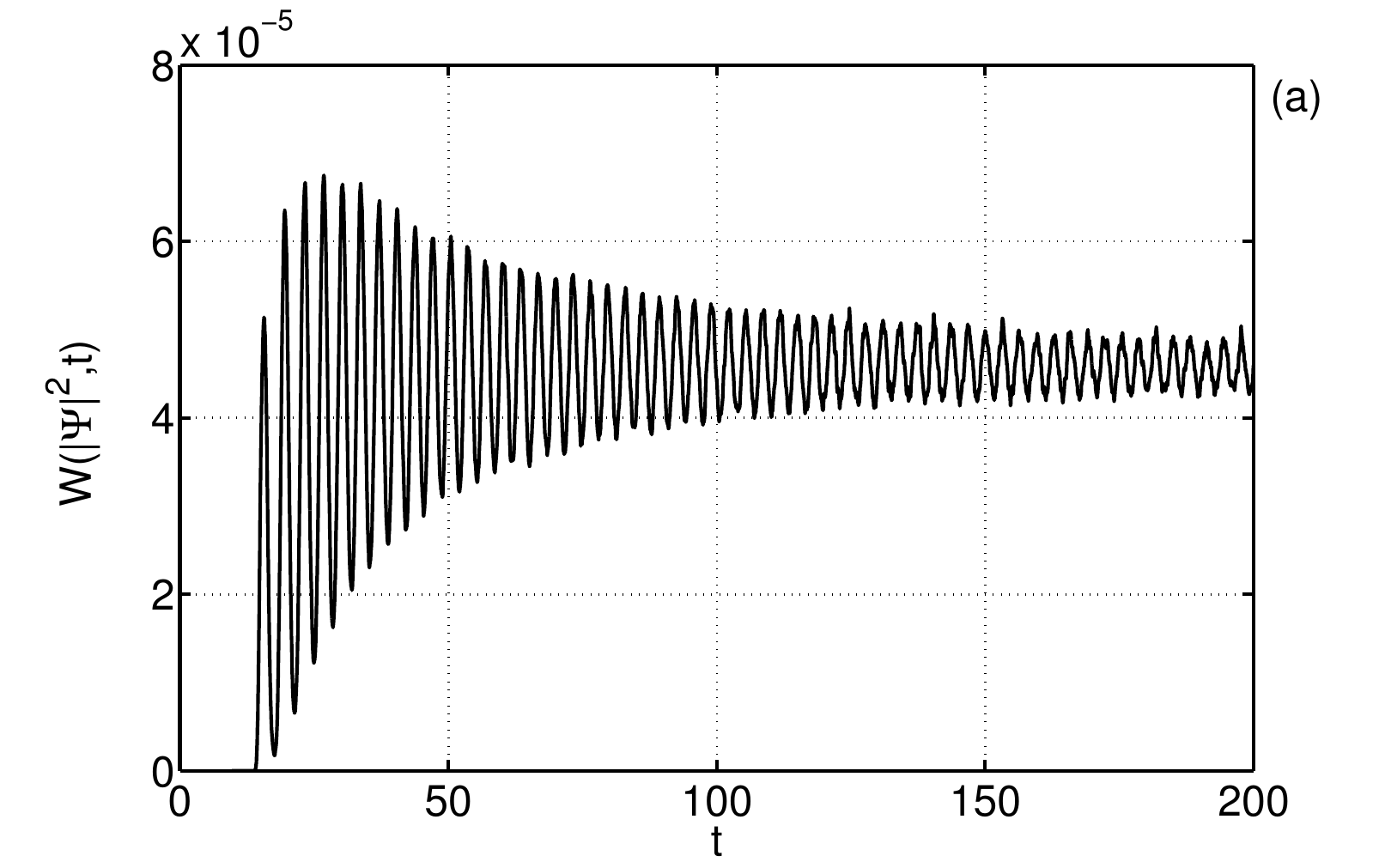}
\includegraphics[width=8.0cm]{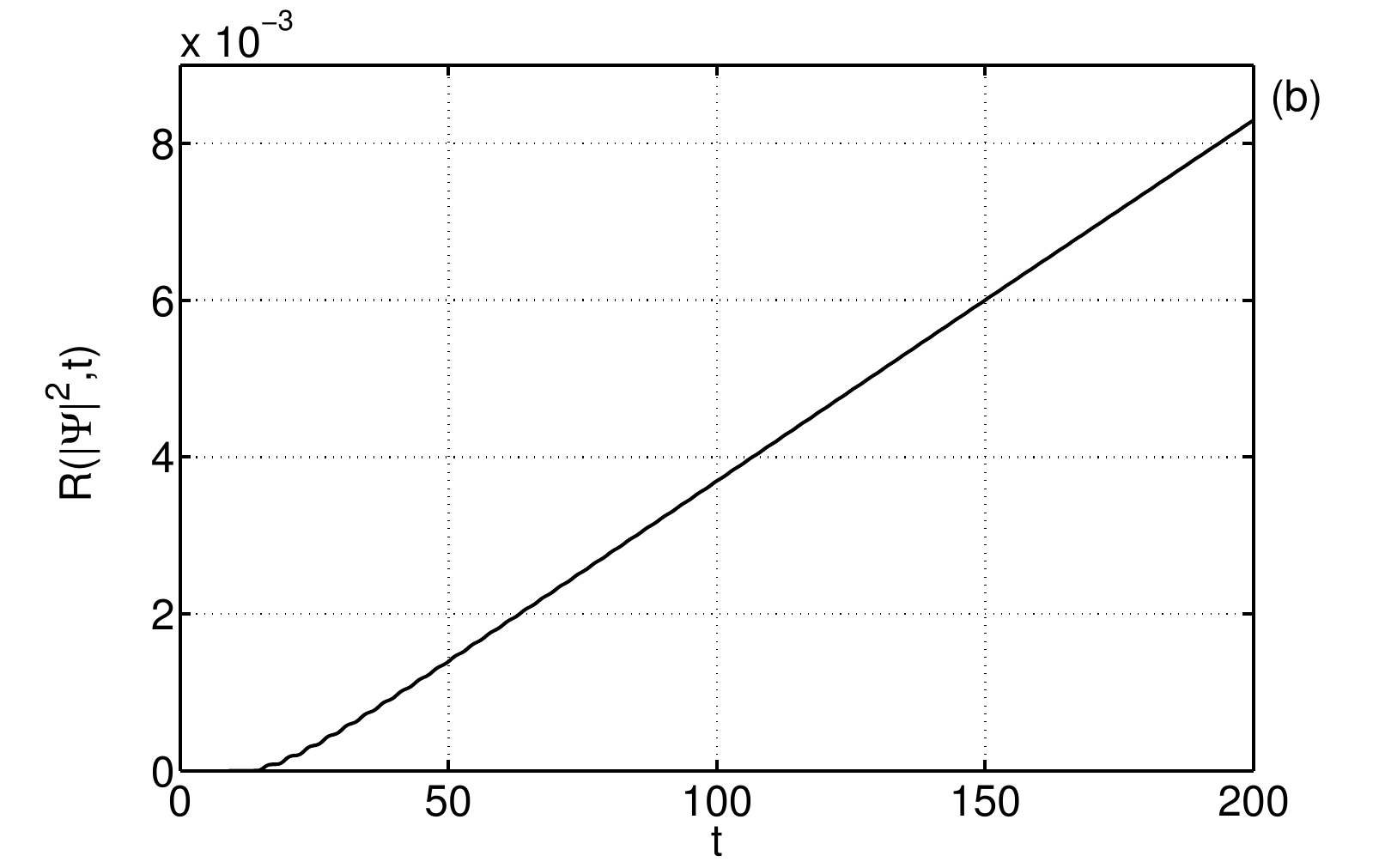}

\caption{\small {\it Time dependence of the probability of occurrence $W(|\Psi|^{2},t)$ (\ref{WY}) (a) and cumulative probability of occurrence $R(|\Psi|^{2},t)$ (\ref{WYcumul}) (b) for waves with amplitudes $|\Psi|^{2}>10$.}}
\label{fig:oscillations_WY_lin03}
\end{figure}

\begin{figure}[H] \centering
\includegraphics[width=8.0cm]{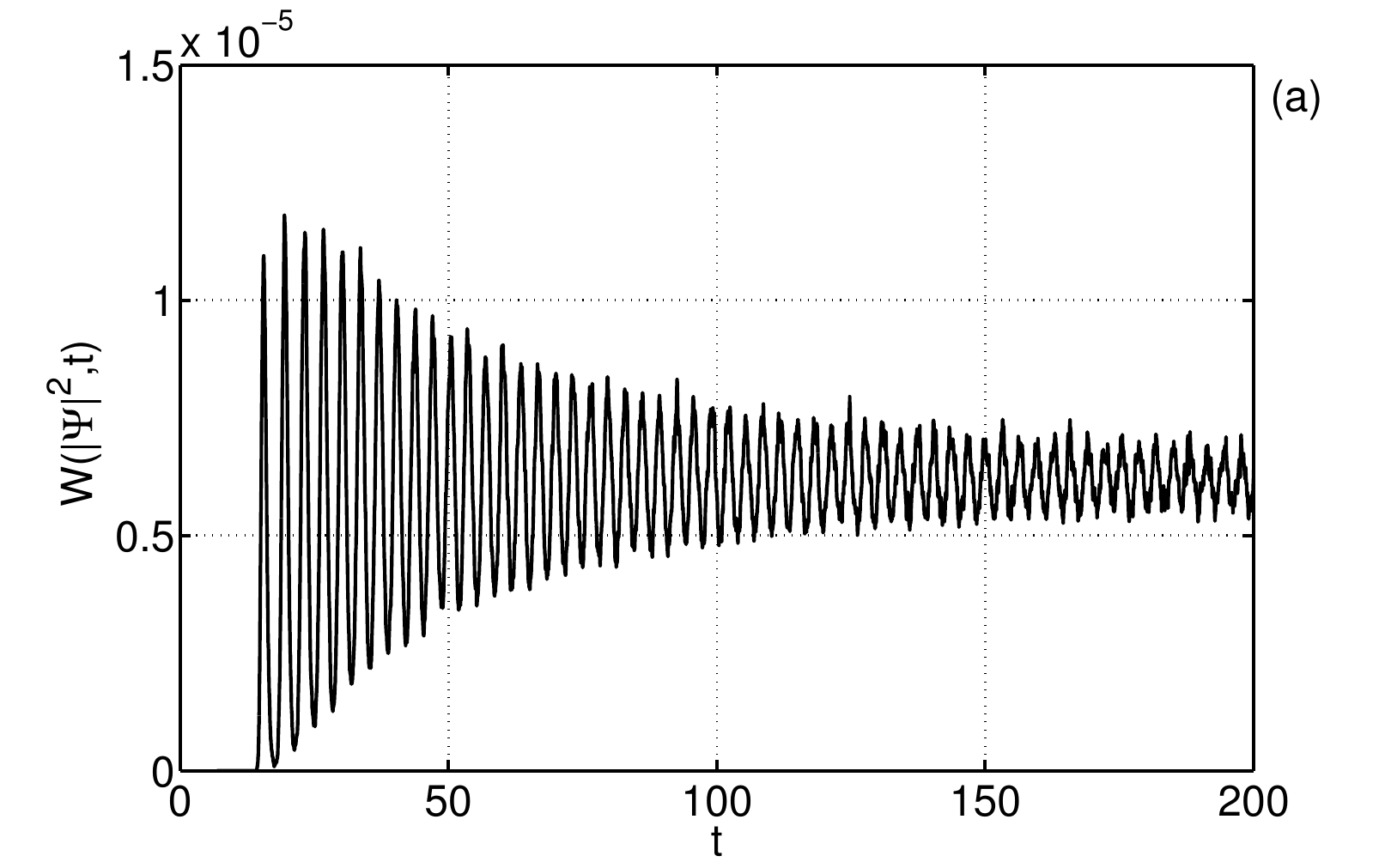}
\includegraphics[width=8.0cm]{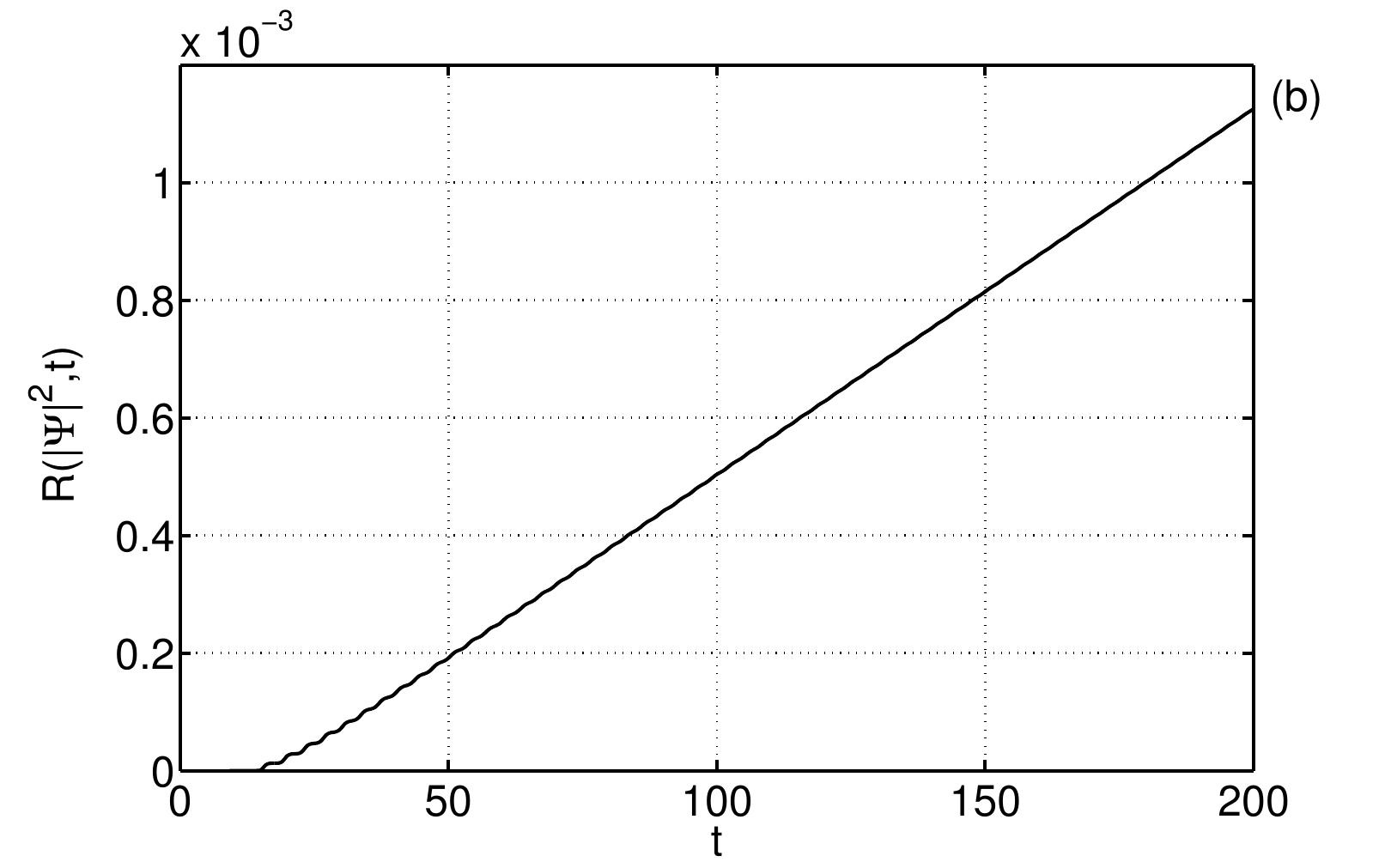}

\caption{\small {\it Time dependence of the probability of occurrence $W(|\Psi|^{2},t)$ (\ref{WY}) (a) and cumulative probability of occurrence $R(|\Psi|^{2},t)$ (\ref{WYcumul}) (b) for waves with amplitudes $|\Psi|^{2}>12$.}}
\label{fig:oscillations_WY_lin05}
\end{figure}


\end{document}